\documentclass[12pt]{article}
\pdfoutput=1
\usepackage{url}
\usepackage{geometry}
\usepackage{setspace}
\usepackage{rotating}
\usepackage{longtable}
\usepackage{graphicx}
\usepackage{ccaption}
\usepackage{booktabs}
\usepackage{bm}
\usepackage{tabularx}
\usepackage{makecell}
\usepackage{placeins}
\usepackage{multirow}
\usepackage{subcaption}
\usepackage{gensymb}
\usepackage{xcolor}
\usepackage{amsmath}

\title{An Early Warning Approach to Monitor COVID-19 Activity with Multiple Digital Traces in Near Real-Time}
\begin{document}

\author{Nicole E. Kogan$^{*,1,2}$ \and Leonardo Clemente$^{*,1}$ \and Parker Liautaud$^{*,3}$ \and Justin Kaashoek$^{1,4}$ \and Nicholas B. Link$^{1,2}$ \and Andre T. Nguyen$^{1,5,6}$ \and Fred S. Lu$^{1,7}$ \and Peter Huybers$^{3}$ \and Bernd Resch$^{8,9}$ \and Clemens Havas$^8$ \and Andreas Petutschnig$^8$ \and Jessica Davis$^{10}$ \and Matteo Chinazzi$^{10}$ \and Backtosch Mustafa$^{1, 12}$ \and William P. Hanage$^2$ \and Alessandro Vespignani$^{10}$ \and Mauricio Santillana$^{1,2,11,\dagger}$}

\date{%
    \today
}

\maketitle

\begin{center}

          {\small $^1$Computational Health Informatics Program, Boston Children's Hospital, Boston, MA \\%
     $^2$Department of Epidemiology, Harvard T.H. Chan School of Public Health, Boston, MA\\
     $^3$Department of Earth and Planetary Sciences, Harvard University, Cambridge, MA\\%
     $^4$School of Engineering and Applied Sciences, Harvard University, Cambridge, MA\\
     $^5$University of Maryland, Baltimore County, Baltimore, MD \\%
     $^6$Booz Allen Hamilton, Columbia, MD\\%
     $^7$Department of Statistics, Stanford University, Stanford, CA\\%
     $^8$Department of Geoinformatics - Z\_GIS, University of Salzburg, Salzburg, Austria\\%
     $^9$Center for Geographic Analysis, Harvard University, Cambridge, MA\\%
     $^{10}$Northeastern University, Boston, MA\\%
     $^{11}$Department of Pediatrics, Harvard Medical School, Boston, MA\\%
     $^{12}$University Medical Center Hamburg-Eppendorf, Hamburg, Germany\\%
     .\\
     $^*$These authors contributed equally to this study.\\%
    
     $^{\dagger}$Correspondance to: Mauricio Santillana (\texttt{msantill@fas.harvard.edu}) \\[2ex]}%
\end{center}

\abstract{

Non-pharmaceutical interventions (NPIs) have been crucial in curbing COVID-19 in the United States (US). Consequently, relaxing NPIs through a phased re-opening of the US amid still-high levels of COVID-19 susceptibility could lead to new epidemic waves. This calls for a COVID-19 early warning system. Here we evaluate multiple digital data streams as early warning indicators of increasing or decreasing state-level US COVID-19 activity between January and June 2020. We estimate the timing of sharp changes in each data stream using a simple Bayesian model that calculates in near real-time the probability of exponential growth or decay. Analysis of COVID-19-related activity on social network microblogs, Internet searches, point-of-care medical software, and a metapopulation mechanistic model, as well as fever anomalies captured by smart thermometer networks, shows exponential growth roughly 2-3 weeks prior to comparable growth in confirmed COVID-19 cases and 3-4 weeks prior to comparable growth in COVID-19 deaths across the US over the last 6 months. We further observe exponential decay in confirmed cases and deaths 5-6 weeks after implementation of NPIs, as measured by anonymized and aggregated human mobility data from mobile phones. Finally, we propose a combined indicator for exponential growth in multiple data streams that may aid in developing an early warning system for future COVID-19 outbreaks. These efforts represent an initial exploratory framework, and both continued study of the predictive power of digital indicators as well as further development of the statistical approach are needed. }


\section{Introduction}
In just over seven months, COVID-19 -- the disease caused by the betacoronavirus SARS-CoV-2 -- has caused over 503,000 deaths worldwide, 125,000 of which are in the US \cite{dong2020jhkdashboard}. In the absence of a vaccine or an effective treatment, authorities have employed non-pharmaceutical interventions (NPIs) to slow epidemic growth, including school and business closures, work-from-home policies, and travel bans. Recently, many US states have begun progressively reopening their economies, despite estimates of cumulative US COVID-19 incidence suggesting that fewer than 10-15\% of the US population has been exposed to SARS-CoV-2 ~\cite{lu2020estimating}. Serological studies also indicate low levels of seroprevalence even in  parts of the US heavily affected by the virus (e.g., 23\% in New York City by May 29, 2020) \cite{rosenberg2020cumulative, okell2020have}. The long development timeline for a vaccine \cite{Corey948} coupled with the possibility that immunity to SARS-CoV-2 may decline over time (as is the case with other coronaviruses) portends the emergence of new epidemic waves \cite{kissler2020projecting}. The availability of a reliable, robust, real-time indicator of emerging COVID-19 outbreaks would aid immensely in appropriately timing a response. \vspace{12pt} 

Despite efforts by the research community to aggregate and make available data streams that are representative of COVID-19 activity, it is not immediately clear which of these data streams is the most dependable for tracking outbreaks. Most metrics for tracking COVID-19, such as confirmed cases, hospitalizations, and deaths, suffer from reporting delays, as well as uncertainties stemming from inefficiencies in the data collection, collation, and dissemination processes \cite{lipsitch2019enhancing}. For example, confirmed cases may be more reflective of testing availability than of disease incidence and, moreover, lag infections by days or weeks \cite{Kaashoek_2020, lu2020estimating}. Previous work has suggested that clinician-provided reports of influenza-like illness (ILI) aggregated by the Centers for Disease Control and Prevention (CDC) may be less sensitive to testing availability than confirmed cases, but these reports suffer from reporting lags of 5-12 days, depend on the thoroughness of clinician reporting, and do not distinguish COVID-19 from other illnesses that may cause similar symptoms.\vspace{12pt} 

Alternatively, forecasting models can assist in long-term planning, but the accuracy of their predictions are limited by the timeliness of data or parameter updates. Specifically, some models demonstrate predictive skill with respect to hospitalizations and deaths \cite{flaxman2020report, kissler2020projecting}, but these predictions are often too late to enable timely NPI implementation. Other models suffer from limited generalizability, with NPI implementation proposed only for a specific city \cite{Austin_Hosp}. The CDC has launched a state-level forecasting initiative aimed at consolidating predictions from multiple models to estimate future COVID-19-attributable deaths, but the use of these predictions in state-level decision-making is still pending \cite{CDC_models}.\vspace{12pt} 

Over the last decade, new methodologies have emerged to track population-level disease spread using data sources not originally conceived for that purpose \cite{salathe2012digital}. These approaches have exploited information from search engines \cite{ginsberg2009detecting, polgreen2008using, santillana2014can, yang2015accurate, mcgough2017forecasting, lu2019improved, santillana2014up2date}, news reports \cite{brownstein2008surveillance, majumder20152014, majumder2016utilizing}, crowd-sourced participatory disease surveillance systems \cite{smolinski2015flu, paolotti2014web}, Twitter microblogs \cite{paul2014twitter, nagar2014case}, electronic health records \cite{viboud2014demonstrating, santillana2016cloud}, Wikipedia traffic \cite{generous2014global}, wearable devices \cite{radin2020harnessing}, smartphone-connected thermometers \cite{miller2018smartphone}, and travel websites \cite{Kogan2019} to estimate disease prevalence in near real-time. Several have already been used to track COVID-19 \cite{lampos2020tracking, lu2020internet}. These data sources are liable to bias, however; for example, Google Search activity is highly sensitive to the intensity of news coverage \cite{santillana2014can, lazer2014parable, bento2020evidence}. Methodologies to mitigate biases in digital data sources commonly involve combining disease history, mechanistic models, and surveys to produce ensemble estimates of disease activity \cite{santillana2015combining, reich2019collaborative}.\vspace{12pt} 

\textbf{Our Contribution}: We propose that several digital data sources may provide earlier indication of epidemic spread than traditional COVID-19 metrics such as confirmed cases or deaths. Six such sources are examined here: (1) Google Trends patterns for a suite of COVID-19-related terms, (2) COVID-19-related Twitter activity, (3) COVID-19-related clinician searches from UpToDate, (4) predictions by GLEAM, a state-of-the-art metapopulation mechanistic model, (5) anonymized and aggregated human mobility data from smartphones, and (6) Kinsa Smart Thermometer measurements. We first evaluate each of these ``proxies'' of COVID-19 activity for their lead or lag relative to traditional measures of COVID-19 activity: confirmed cases, deaths attributed, and ILI. We then propose the use of a metric combining these data sources into a multi-proxy estimate of the probability of an impending COVID-19 outbreak. Finally, we develop probabilistic estimates of when such a COVID-19 outbreak will occur conditional on proxy behaviors. Consistent behavior among proxies increases the confidence that they capture a real change in the trajectory of COVID-19.

\section{Results}

\begin{figure}
    \centering
    \vspace{-0.7in}
    \includegraphics[width=.9\textwidth]{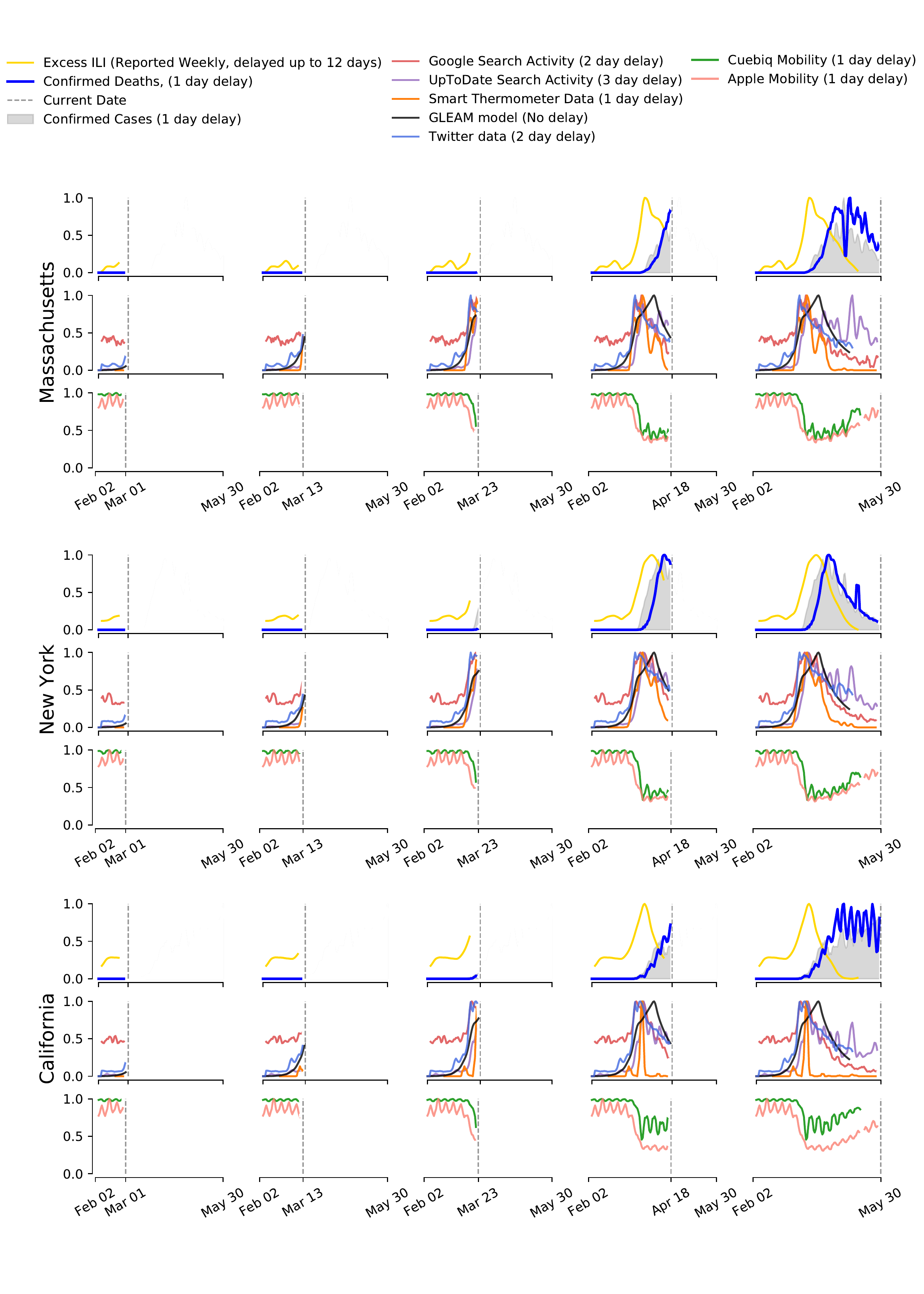}
    \caption{Visualization of the evolution of each COVID-19 proxy in Massachusetts, New York, and California. Columns depict progressively increasing time periods over which proxies become available (vertical dashed line indicates the latest date in the time period) in order to illustrate how the availability of different proxies informs upon the evolution of COVID-19. Time series were normalized between 0 and 1 and smoothed using a simple moving average for purposes of visualization. The legend at top shows each data stream alongside typical delays between measurement and reporting.}
    \label{fig:proxies_visualization}
\end{figure}

\noindent \textbf{Visualizing the behavior of COVID-19-tracking data sources: motivation for designing an early-warning system.} Figure \ref{fig:proxies_visualization} displays the temporal evolution of all available signals considered in this study for three US states - Massachusetts (MA), New York (NY), and California (CA) - over five lengthening time intervals. These states illustrate different epidemic trajectories within the US, with NY among the worst affected states to date and CA experiencing a more gradual increase in cases than both MA and NY. \vspace{12pt} 

The top row of Figure \ref{fig:proxies_visualization} for each state displays normalized COVID-19 activity as captured by daily reported confirmed cases, deaths, and ``Excess ILI'' (hospitalizations were discounted due to data sparseness). Excess ILI refers to hospital visits due to influenza-like illness in excess of what is expected from a normal flu season \cite{lu2020estimating}, which we attribute to COVID-19 in 2020. ILI data were taken from the CDC's US Outpatient Influenza-like Illness Surveillance Network (ILINet). The middle row for each state displays time series for five proxies of COVID-19 activity. The bottom row for each state displays state-level anonymized and aggregated human mobility data as collected by mobile phones; mobility data is viewed as a proxy for adherence to social distancing recommendations. Similar visualizations for all states are shown in Figures S1 to S17 in the Supplementary Materials.\vspace{12pt} 

Figure \ref{fig:proxies_visualization} demonstrates that for MA, NY, and CA, COVID-19-related clinicians' and general population's Internet activity, smart thermometers, and GLEAM model predictions exhibit early increases that lead increases of confirmed cases and deaths due to COVID-19. Analogously, decreases in other proxies - especially in mobility - mirror later decreases in COVID-19-attributable confirmed cases and deaths for the three states represented. This is not universally observable, however, as some states such as North Carolina, Arizona, Florida, and Texas have not seen decreases in COVID-19 activity. \vspace{12pt} 

\noindent \textbf{Quantifying the timing of growth in proxies of COVID-19 activity.} To quantify the relative leads and lags in our collection of disease proxies, we formulated a change of behavior ``event'' for each proxy and compared it to three ``gold standards'' of COVID-19 activity: confirmed cases, deaths attributed, and Excess ILI. In keeping with classical disease dynamics, we defined an event as any initiation of exponential growth (``uptrend''). Using a Bayesian approach, we obtained a joint posterior probability distribution for parameter values in a function $y(t) \sim \beta \; e^{\gamma (t-t_{0})} + \epsilon(t)$ over a time window of 14 days, evaluating $\beta$, $\gamma$, and $\sigma^{2}$ (the variance of $\epsilon$). A $p$-value was then calculated per proxy per day, representing the posterior probability that $\gamma$ is greater than zero. As the $p$-values decrease, we grow more confident that a given time series is exhibiting sustained growth. When the $p$-value decreases below 0.05, we define this as an individual proxy's ``uptrend'' event. \vspace{12pt}

\begin{figure}
    \centering
    \includegraphics[width=.72\textwidth]{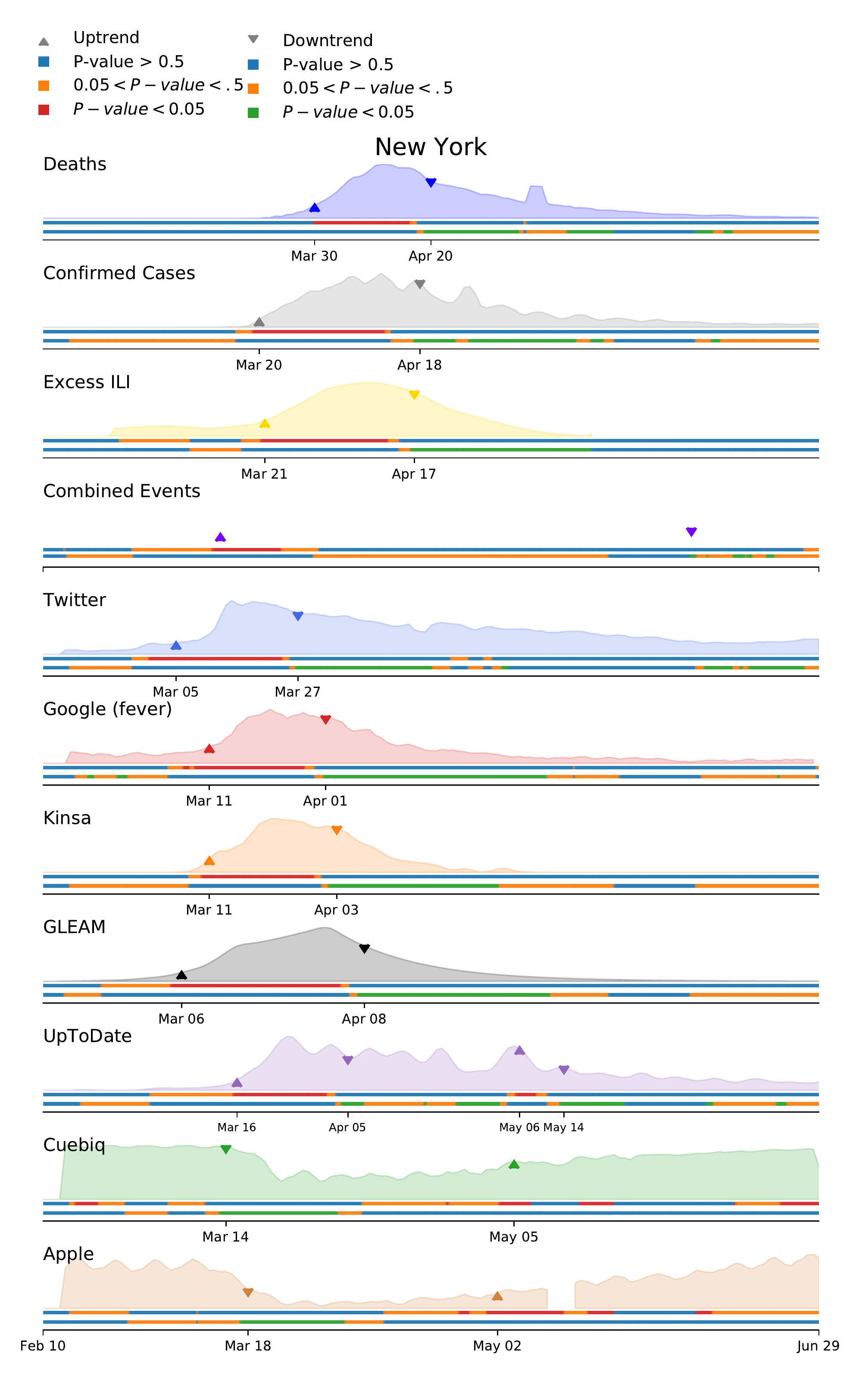}
    \caption{\small {Visualization of the event detection procedure applied to COVID-19 proxies. An event is detected by setting a threshold over the $p$-value of the exponential coefficient $\gamma$ (in this case, $p$-value $< 0.05$). Under each curve, the $p$-values are shown as blue-to-red colorbars for uptrends and blue-to-green colorbars for down-trends; red and green indicate periods with $p$-value $< 0.05$, while orange indicates periods with $p$-value between 0.05 and 0.5. Triangular markers are used to signal the date when an uptrend or a downtrend is detected. The time series are adjusted to account for expected reporting delays in the source of information.}}
    \label{fig:EventDetection}
\end{figure}

The sequences of proxy-specific uptrends for an example state, New York (NY), are depicted in Figure \ref{fig:EventDetection}. Upward-pointing triangles denote the date on which a growth event is identifiable. For the example state, COVID-19-related Twitter posts gave the earliest indication of increasing COVID-19 activity, exhibiting an uptrend around March 2. This was closely followed by uptrends in GLEAM-modeled infections, Google Searches for ``fever'', fever incidence, and COVID-19-related searches by clinicians.\vspace{12pt} 

\begin{figure}[t!]
    \centering
    \begin{subfigure}[t]{0.6\textwidth}
    \centering
        \hspace*{0.8cm}\includegraphics[width=\textwidth]{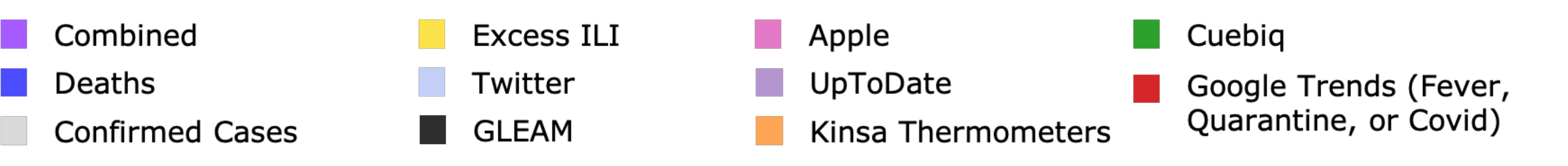}
        \label{fig:legend}
    \end{subfigure}
    \vspace{12pt} 
    \begin{subfigure}[t]{1.0\textwidth}
        \includegraphics[width=\textwidth]{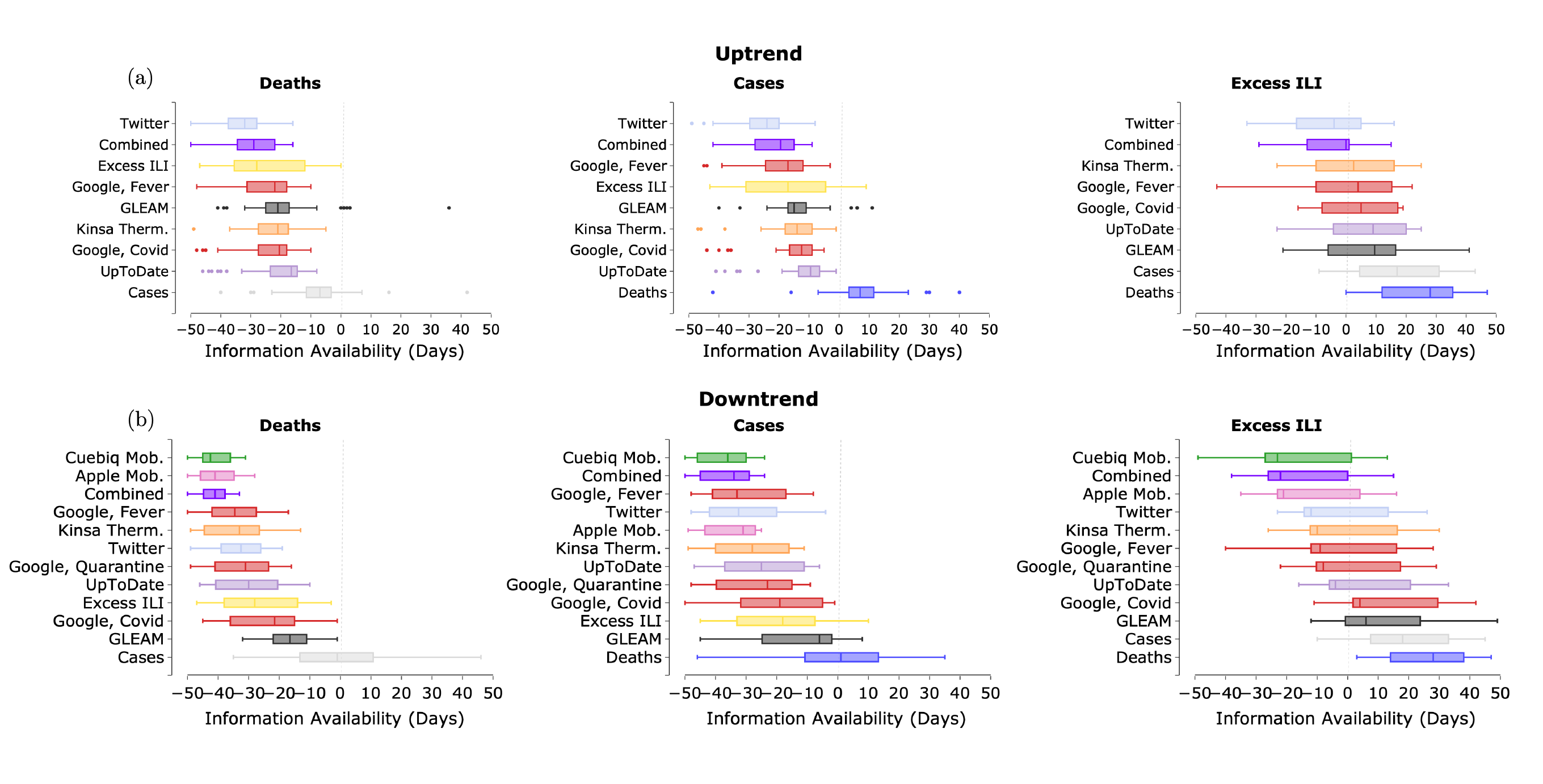}
    \end{subfigure} 
    \caption{Event detection results for pairwise comparisons between COVID-19 proxies and gold standards for US states with available data. (a) Boxplots showing proxy-specific uptrends, or intervals of significant exponential growth relative to deaths, confirmed COVID-19 cases, and Excess ILI. (b) Boxplots showing proxy-specific downtrends. Boxplots indicate the median (central vertical line), interquartile range (vertical lines flanking the median), extrema (whiskers), and outliers (dots); differences between input variable (y-axis) and response variable (title) exceeding 50 days are omitted. Negative differences indicate the input variable event activation preceded the response variable event activation. Deaths, cases, and Excess ILI, as well as the combined measure defined in Figure \ref{fig:EventDetection}, are also included for purposes of intercomparing gold standards. Boxplots are sorted according to median value and shifted to offset delays in real-time availability. Only the event activations within the first wave are considered; the box plots therefore do not account for subsequent minor activations.}
  \label{fig:violins_parker}
\end{figure}

The activation order of COVID-19 early warning indicators in NY is characterized by earlier growth in proxies reflecting public sentiment than in more clinically-specific proxies. This ordering is broadly repeated across states (Figure \ref{fig:violins_parker}a). COVID-19-related Twitter posts and Google Searches for ``fever'' were among the earliest proxies to activate, with Twitter activating first for 35 states and Google activating first for 7 states. UpToDate showed the latest activation among proxies, activating last in 37 states albeit still predating  an uptrend in confirmed cases (Supplementary Figure S67). This analysis was conducted for all states excepting those with data unavailable due to reporting delays; event information was missing for deaths in 2 states, Kinsa in 1 state, and Excess ILI in 5 states.\vspace{12pt}  

Although data streams that directly measure public sentiment (i.e., Google and Twitter) are sensitive to the intensity of news reporting, we note that median growth in Google Searches for ``fever'' occur within 3 days of median growth in fever incidence (as measured by Kinsa), suggesting that many searches may be driven by newly-ill people (Figure~\ref{fig:violins_parker}a). We additionally observed that the median lags between deaths and either fever incidences or Google Searches for ``fever'' were respectively 22 days and 21 days, broadly consistent with previous estimates that the average delay between the onset of symptoms and death is 20 days \cite{linton2020incubation}. Detailed time-series and event activation dates are displayed in Figures S18 to S66 in the Supplemental Materials.\vspace{12pt}

Consensus among proxies was recorded through a ``combined'' growth event that acts as a consolidated metric for all uptrend events (Figures \ref{fig:EventDetection}, \ref{fig:violins_parker}). As described in more detail in Section 4, a harmonic mean was taken across the $p$-values associated with each of the indicators. The harmonic mean was used because it does not require $p$-values across different proxies to be independent \cite{wilson_201pnas}. Similar to the case for individual proxies, we defined detection of a growth event to occur when the harmonic mean $p$-value (HMP) decreases below 0.05.\vspace{12pt}. 

\noindent \textbf{Quantifying the timing of decay in proxies.}
An examination analogous to that made for the uptrend events was made to identify the timing of exponential decay (``downtrend'') in proxies and gold standard time series. On each day of the time series and for each proxy, a $p$-value is calculated to represent the posterior probability that $\gamma$ is less than zero. An individual proxy's downtrend was defined to occur on days when the associated $p$-value decreases below 0.05. A sample sequence of downtrends is also depicted in Figure \ref{fig:EventDetection}, where downward-pointing triangles denote the date on which a decay event is identifiable. For the example state, Cuebiq and Apple mobility data gave the earliest indication of decreasing COVID-19 activity, exhibiting downtrends around March 15. \vspace{12pt} 

The added value of extending our analysis to include decay events is the ability to characterize downstream effects of NPIs. Specifically, opportunities to rapidly assess NPI effectiveness may arise if NPI influence on transmission rates is recorded in proxy time series before it is recorded in confirmed case or death time series. We used two smartphone-based metrics of human mobility that are indicators of population-wide propensity to travel within a given US county as provided by the location analytics company Cuebiq and by Apple. These mobility indicators are used as proxies for adherence to social distancing policies and are described in detail in section \ref{data_and_methods}. Apple Mobility and the Cuebiq Mobility Index (CMI) are imperfect insofar as they do not account for several important transmission factors, including importing of cases from other states. Although local travel distances are incomplete proxies for the scale at which NPIs are performed, reductions in mobility have been shown to lead subsequent reduction in fever incidence by an average of 6.5 days - an interval approximately equal to the incubation period of COVID-19 - across hundreds of US counties \cite{liautaud_2020arxiv}, suggesting that they capture NPI-induced reductions in transmission rates. Figure \ref{fig:violins_parker}b supports this observation, with median fever incidence lagging median CMI and Apple Mobility activation by an average of 8.5 days and 5.5 days, respectively. Our use of two distinct mobility metrics is intended to reduce the influence of systematic biases arising from the methodology of either metric. \vspace{12pt} 

The timing of the first downtrend is consistent between Apple Mobility and CMI (maximum difference of 4 days in median activation across all states with available data), with median downtrend activation for CMI preceding median activation of all other proxies and gold standard time series (Figure \ref{fig:violins_parker}b). Median decay in these indices predated median decay in deaths and confirmed cases by a median of 6 and 5 weeks, respectively; CMI was first to activate in 60\% of states (refer to Figure S68). GLEAM, Google Searches for ``covid'', and UpToDate were among the latest proxies to activate across states. Median downtrend activation for Google Searches for ``quarantine'' - included as a surrogate measure of mobility - lagged CMI and Apple Mobility median downtrend activation by an average of 12 days and 10.5 days, respectively. Statistically significant events were not detected, or data were not available, for GLEAM in 22 states, Excess ILI in 5 states, Apple mobility in 1 state, deaths in 2 states, and confirmed cases in 7 states. \vspace{12pt} 

To complement our lead-lag analysis, we conducted a diagnostic post-hoc analysis using correlation coefficients between lagged time-series, described in further detail in Supplemental Materials (Supplementary Figures S70-S76).\vspace{12pt} 

\noindent\textbf{Applying early signals by proxies to the prediction of future outbreak timing.} We hypothesize that an early warning system for COVID-19 could be derived from uptrend event dates across a network of proxies. For each state, $\Delta x_{i,t}$ is defined as the number of days since an uptrend event for proxy $i$, where $t$ is the current date. A posterior probability distribution of an uptrend in confirmed COVID-19 cases is then estimated for each state conditional on the collection of proxies, $p(y|\Delta x_{1,t}, ..., \Delta x_{n,t})$, where each proxy is treated as an independent expert predicting the probability of a COVID-19 event. In this case, $n$ is the number of proxies. See Section \ref{data_and_methods} for a more detailed explanation. A similar analysis is also feasible for downtrends. This method is introduced to better formalize predictions of growth in gold standard indicators using a network of proxies, but further evaluation is required, including relative to subsequent ``waves'' of COVID-19 cases.  

\begin{figure}[t!]
    \centering
    \begin{subfigure}[t]
    {.45\textwidth}
        \includegraphics[width=\textwidth]{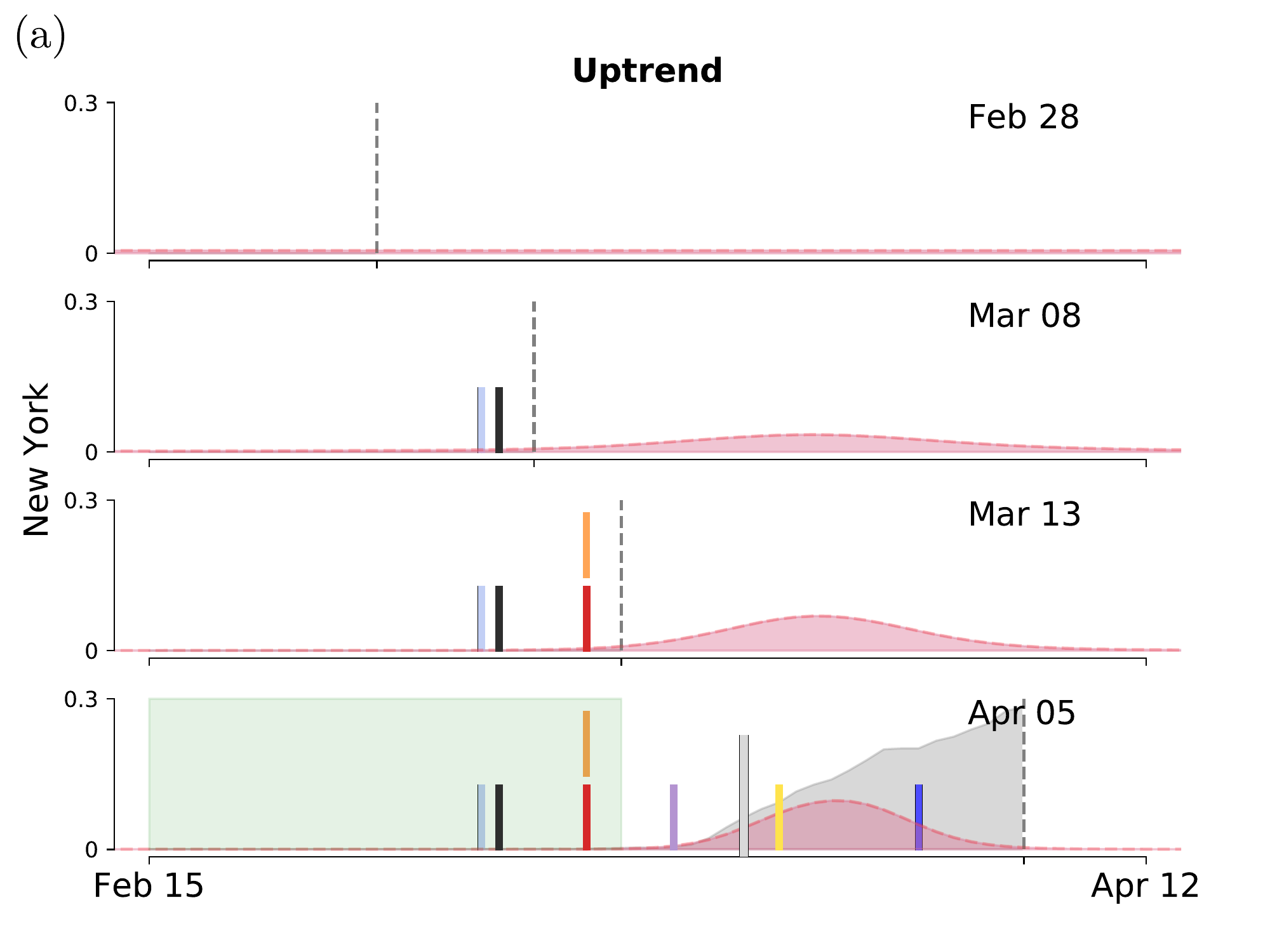}
    \end{subfigure}  
    \begin{subfigure}[t]
    {.45\textwidth}
        \includegraphics[width=\textwidth]{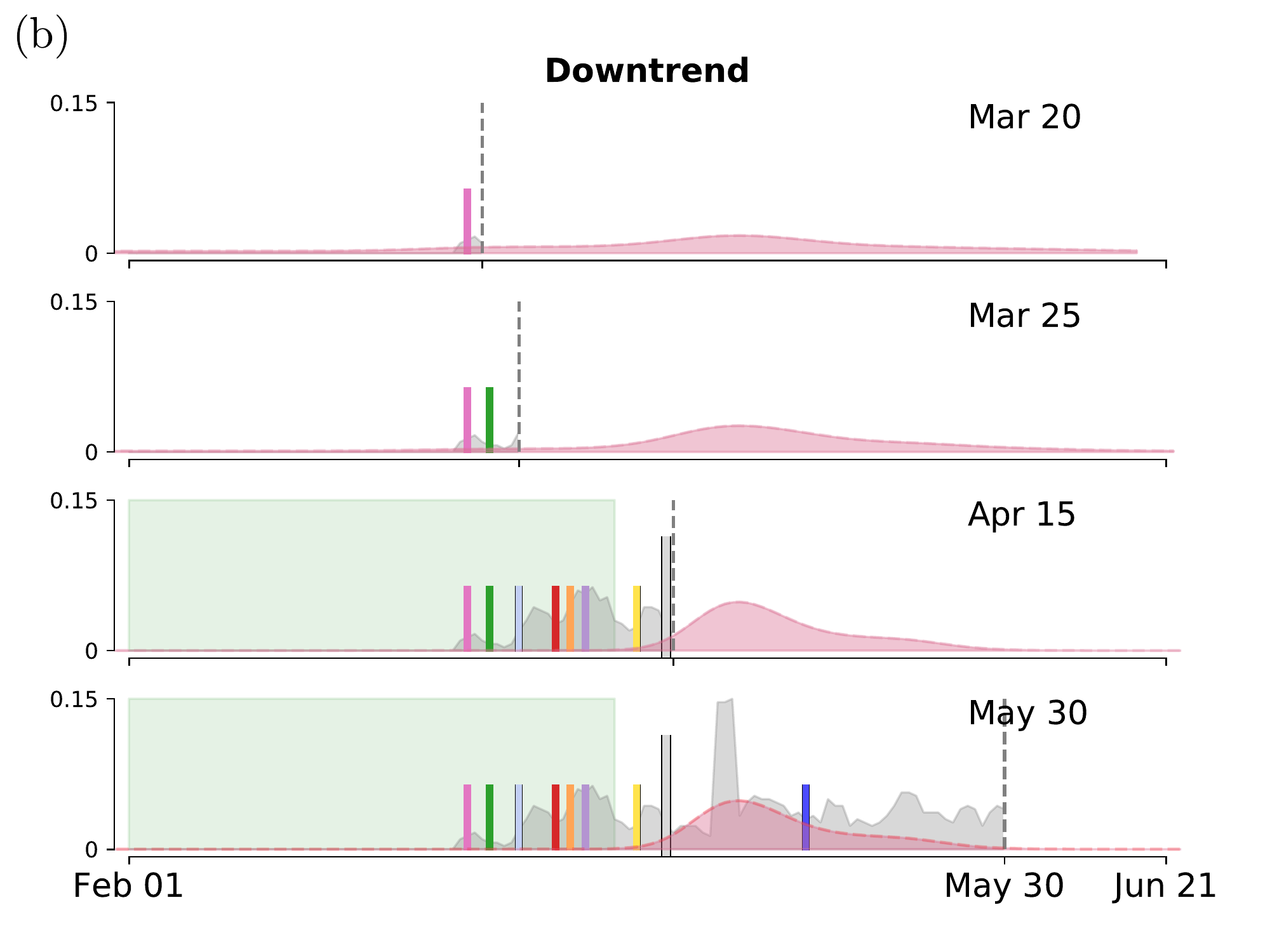}
    \end{subfigure} ~
    \begin{subfigure}[t]
    {1\textwidth}
        \includegraphics[width=\textwidth]{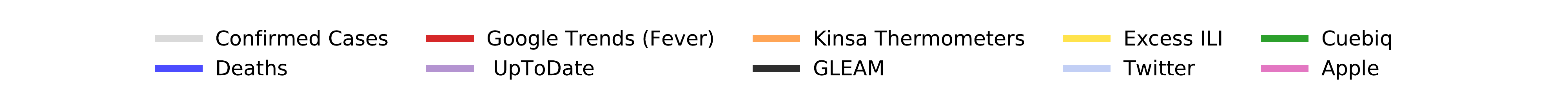}
    \end{subfigure}
    \caption{Illustration of the evolving probability distribution (in red) for the time-to-event estimation as applied to New York. The probability distributions are calculated from information up to a specified time horizon (vertical dashed line). Events that signaled an exponential increase or decrease more than 7 days prior to the true event (long vertical solid line) are contained within a green background. (a) shows the estimated uptrend posterior probability distribution and uptrend events. (b) shows the estimated downtrend posterior probability distribution and downtrend events. }
  \label{fig:probability_evol}
\end{figure}

Figure \ref{fig:probability_evol}a shows uptrend events from proxies (vertical solid lines) and the predicted uptrend probability distribution for confirmed COVID-19 cases (in red) overlayed on confirmed COVID-19 cases (gray). As more proxy-derived uptrend events are observed, the probability mass of the predicted uptrend event becomes increasingly concentrated in the vicinity of identified exponential growth in confirmed COVID-19 cases (long vertical solid line). In the case of NY, exponential growth is identified in confirmed COVID-19 cases in the earlier part of the prediction distribution, though for most states it occurs near the center as follows from how the prediction is estimated. The right panel of Figure \ref{fig:probability_evol}b similarly shows downtrend events in proxies and the estimated downtrend posterior probability distribution for decay in daily reports of confirmed COVID-19 cases. The downtrend posterior probability distribution has a greater variance than the uptrend posterior probability distribution, with the true downtrend event again occurring earlier in this high variance distribution. A visualization of the probability distribution for all the states is included in the Supplementary Materials (Figures S68 and S69).\vspace{12pt} 

\section{Discussion}

Here we have assessed the utility of various digital data streams, individually and collectively, as components of a near real-time COVID-19 early warning system. Specifically, we focused on identifying early signals of impending outbreak characterized by significant exponential growth and subsiding outbreak characterized by exponential decay. We found that COVID-19-related activity on Twitter showed significant  growth 2-3 weeks before such growth occurred in confirmed cases and 3-4 weeks before such growth occurred in reported deaths. We also observed that for exponential decay, NPIs - as represented by reductions in human mobility - predated decreases in confirmed cases and deaths by 5-6 weeks. Clinicians' search activity, fever data, estimates from the GLEAM metapopulation mechanistic epidemiological model, and Google Searches for COVID-19-related terms were similarly found to anticipate changes in COVID-19 activity. We also developed a consensus indicator of COVID-19 activity using the harmonic mean of all proxies. This combined indicator predated an increase in COVID-19 cases by a median of 19.5 days, an increase in COVID-19 deaths by a median of 29 days, and was synchronous with Excess ILI. Such a combined indicator may provide timely information, like a ``thermostat'' in a heating or cooling system, to guide intermittent activation, intensification, or relaxation of public health interventions as the COVID-19 pandemic evolves. \vspace{12pt} 

The most reliable metric for tracking the spread of COVID-19 remains unclear, and all metrics discussed in this study feature important limitations. For example, a recent study has shown that confirmed US cases of COVID-19 may not necessarily track the evolution of the disease considering limited testing frequency at early stages of the pandemic \cite{Kaashoek_2020}. While deaths may seem a more accurate tracker of disease evolution, they are limited in their real-time availability, as they tend to lag cases by nearly 20 days \cite{linton2020incubation}. Influenza-like illness (ILI) activity, anomalies in which may partly reflect COVID-19 \cite{lu2020estimating}, similarly suffers from a lag in availability because reports are released with a 5-12 day lag; for simplicity, we approximated this lag as 10 days in our analysis. Furthermore, a decrease in ILI reporting is frequently observed after flu season (which ended April 4 2020 and will begin again in October 2020), rendering ILI-based analyses useful only when surveillance systems are fully operational. Ref. \cite{lu2020estimating} supports this conjecture, reporting a rapid decrease in the number of ILI patients reported in late March 2020 despite the number of reporting providers remaining largely unchanged. This decrease may also be attributed to patients foregoing treatment for milder, non-COVID-19-attributed ILI. Hospitalizations, though possibly less biased than confirmed case numbers, were ultimately omitted due to sparseness and poor quality of data.\vspace{12pt} 

The near real-time availability of digital data streams can facilitate tracking of COVID-19 activity by public health officials. Increases in discussions of disease terminology on Twitter and Google, for example, may be early signals of increase in COVID-19 activity (Figures \ref{fig:EventDetection}, \ref{fig:violins_parker}). These data streams have been used in the past to track other infectious diseases in the US \cite{paul2014twitter, yang2015accurate, gluskin2014denguetrack}. Although Twitter and Google Trends both show growth and decay well ahead of confirmed cases, deaths, and ILI, it is unclear if their activity is in fact representative of disease prevalence. This activity may instead reflect the intensity of news coverage \cite{santillana2014can, lazer2014parable, bento2020evidence} and could perhaps be triggered by ``panic searches'' following the identification of several cases. Such false positives in social and search activity may be reduced by confirmatory use of UpToDate, whose clinician-restricted access and clinical focus limit the likelihood that searches are falsely inflated \cite{santillana2014up2date}. Kinsa data may be used in a similar confirmatory capacity as they directly report user symptoms. However, the number of users and their demographics, as well as several aspects of the incidence estimation procedure, are not disclosed by Kinsa \cite{covid_ineq}, limiting our ability to account for possible sources of bias in their data.\vspace{12pt} 

Given the near-ubiquity of smartphones in the US \cite{smartphone_ub}, smartphone-derived mobility data may reflect aspects of the local population response to COVID-19. We found that decreasing mobility - as measured by Apple and Cuebiq - predated decreases in deaths and cases by 6 and 5 weeks, respectively. Our results may be compared to documented findings that decreases in mobility preceded decreases in COVID-19 cases by up to 1 week in China \cite{vespignani_mobility, scarpino_mobility}, by up to 2 weeks in Brazil \cite{Ganem2020}, by up to 3 weeks in certain US states \cite{cdc_mobility}, and by up to 3 weeks globally \cite{batiashvili_mobility}. This variability may be attributed to differences in swiftness and strictness of implementing mobility restrictions, as well as discrepancies in definitions of ``decrease.''\vspace{12pt}  

In contrast to the aforementioned digital data streams, GLEAM assimilates many epidemiological parameters (e.g., literature-derived incubation period and generation time \cite{backer2020incubation, kissler2020projecting, verity_2020lancet}) essential for describing COVID-19. Coupled with Internet sources, GLEAM can provide a more robust outbreak detection method due to its different framework and, consequently, at least partially uncorrelated errors with the Internet-derived sources. Estimates produced by this model suggest a median increase in cases and deaths of 15 and 22 days later, respectively (Figure \ref{fig:violins_parker}a). However, some parameters are excluded from the model due to lack of availability (e.g., age-related COVID-19 susceptibility). These excluded parameters - coupled with the need to regularly update existing parameters - may lead to sub-optimal downstream performance by the model.\vspace{12pt} 

The analysis we presented focuses largely on temporally analyzing different data streams that are aggregated to the state-level. This level of aggregation is able to provide a coarse overview of regional differences within the US. Smart thermometer, Twitter, and mobility data may help us replicate our analysis at finer spatial resolutions, making them suitable for capturing both regional and local effects. It follows that a promising future research avenue is the detection of local COVID-19 clusters (``hotspots'') through more spatially-revolved approaches \cite{Ord1995}. Such an approach would better inform regarding at-risk populations and, therefore, permit for more targeted NPIs. Whereas the data streams that we analyze do not capture important population dynamics, integrated spatial, temporal, and semantic analysis of web data \cite{resch2015urbanemotions} could give a more nuanced understanding of public reaction, such as estimating changes in the public emotional response to epidemics \cite{resch2016twitteremotions}.\vspace{12pt} 

Using an exponential model to characterize the increase (and decrease) in activity of a COVID-19 proxy offers various advantages in event detection.Our current procedure is capable of estimating the value of $\gamma$ with a measure on the confidence that the $\gamma > 0$ or $\gamma < 0$. In this work, we provide event dates based on a confidence of 95\% ($p$-value $< 0.05$). The degree of confidence can be adjusted to provide earlier event dates (at the cost of less confidence and, consequently, more frequent false positives). $p$-values are combined into a single metric using a harmonic mean, but a more sensitive design  may be realized by assuming independence between the proxies and using Fisher's method. Although this would generally lead to a lower combined $p$-value given decreases in any individual proxy $p$-value (i.e., higher sensitivity), assuming independence would make such an approach prone to false positives (i.e., lower specificity) than the HMP, which makes no independence assumption. The choice of method for combining proxy indicators requires a trade-off between specificity and sensitivity.\vspace{12pt} 

The ability to detect future epidemic changes depends on the stability of historical patterns observed in multiple measures of COVID-19 activity. We posit that using multiple measures in our COVID-19 early warning system leads to improved performance and robustness to measure-specific flaws. The probabilistic framework we developed (Figure \ref{fig:probability_evol}) also gives decision-makers the freedom to decide how conservative they want to be in interpreting and consolidating different measures of COVID-19 activity (i.e., by revising the $p$-value required for event detection). Although we can expect COVID-19 activity to increase in the future given continued opportunities for transmission, the human population in which said transmission occurs may not remain identical in terms of behavior, immune status, or risk. For example, death statistics in the early pandemic have been driven by disease activity in nursing homes and long term care facilities \cite{nurse_homes}, but the effect of future COVID-19 waves in these settings may be attenuated by better safeguards implemented after the first wave. Our approach combines multiple measures of COVID-19 such that these changes in population dynamics - if reflected in any measure - would presumably be reflected in the event-specific probability distribution.\vspace{12pt} 

\section{Data and Methods}
\label{data_and_methods}
For our study, we collected the following daily-reported data streams: 1) official COVID-19 reports from three different organizations, 2) ILI cases, as reported weekly by the ILINet, 3) COVID-19-related search term activity from UpToDate and Google Trends, 4) Twitter microblogs, 5) fever incidence as recorded by a network of digital thermometers distributed by Kinsa, and 6) human mobility data, as reported by Cuebiq and Apple.\vspace{12pt} 

\textbf{COVID-19 Case Reports}: Every state in the US is providing daily updates about its COVID-19 situation as a function of testing. Subject to state-to-state and organizational variability in data collection, these reports include information about the daily number of positive, negative, pending, and total COVID-19 tests, hospitalizations, ICU visits, and deaths. Daily efforts in collecting data by research and news organizations have resulted in several data networks, from which official health reports have been made available to the public. The three predominant data networks are: the John Hopkins Resource Center, the CovidTracking project, and the New York Times Repository \cite{dong2020jhkdashboard, meyer2020covidtrackingproject, nytcovidrepo}. We obtained daily COVID-19 testing summaries from all three repositories with the purpose of analyzing the variability and quality in the data networks.\vspace{12pt} 

\textbf{ILINet}: Influenza-like illnesses (ILI) are characterized by fever and either a cough or sore throat. An overlap in symptoms of COVID-19 and ILI has been observed, and it has further been shown that ILI signals can be useful in the estimation of COVID-19 incidence when testing data is unavailable or unreliable \cite{lu2020estimating}.\vspace{12pt} 

ILINet is a sentinel system created and maintained by the US CDC~\cite{cdc2020fluview, cdc2020iss} that aggregates information from clinicians' reports on patients seeking medical attention for ILI symptoms. ILINet provides weekly estimates of ILI activity with a lag of 5-12 days; because detailed delay information is unavailable, we arbitrarily apply a lag of 10 days throughout this work. At the national-level, ILI activity is estimated via a population-weighted average of state-level ILI data. ILINet data are unavailable for Florida.\vspace{12pt}  

The CDC also releases data on laboratory test results for influenza types A and B, shared by labs collaborating with the World Health Organization (WHO) and the National Respiratory and Enteric Virus Surveillance System (NREVSS). Both ILI activity and virology data are available from the CDC FluView dashboard~\cite{cdc2020fluview}.\vspace{12pt} 

We followed the methodology of Ref.\cite{lu2020estimating} to estimate unusual ILI activity, a potential signal of a new outbreak such as COVID-19. In particular, we employed the divergence-based methods, which treat COVID-19 as an intervention and try to measure the impact of COVID-19 on ILI activity by constructing two control time series representing the counterfactual 2019-2020 influenza season had the COVID-19 outbreak not occurred.\vspace{12pt}  

The first control time series is based on an epidemiological model, specifically the Incidence Decay and Exponential Adjustment (IDEA) model~\cite{fisman2013idea}. IDEA models disease prevalence over time while accounting for factors as control activities that may dampen the spread of a disease. The model is written as follows:
$$I(t) = \left(\frac{R_0}{(1+d)^t}\right)^t$$
where $I(t)$ represents the incident case count at serial interval time step $t$. $R_0$ represents the basic reproduction number, and $d$ is a discount factor modeling reductions in the effective reproduction number over time.\vspace{12pt} 

In line with the approach of Ref.\cite{lu2020estimating}, we fit the IDEA model to ILI case counts from the start of the 2019-2020 flu season to the last week of February 2020, where the start of flu season is defined as the first occurrence of two consecutive weeks with an ILI activity above $2\%$. The serial interval used was half a week, consistent with the influenza serial interval estimates from~\cite{10.1093/aje/kwu209}.\vspace{12pt}

The second control time series used the CDC's virological influenza surveillance data. For any week $t$ the following was computed:
$$ F_t = \frac{F_t^+ \cdot I_t}{N_t}$$
where $F_t^+$, $N_t$, $I_t$, and $F_t$ denote positive flu tests, total specimens, ILI visit counts, and the true underlying flu counts, respectively. This can be interpreted as the extrapolation of the positive test percentage to all ILI patients. Least squares regression (fit on pre-COVID-19 data) is then used to map $F_t$ to an estimate of ILI activity.\vspace{12pt}

The differences between the observed ILI activity time series and these counterfactual control time series can then be used as signals of COVID-19 activity. In particular, we used the difference between observed ILI activity and the virology-based counterfactual control time series to produce Excess ILI. The supplementary materials show \textbf{Excess ILI} computed using both the virology time series and IDEA model-based counterfactuals for all states.\vspace{12pt}  

\textbf{UpToDate Trends}: UpToDate is a private-access search database - part of Wolters Kluwer, Health - with clinical knowledge about diseases and their treatments. It is used by physicians around the world and the majority of academic medical centers in the US as a clinical decision support resource given the stringent standards on information within the database (in comparison to Google Trends, information provided within the database is heavily edited and authored by experienced clinicians) \cite{santillana2014up2date}.\vspace{12pt}  

Recently, UpToDate has made available a visualization tool on their website in which they compare their search volumes of COVID-19 related terms to John Hopkins University official health reports \cite{uptodateSearchIntensity}. The visualization shows that UpToDate trends may have the potential to track confirmed cases of COVID-19. From this tool, we obtained UpToDate's COVID-19-related search frequencies for every US state. These search frequencies consist only of one time series described as ``normalized search intensity'' for selected COVID-19-related terms, where normalization is calculated as the total number of COVID-19-related search terms divided by the total number of searches within a location. At the time of analysis, the 
website visualization appeared to update with a 3 day delay; however, UpToDate is since operationally capable of producing time series with delays of 1 day. More details are available at \texttt{https://covid19map.uptodate.com/}.\vspace{12pt}  

\textbf{Google Trends}: Google Search volumes have been shown to track successfully with various diseases such as influenza \cite{lu2019improved, lu2018boston}, Dengue \cite{gluskin2014denguetrack}, and Zika \cite{mcgough2017forecasting, teng2017dynamicZika}, among others \cite{carneiro2009google}. In recent months, Google has even created a ``Coronavirus Search Trends'' \cite{google_coronavirus} page that tracks trending pandemic-related searches. We obtained such daily COVID-19-related search trends through the Google Trends Application Programming Interface (API). The original search terms queried using the Google Trends API were composed of: a) official symptoms of COVID-19, as reported by the WHO, b) a list of COVID-19 related terms shown to have the potential to track confirmed cases \cite{lampos2020tracking}, and c) a list of search terms previously used to successfully track ILI activity \cite{lu2019improved}. The list of terms can be seen in Table \ref{tab:google_term_list}. For purposes of the analysis, we narrowed down the list of Google Search terms to those we felt to be most representative of the pandemic to date: ``fever'', ``covid'', and ``quarantine.'' Given that lexicon, however, is naturally subject to change as the pandemic progresses, other terms may become more suitable downstream.\vspace{12pt} 

\renewcommand{\arraystretch}{1}%
\begin{table}[h!]
\centering
\caption{Search term list for Google Trends.}
\label{tab:google_term_list}
\begin{tabular}{@{} p{15cm}p{3cm}|p{3cm}|p{3cm} @{}}
\toprule
anosmia, chest pain, chest tightness, cold, cold symptoms, cold with fever, contagious flu, cough, cough and fever, cough fever, covid, covid nhs, covid symptoms, covid-19, covid-19 who, dry cough, feeling exhausted, feeling tired, fever, fever cough, flu and bronchitis, flu complications, how long are you contagious, how long does covid last, how to get over the flu, how to get rid of flu, how to get rid of the flu, how to reduce fever, influenza, influenza b symptoms, isolation, joints aching, loss of smell, loss smell, loss taste, nose bleed, oseltamivir, painful cough, pneumonia, pneumonia, pregnant and have the flu, quarantine, remedies for the flu, respiratory flu, robitussin, robitussin cf, robitussin cough, rsv, runny nose, sars-cov 2, sars-cov-2 , sore throat, stay home, strep, strep throat, symptoms of bronchitis, symptoms of flu, symptoms of influenza, symptoms of influenza b, symptoms of pneumonia, symptoms of rsv, tamiflu dosage, tamiflu dose, tamiflu drug, tamiflu generic, tamiflu side effects, tamiflu suspension, tamiflu while pregnant, tamiflu wiki, tessalon \\ \bottomrule
\end{tabular}
\end{table}

\textbf{Twitter API}: We developed a geocrawler software to collect as much georeferenced social media data as possible in a reasonable time. This software requests data from Twitter's APIs. Twitter provides two types of APIs to collect tweets: REST and streaming \cite{TwitterDeveloper}. The REST API offers various endpoints to use Twitter functionalities, including the “search/tweets” endpoint that enables, with limitations, the collection of tweets from the last seven days. These limitations complicate the collection process, necessitating a complementary strategy to manage the fast-moving time window of the API in order to harvest all offered tweets with a minimal number of requests. In contrast, the streaming API provides a real-time data stream that can be filtered using multiple parameters.\vspace{12pt}  

Our software focuses on requesting tweets featuring location information either as a point coordinate from the positioning device of the mobile device used for tweeting or a rectangular outline based on a geocoded place name, using APIs. The combination of APIs makes crawling robust against interruptions or backend issues that would lead to missing data. For example, if data from the streaming API cannot be stored in time, the missing data can be retrieved via the redundant REST API.\vspace{12pt} 

All collected tweets are located within the US. To limit the dataset to COVID-19-relevant tweets, we performed simple keyword-based filtering using the keywords listed in table \ref{tab:twitter_term_list}. This method was chosen for reasons of performance, delivery of results in near real-time, and its simplicity. While a machine learning-based semantic clustering method like Guided Latent Dirichlet Allocation (LDA) may deliver more comprehensive results (e.g., through identifying co-occurring and unknown terms \cite{Resch2018}), controlling the ratio between false positives and false negatives requires extensive experimental work and expert knowledge.\vspace{12pt} 

\begin{table}[h!]
\centering
\caption{Search term list for Twitter.}
\label{tab:twitter_term_list}
\begin{tabular}{@{} p{15cm}p{3cm}|p{3cm}|p{3cm} @{}}
\toprule
 covid, corona, epidemic, flu, influenza, face mask, spread, virus, infection, fever, panic buying, state of emergency, masks, quarantine, sars, 2019-ncov \\ \bottomrule
\end{tabular}
\end{table}

\textbf{Kinsa Smart Thermometer Data}:
County-level estimates of US fever incidence are provided by Kinsa Insights using data from a network of volunteers who have agreed to regularly record and share their temperatures (\texttt{https://healthweather.us/}). Estimates from past years have been shown to correlate closely with reported ILI incidence from the US CDC across regions and age groups \cite{miller_2018cid}. Such historical data, coupled with county-specific characteristics (e.g., climate and population size), are used to establish the ``expected'', or forecast, number of fevers as a function of time \cite{miller_2018cid, dalziel_2018}. An ``excess fevers'' time series presumed to represent COVID-19 cases is approximated as the difference between the observed fever incidence and the forecast, with values truncated at zero so that negative excess fevers are not possible. County-level data is aggregated up to the state-level by a population-weighted average. A limitation of tracking febrility as a proxy for COVID-19 is that it is not a symptom exclusive to COVID-19, nor is it present in all COVID-19 patients. In a study with more than 5000 patients from New York who were hospitalized with COVID-19, only 30\% presented with fever ($>$100.4F/38C) at triage \cite{10.1001/jama.2020.6775}.\vspace{12pt} 

\textbf{Cuebiq Mobility Index}:
Data are provided by the location analytics company Cuebiq which collects first-party location information from smartphone users who opted in to anonymously provide their data through a General Data Protection Regulation-compliant framework; this first-party data is not linked to any third-party data. Cuebiq further anonymizes, then aggregates location data into an index, $M$, defined as the base-10 logarithm of median distance traveled per user per day; ``distance'' is measured as the diagonal distance across a bounding box that contains all GPS points recorded for a particular user on a particular day. A county index of 3.0, for example, indicates that a median county user has traveled 1,000m. We were provided with county-level data - derived from these privacy preservation steps - which we then aggregated up to the state level by a population-weighted average. \vspace{12pt} 

\textbf{Apple Mobility}: 
Apple mobility data is generated by counting the number of requests made to Apple Maps for directions in select countries/regions, sub-regions, and cities. Data that is sent from users’ devices to the Maps service is associated with random, rotating identifiers so Apple does not have a profile of users' movements and searches. The availability of data in a particular country/region, sub-region, or city is based on a number of factors, including minimum thresholds for direction requests per day. More details are available at \texttt{https://www.apple.com/covid19/mobility}.\vspace{12pt} 

\textbf{Global Epidemic and Mobility Model (GLEAM):} GLEAM is a spatially structured epidemic model that integrates population and mobility data with an individual-based stochastic infectious disease dynamic to track the global spread of a disease \cite{vespignani_mobility,balcan2010modeling,gomes2014pastore,ZhangE4334}. The model divides the world into more than 3,200 subpopulations, with mobility data between subpopulations including air travel and commuting behavior. Air travel data are obtained from origin-destination traffic flows from the Official Aviation Guide (OAG) and the IATA databases\cite{IATA,OAG}, while short-range mobility flows as commuting behavior are derived from the analysis and modeling of data collected from the statistics offices for 30 countries on five continents~\cite{balcan2010modeling}. Infectious disease dynamics are modeled within each subpopulation using a compartmental representation of the disease where each individual can occupy a single disease state: Susceptible ($S$), Latent ($L$), Infectious ($I$) and Removed ($R$). The infection process is modeled by using age-stratified contact patterns at the state level~\cite{mistry2020inferring}. These contact patterns incorporate interactions that occur in four settings: school, household, workplace, and the general community. Latent individuals progress to the infectious stage with a rate inversely proportional to the latent period. Infectious individuals progress into the removed stage with a rate inversely proportional to the infectious period. The sum of the average latent and infectious periods defines the generation time. Removed individuals represent those who can no longer infect others, as they were either isolated, hospitalized, died, or have recovered. To take into account mitigation policies adopted widely across the US, we incorporated a reduction in contacts in the school, workplace, and community settings (which is reflected in the contact matrices used for each state). Details on the timeline of specific mitigation policies implemented are described in Ref.~\cite{gleam_description}. A full discussion of the model for COVID-19 is reported in Ref.~\cite{vespignani_mobility}. \vspace{12pt} 

\textbf{Growth and decay event detection}:
Here we describe a simple Bayesian method that estimates the probability of exponential growth in a time-series that features uncertain error variance. We model a proxy time-series as following exponential growth,
\begin{align}
\label{eq1}
    y(t) = \beta \exp(\gamma (t-t_o)) + \epsilon(t),
\end{align}
over successive 14 day intervals. Before inference, proxies are adjusted to have a common minimum value of 1. The error, $\epsilon$, is assumed Gaussian with zero mean and standard deviation, $\sigma$. We assess the probability that $\gamma$ is greater than zero over each successive window. The joint distribution of $\beta$ and $\gamma$, conditional on $\sigma$, is proportional to $ p(\bm{y}, \sigma| \beta, \gamma) \times p(\beta) \times p(\gamma)$. Prior distributions, $p(\beta)$ and $p(\gamma)$, are specified as uniform and uninformative, and samples are obtained using the Metropolis-Hastings algorithm \cite{gelman_2013book} with $5\times10^{3}$ posterior draws.The first 500 samples are discarded to remove the influence of initial parameter values and, to reduce autocorrelation between draws, only every fifth sample is retained. The conditional posterior distribution for $\sigma_{n}$ is inverse-Gamma and is obtained using Gibbs sampling \cite{gelman_2006priors, fink_1997compendium}
\begin{equation}
    p(\sigma | \bm{y}, \gamma, \beta) \sim \Gamma^{-1} \left( \frac{N}{2} + \alpha, \beta + \frac{\sum_{i=1}^{N} \epsilon^{2}}{2} \right),
\end{equation}
where $\Gamma^{-1}$ is the inverse-Gamma distribution, $\bm{y}$ is the vector of observations, and $N$ is the number of observations. Terms $\alpha$ and $\beta$ are, respectively, specified to equal 4 and 1. On any given day, a $p$-value for rejecting the null hypothesis of no exponential growth is obtained as the fraction of posterior draws with $\gamma \leq 0$. The procedure is repeated on successive days to obtain a time-series of $p$-values.\vspace{12pt} 

Our current approach has some important limitations. The mean value in a time series is not inferred, and a highly simplified treatment of errors neglects the possibility of autocorrelation and heteroscedasticity. A more complete treatment might employ a more sophisticated sampling strategy and jointly infer (rather than impose) a mean proxy value, non-zero error mean, error autoregressive parameter(s), and heteroscedasticity across each 14-day window. Extensions to the current work will include these considerations.\vspace{12pt}

\textbf{Multi-Proxy $p$-Value:}
$p$-values estimated across multiple proxies are combined into a single metric representing the family-wise probability that $\gamma>0$. Because proxies cannot be assumed independent, we use the harmonic mean $p$-value \cite{wilson_201pnas}, 
\begin{equation}
\label{eq:harmonic_mean}
    \overset{\circ}{p} = \frac{\sum_{i=0}^{k}  w_{i}}{\sum_{i=0}^{k} w_{i}p_{i}^{-1}},
\end{equation}
where $w_{i}$ are weights that sum to 1 and, for the purposes of our analyses, are treated as equal.\vspace{12pt}  

\textbf{Time to Outbreak Estimation}: 
A time to outbreak estimation strategy can be formulated to provide probability estimates for when the next outbreak will occur given early indicators. We propose a strategy based on the timing of detected events among input data sources with respect to the eventual COVID-19 outbreak event in each state, as defined in the preceding section. We first modeled the behavior of the data sources in each state as a function of the state's underlying COVID-19 case trajectory over the time period studied. Specifically, we modeled the detected events in each data source as conditionally independent given the state-level latent COVID-19 dynamics. This follows from the assumption that exponential behavior in each data source is a causal effect of a COVID-19 outbreak and that other correlations unrelated to COVID-19 are mostly minor in comparison.\vspace{12pt} 

The time intervals between each event and the eventual outbreak event were then pooled across states to form an empirical conditional distribution for each dataset. Since observations were sparse relative to the time window, we used kernel density estimation with Gaussian kernels to smooth the empirical distributions, where bandwidth selection for the kernel density estimation was performed using Scott's Rule \cite{scott2015multivariate}.\vspace{12pt} 

Thus, for any given dataset, a detected event implies a distribution of likely timings for the COVID-19 outbreak event. We define $\Delta x_{it}$ as the number of days since an uptrend event for signal $i$ where $t$ is the current date. Within a state, the relative intervals of the events for each data source $\Delta x_{it}$ specify a posterior distribution over the probability of a COVID-19 event in $y$ days from current date $t$:
$$p(y | \Delta x_{1t}, \ldots, \Delta x_{nt}) \propto p(y) \prod_{i=1}^n p(\Delta x_{it} | y)$$
where we decomposed the joint likelihood
$$p(\Delta x_{1t}, \ldots, \Delta x_{nt} | y) =   \prod_{i=1}^n p(\Delta x_{it} | y)$$ using conditional independence.\vspace{12pt} 

A uniform prior $p(y)$ over the entire range of possible delays (a period of 180 days) was assumed, and additive smoothing was used when combining the probabilities. Because we modeled $y$ at the daily-level, the distributions are discrete, allowing evaluation of the posterior likelihood explicitly using marginalization. This process was repeated for each state. Such an approach can be viewed as pooling predictions from a set of ``experts'' when they have conditionally independent likelihoods given the truth \cite{punska1999bayesian}, with each expert corresponding to a data source. We note as a limitation that the assumption of conditionally independent expert likelihoods given the truth is unlikely to hold perfectly as, for example, an increase in measured fevers could be correlated with an increase in fever-related Google Searches even when the underlying COVID-19 infection dynamics are similar. Such dependencies may manifest heterogeneously as correlations among locations with similar COVID-19 outbreak timings, but are likely to be small since most of our inputs represent disparate data sources.\vspace{12pt} 

For the purposes of this model, we excluded mobility data when modeling the uptrend because it is a direct consequence of government intervention rather than of COVID-19 activity. When no event is detected for a data source, that data source's expert's prediction is zero across all possible timings, which translates with smoothing to a uniform distribution.\vspace{12pt}

\clearpage
\bibliographystyle{unsrt}
\bibliography{refs}

\section*{Funding}

MS was partially supported by the National Institute of General Medical Sciences of the National Institutes of Health under Award Number R01GM130668. The content is solely the responsibility of the authors and does not necessarily represent the official views of the National Institutes of Health.

\end{document}


\author{Nicole E. Kogan$^{*,1,2}$ \and Leonardo Clemente$^{*,1}$ \and Parker Liautaud$^{*,3}$ \and Justin Kaashoek$^{1,4}$ \and Nicholas B. Link$^{1,2}$ \and Andre T. Nguyen$^{1,5,6}$ \and Fred S. Lu$^{1,7}$ \and Peter Huybers$^{3}$ \and Bernd Resch$^{8,9}$ \and Clemens Havas$^8$ \and Andreas Petutschnig$^8$ \and Jessica Davis$^{10}$ \and Matteo Chinazzi$^{10}$ \and Backtosch Mustafa$^{1, 12}$ \and William P. Hanage$^2$ \and Alessandro Vespignani$^{10}$ \and Mauricio Santillana$^{1,2,11,\dagger}$}

\date{%
    \today
}

\maketitle

\begin{center}

          {\small $^1$Computational Health Informatics Program, Boston Children's Hospital, Boston, MA \\%
     $^2$Department of Epidemiology, Harvard T.H. Chan School of Public Health, Boston, MA\\
     $^3$Department of Earth and Planetary Sciences, Harvard University, Cambridge, MA\\%
     $^4$School of Engineering and Applied Sciences, Harvard University, Cambridge, MA\\
     $^5$University of Maryland, Baltimore County, Baltimore, MD \\%
     $^6$Booz Allen Hamilton, Columbia, MD\\%
     $^7$Department of Statistics, Stanford University, Stanford, CA\\%
     $^8$Department of Geoinformatics - Z\_GIS, University of Salzburg, Salzburg, Austria\\%
     $^9$Center for Geographic Analysis, Harvard University, Cambridge, MA\\%
     $^{10}$Northeastern University, Boston, MA\\%
     $^{11}$Department of Pediatrics, Harvard Medical School, Boston, MA\\%
     $^{12}$University Medical Center Hamburg-Eppendorf, Hamburg, Germany\\%
     .\\
     $^*$These authors contributed equally to this study.\\%
    
     $^{\dagger}$Correspondance to: Mauricio Santillana (\texttt{msantill@fas.harvard.edu}) \\[2ex]}%

\vspace{1cm}
{\LARGE\textbf{Supplementary Materials}}\\

\renewcommand\thefigure{S\arabic{figure}}

\vspace{1cm}
\textbf{Time evolution for the COVID-19 proxies for each state.}
\end{center}

 \begin{figure}
    \centering
    \includegraphics[width=.75\textwidth]{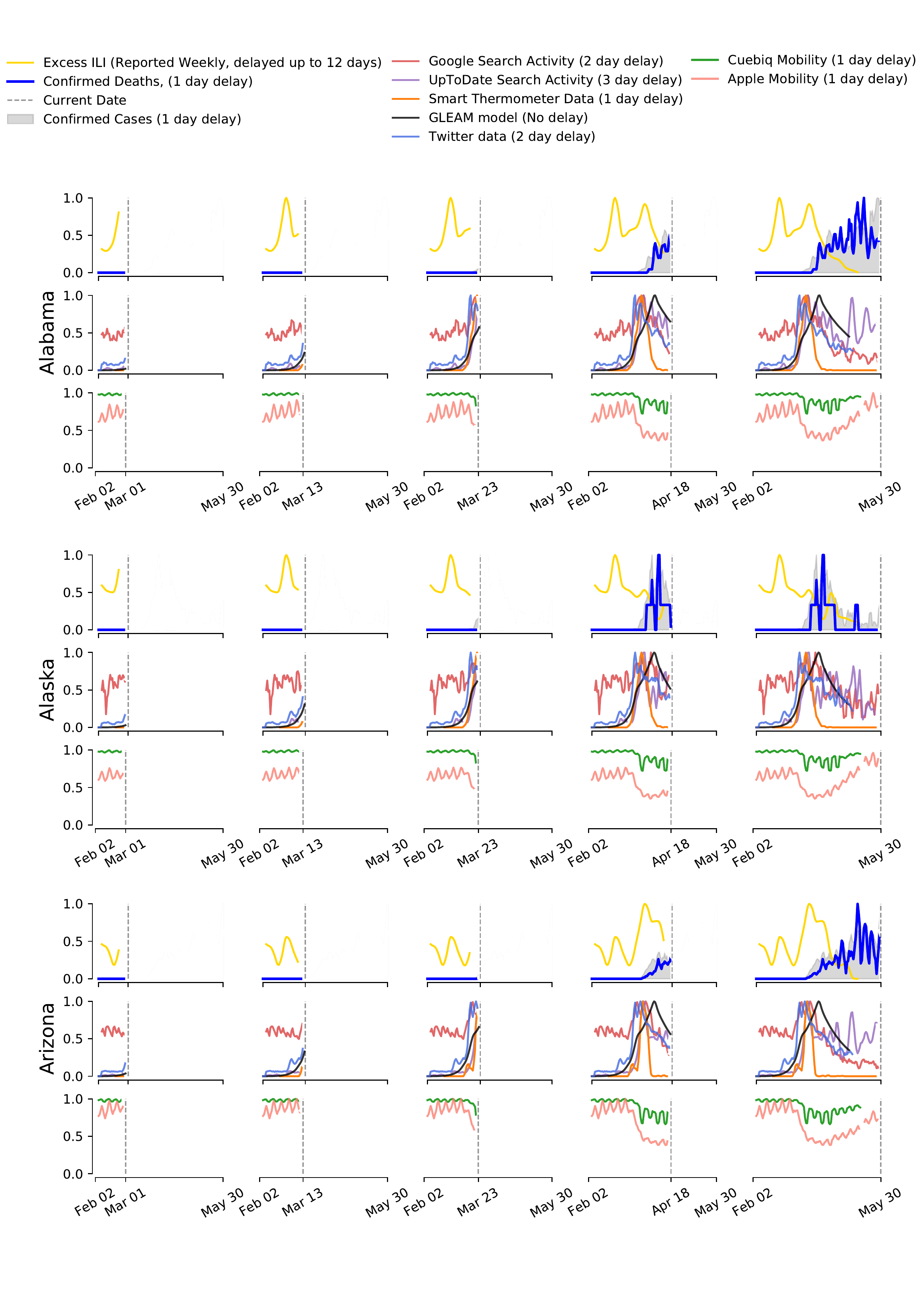}
    \label{fig:EventDetection} \caption{}
    \end{figure}
    
 \begin{figure}
    \centering
    \includegraphics[width=.75\textwidth]{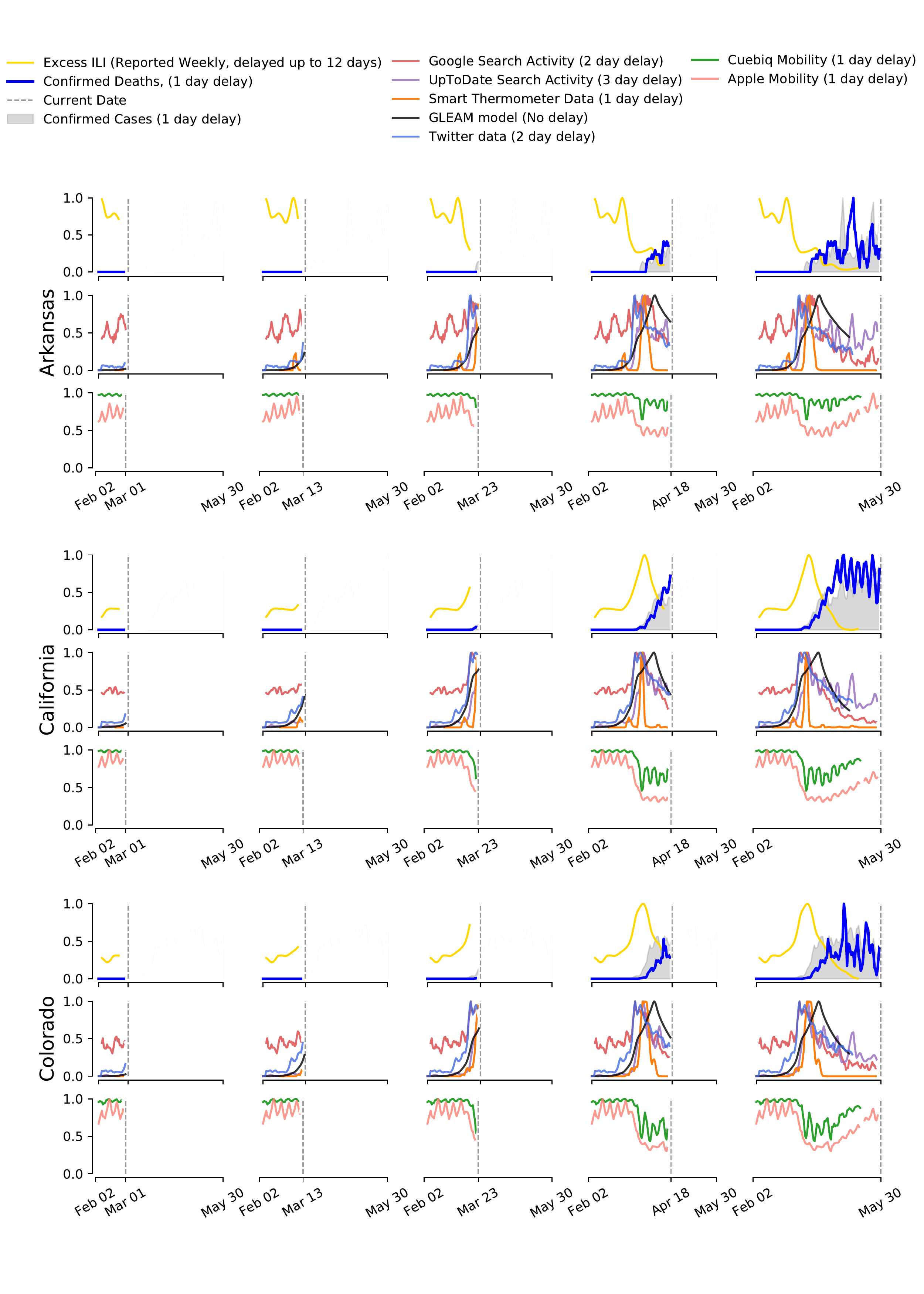}
    \label{fig:EventDetection} \caption{}
    \end{figure}
    
 \begin{figure}
    \centering
    \includegraphics[width=.75\textwidth]{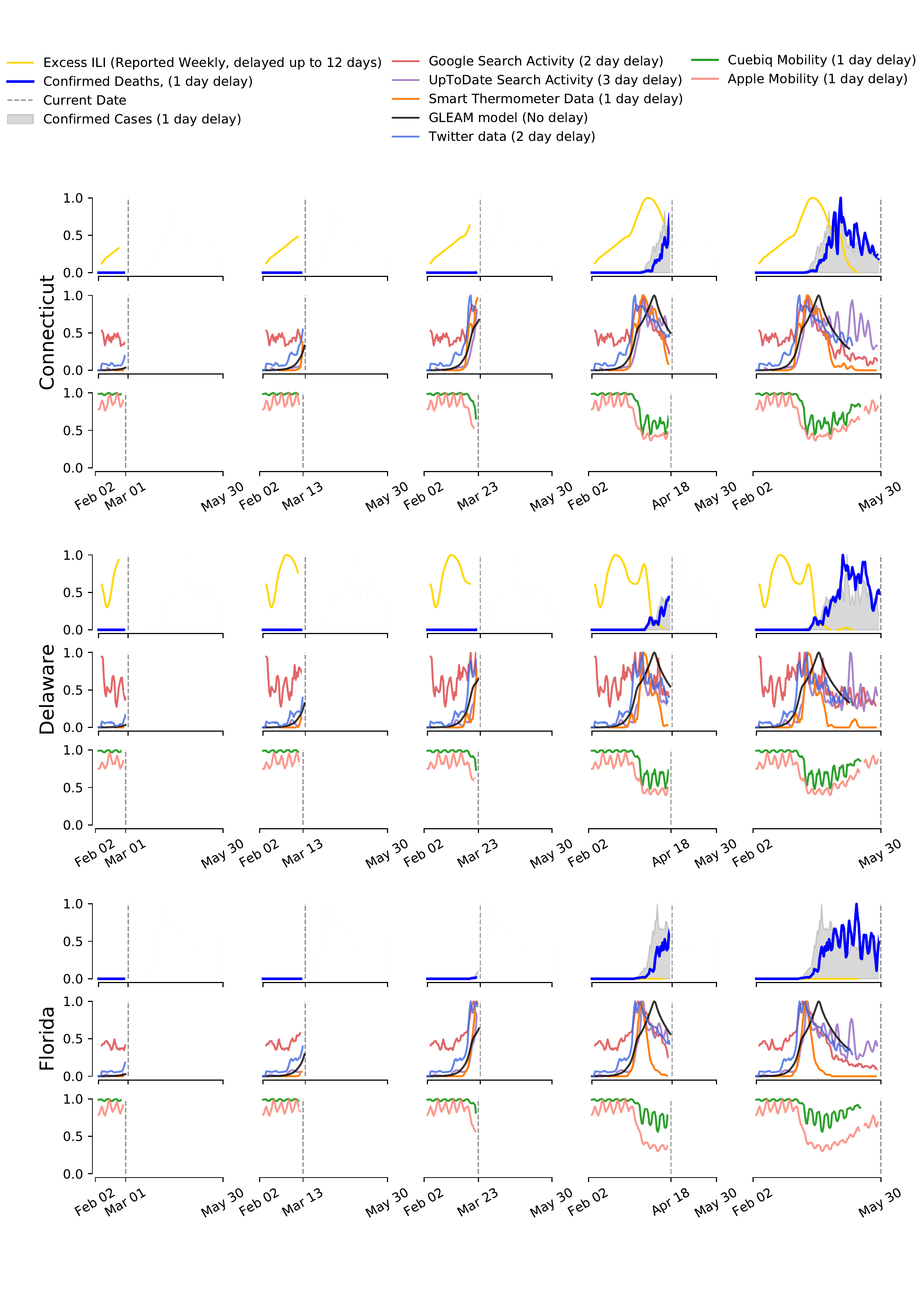}
    \label{fig:EventDetection} \caption{}
    \end{figure}
    
 \begin{figure}
    \centering
    \includegraphics[width=.75\textwidth]{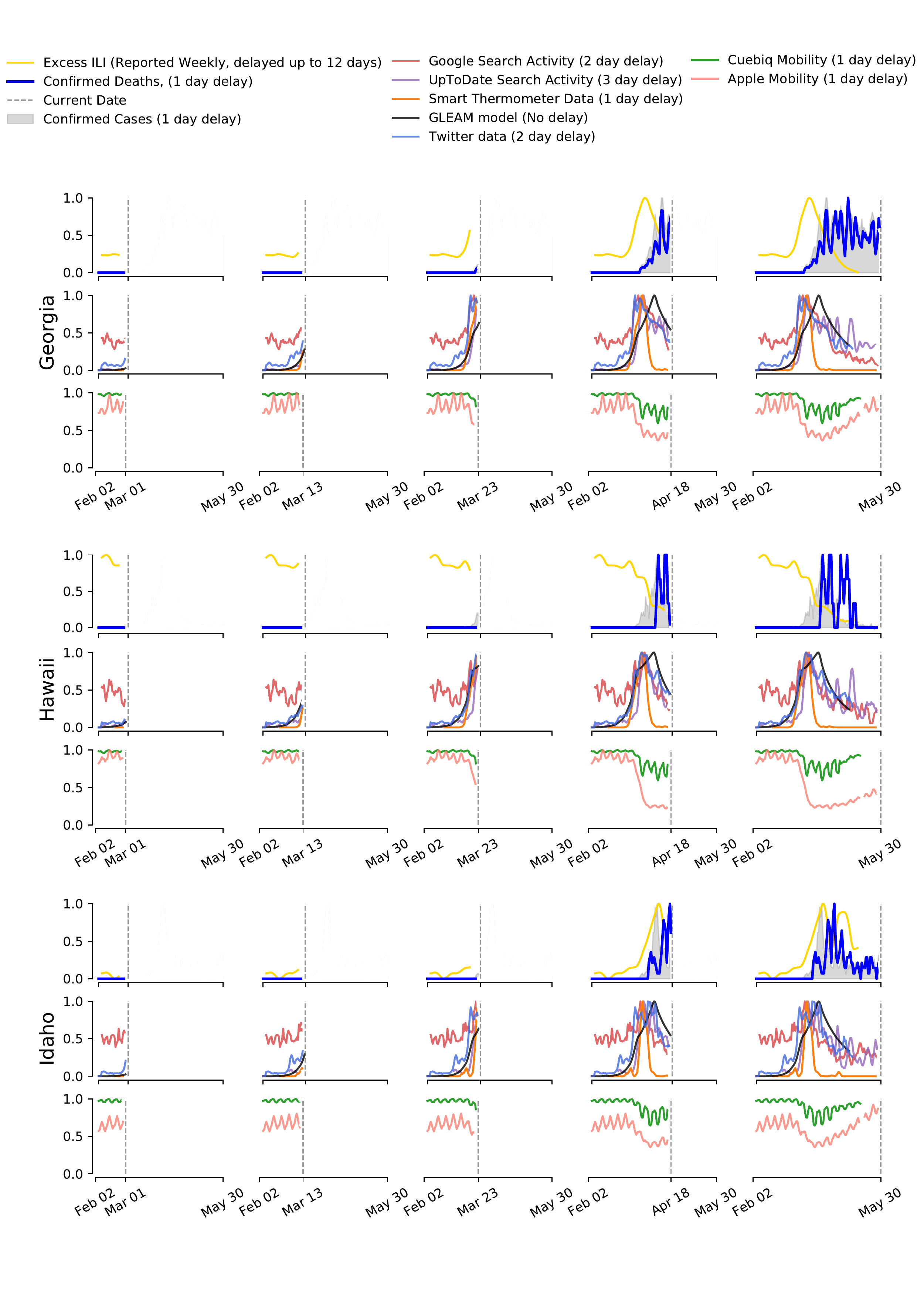}
    \label{fig:EventDetection} \caption{}
    \end{figure}
    
 \begin{figure}
    \centering
    \includegraphics[width=.75\textwidth]{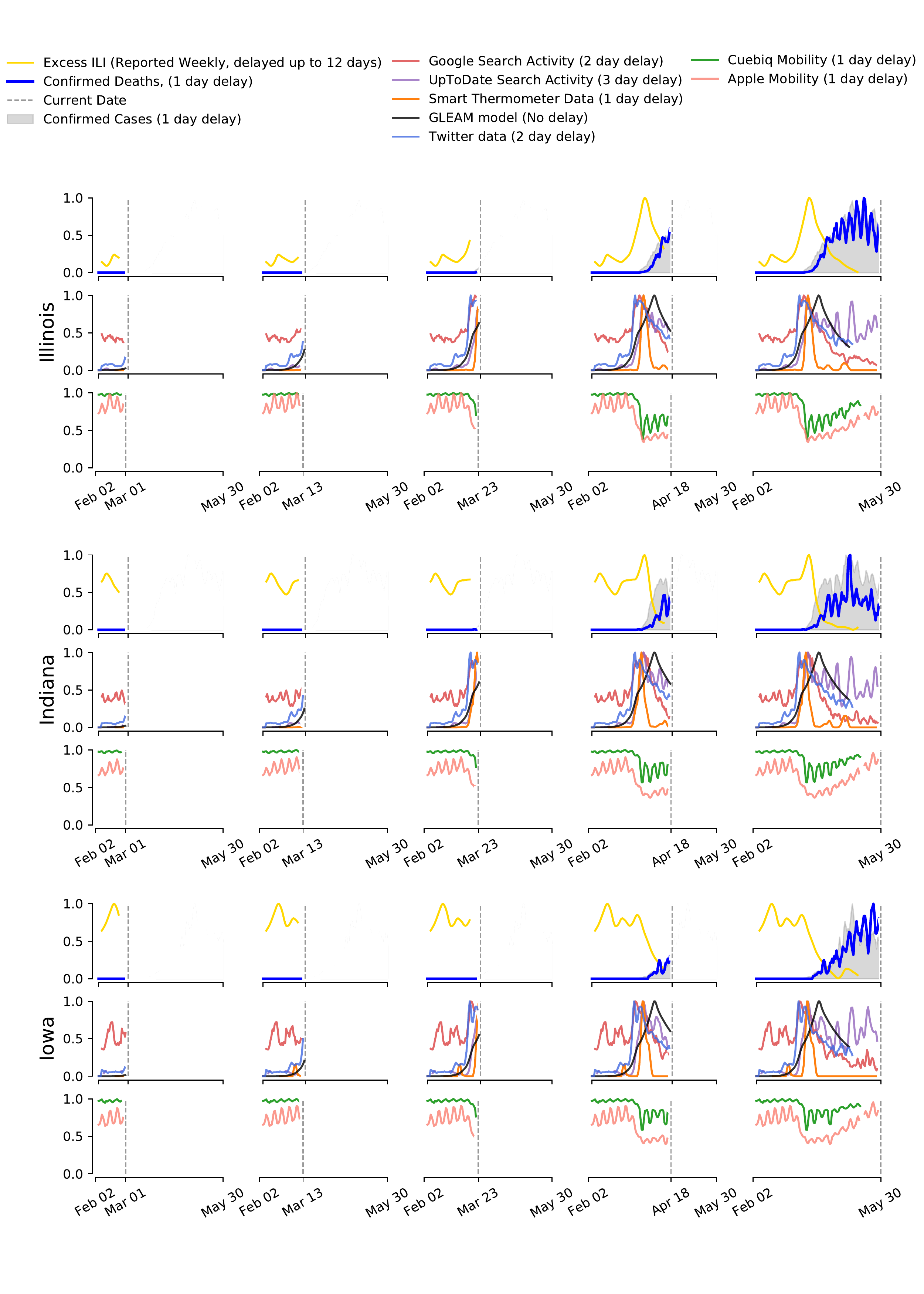}
    \label{fig:EventDetection} \caption{}
    \end{figure}
    
 \begin{figure}
    \centering
    \includegraphics[width=.75\textwidth]{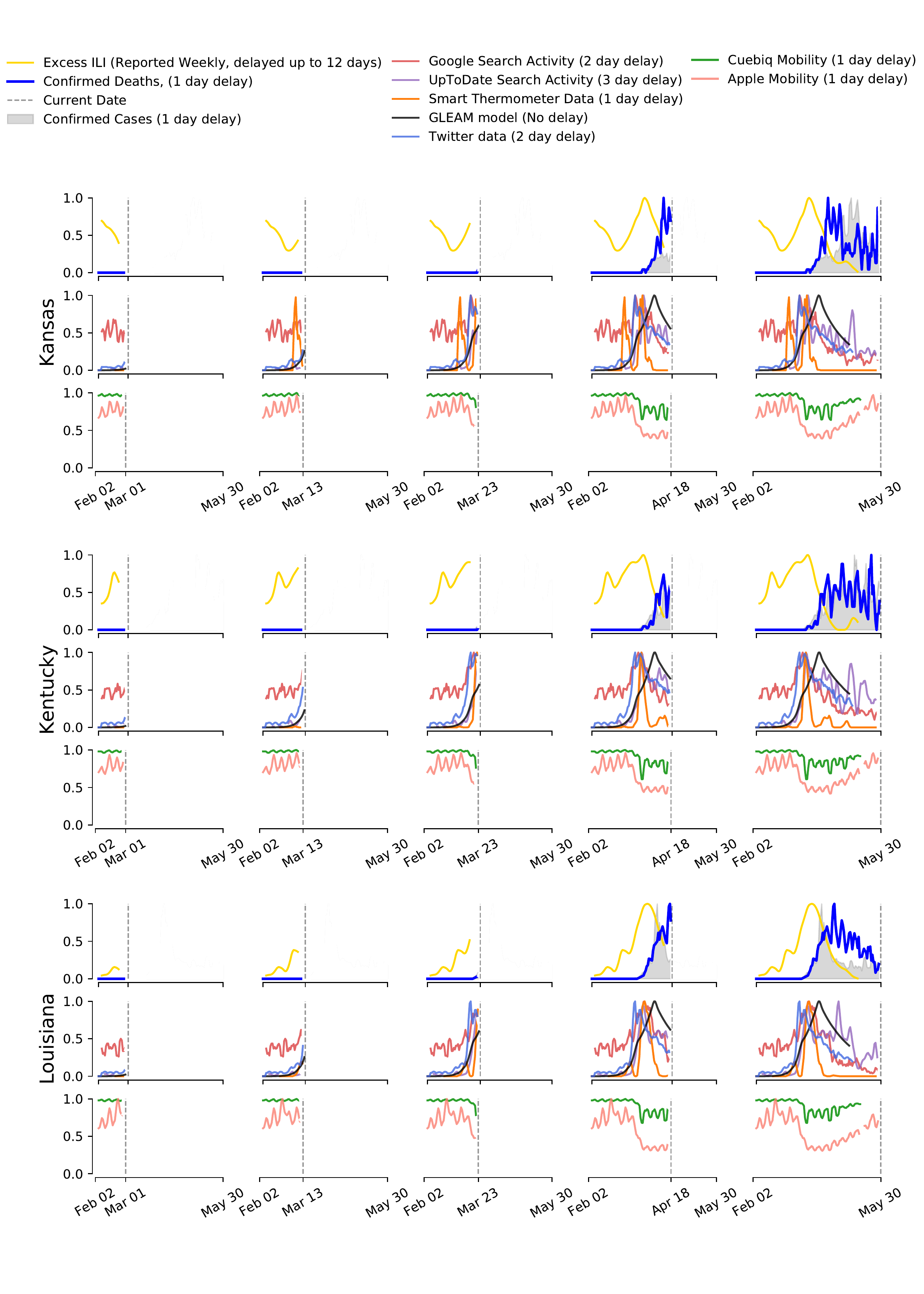}
    \label{fig:EventDetection} \caption{}
    \end{figure}
    
 \begin{figure}
    \centering
    \includegraphics[width=.75\textwidth]{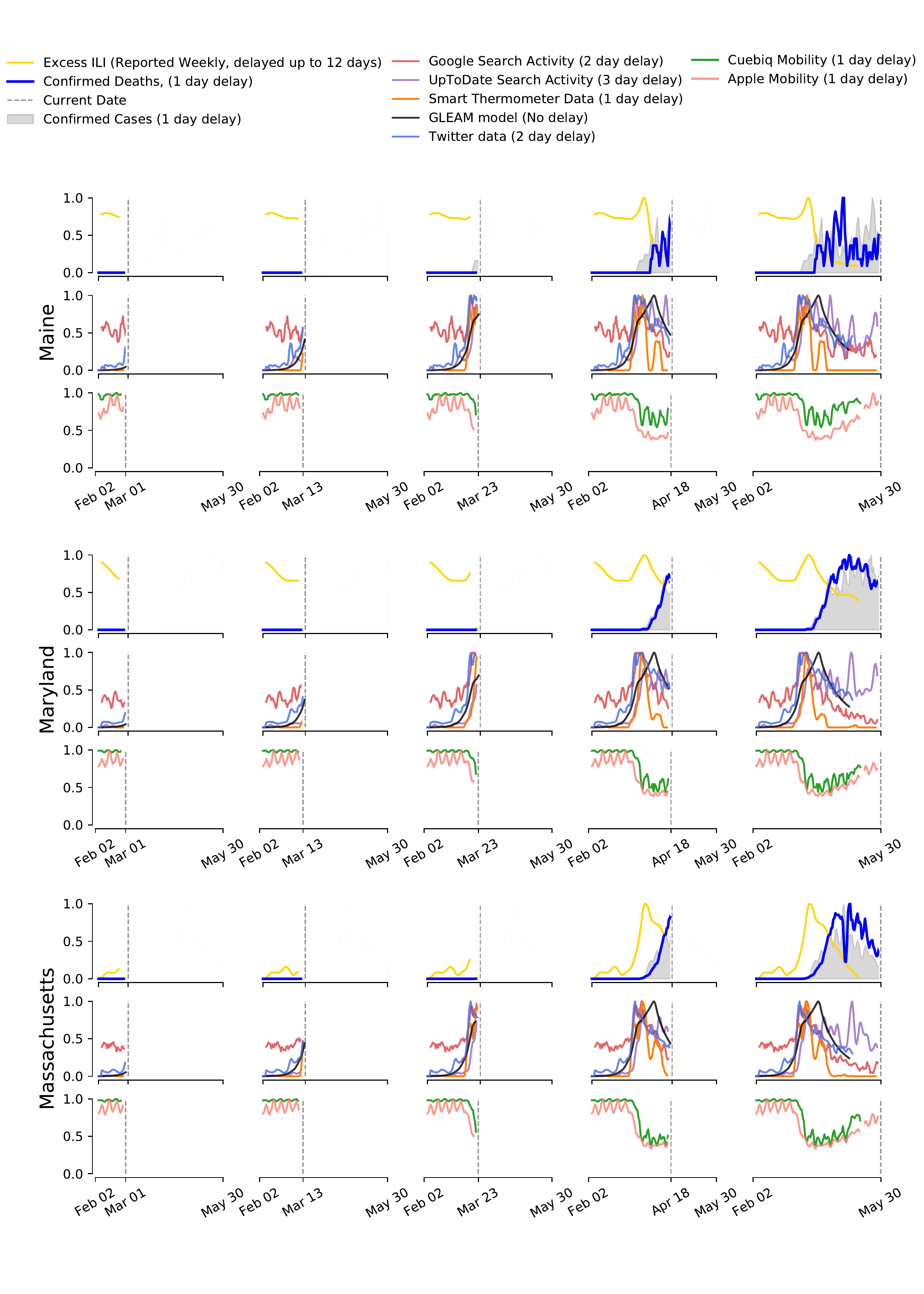}
    \label{fig:EventDetection} \caption{}
    \end{figure}
    
 \begin{figure}
    \centering
    \includegraphics[width=.75\textwidth]{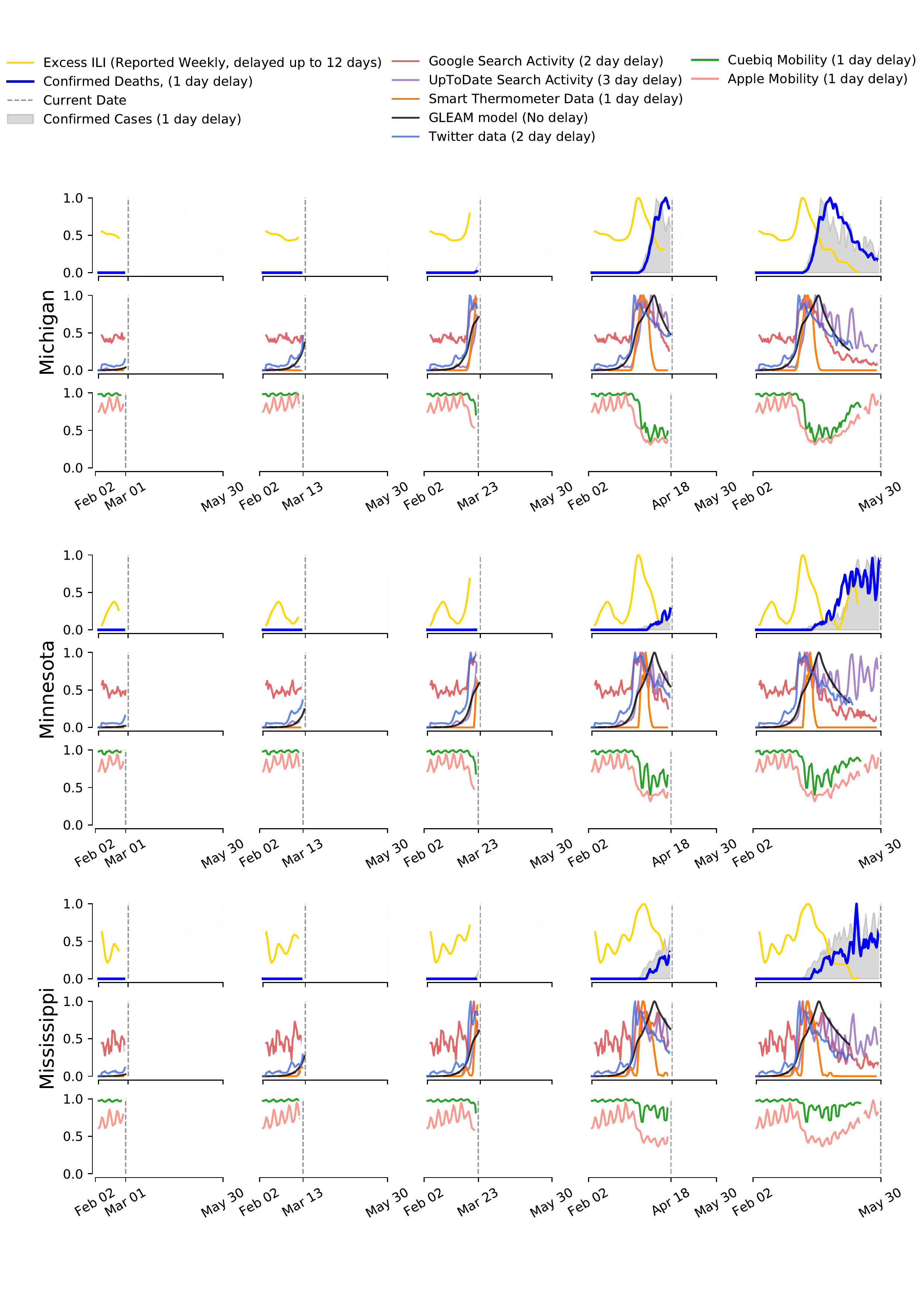}
    \label{fig:EventDetection} \caption{}
    \end{figure}
    
 \begin{figure}
    \centering
    \includegraphics[width=.75\textwidth]{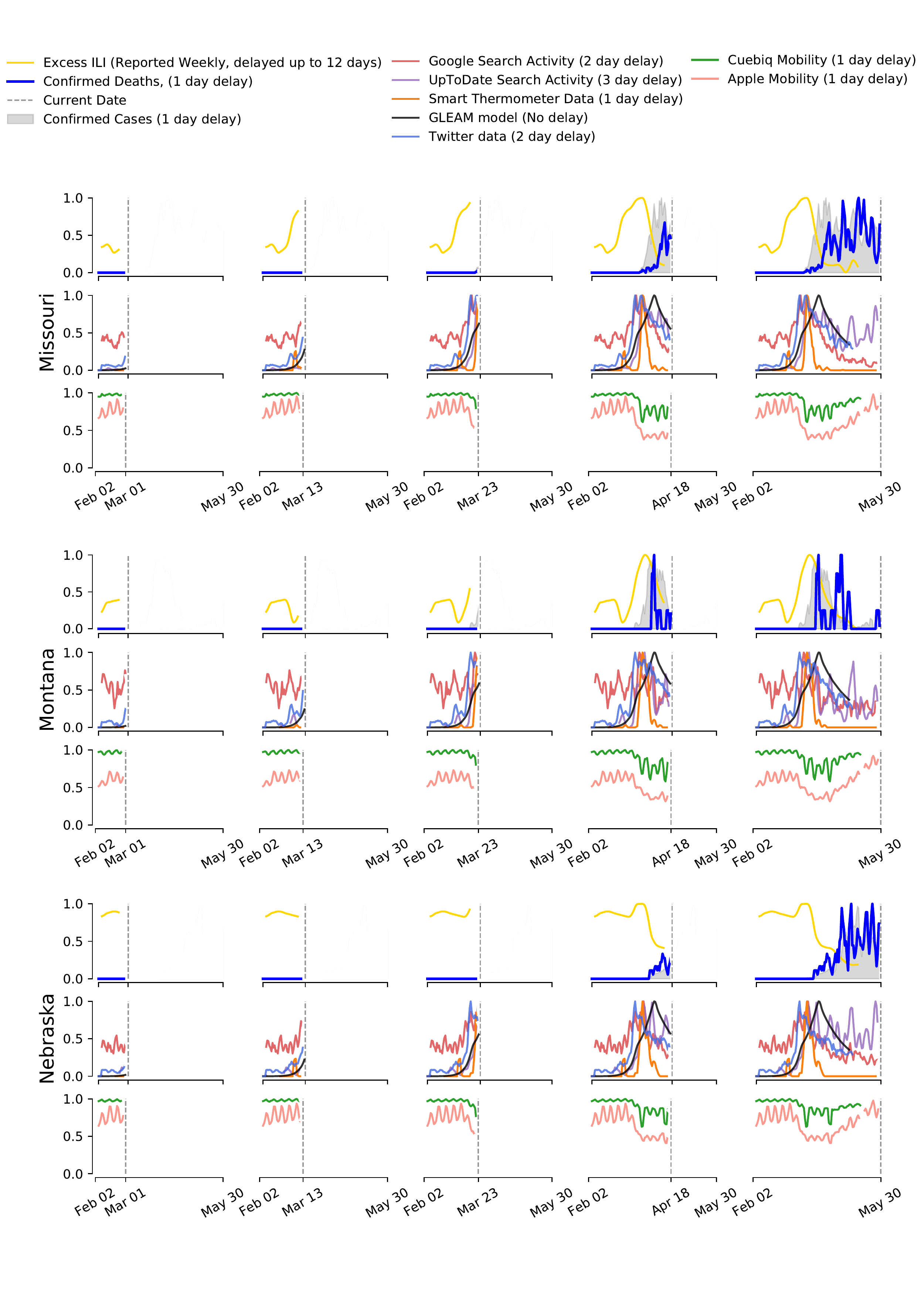}
    \label{fig:EventDetection} \caption{}
    \end{figure}
    
 \begin{figure}
    \centering
    \includegraphics[width=.75\textwidth]{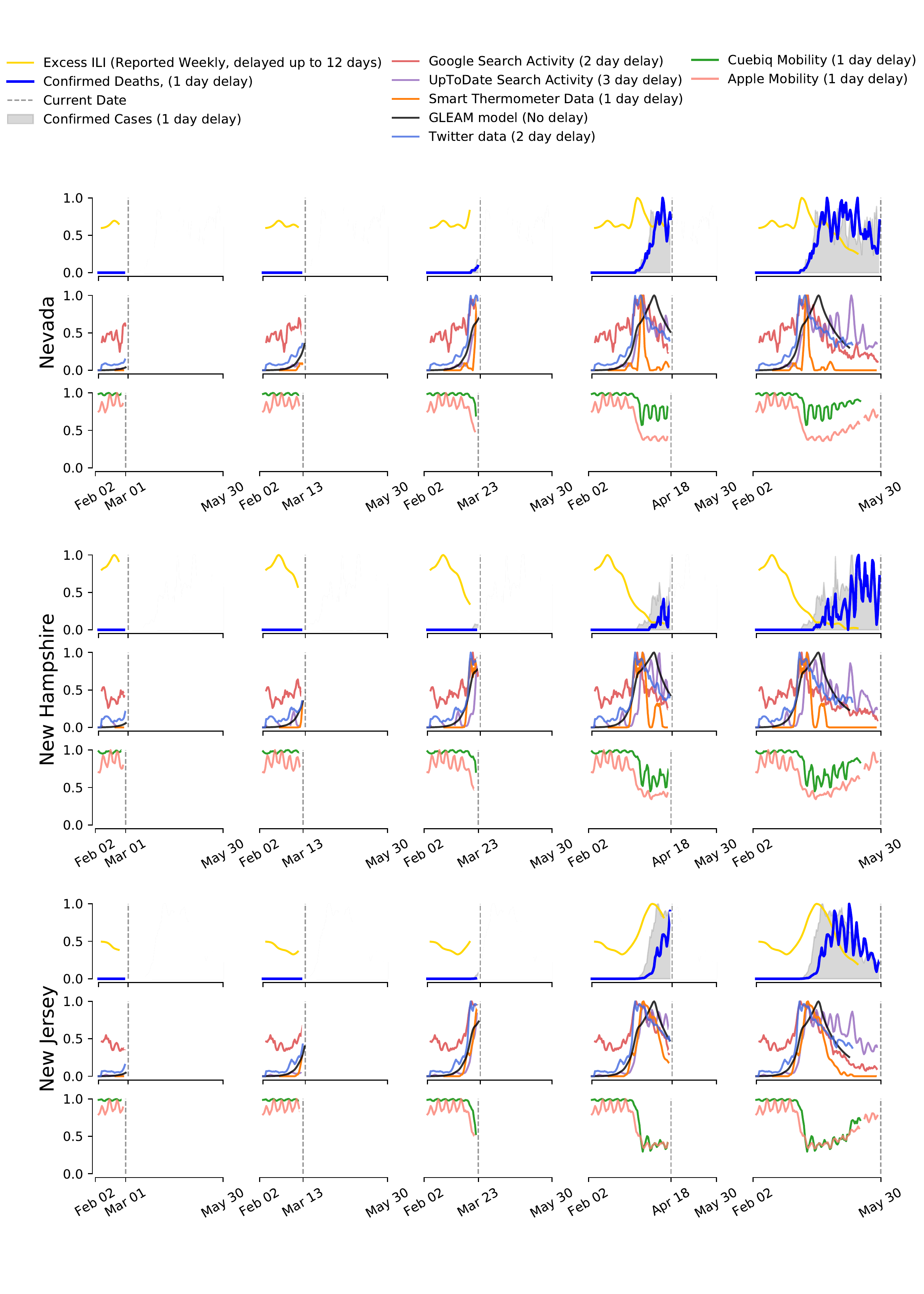}
    \label{fig:EventDetection} \caption{}
    \end{figure}
    
 \begin{figure}
    \centering
    \includegraphics[width=.75\textwidth]{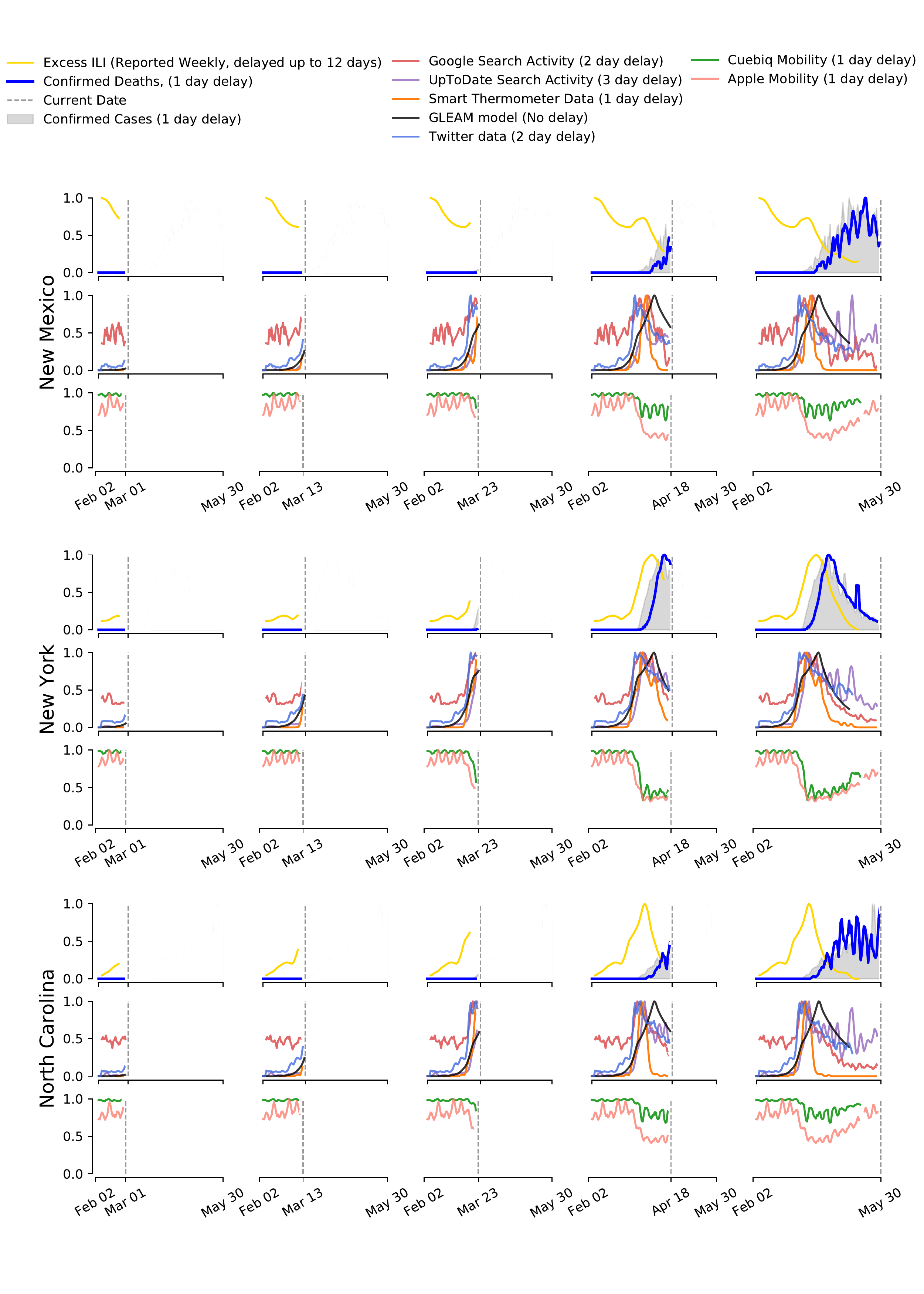}
    \label{fig:EventDetection} \caption{}
    \end{figure}
    
 \begin{figure}
    \centering
    \includegraphics[width=.75\textwidth]{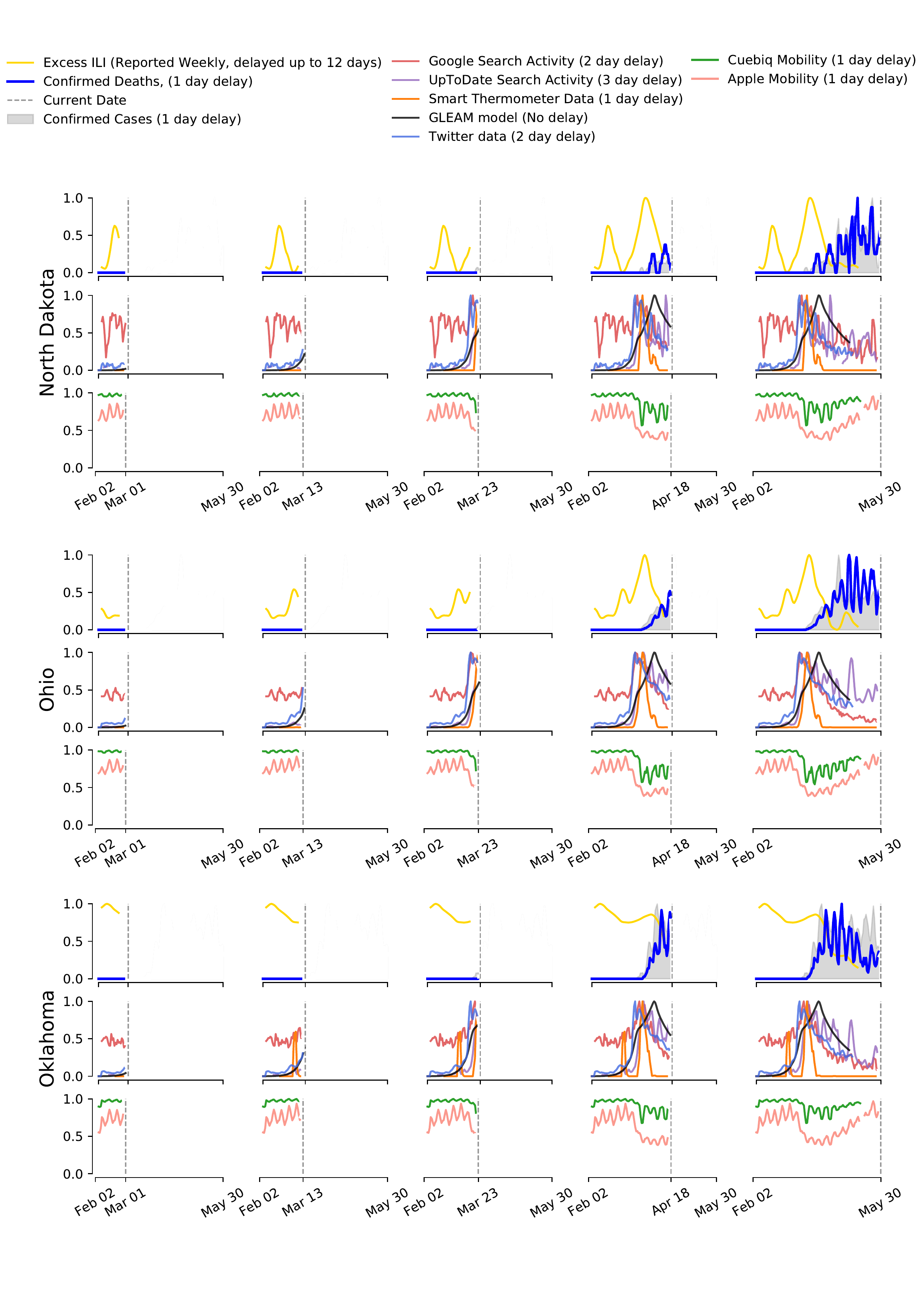}
    \label{fig:EventDetection} \caption{}
    \end{figure}
    
 \begin{figure}
    \centering
    \includegraphics[width=.75\textwidth]{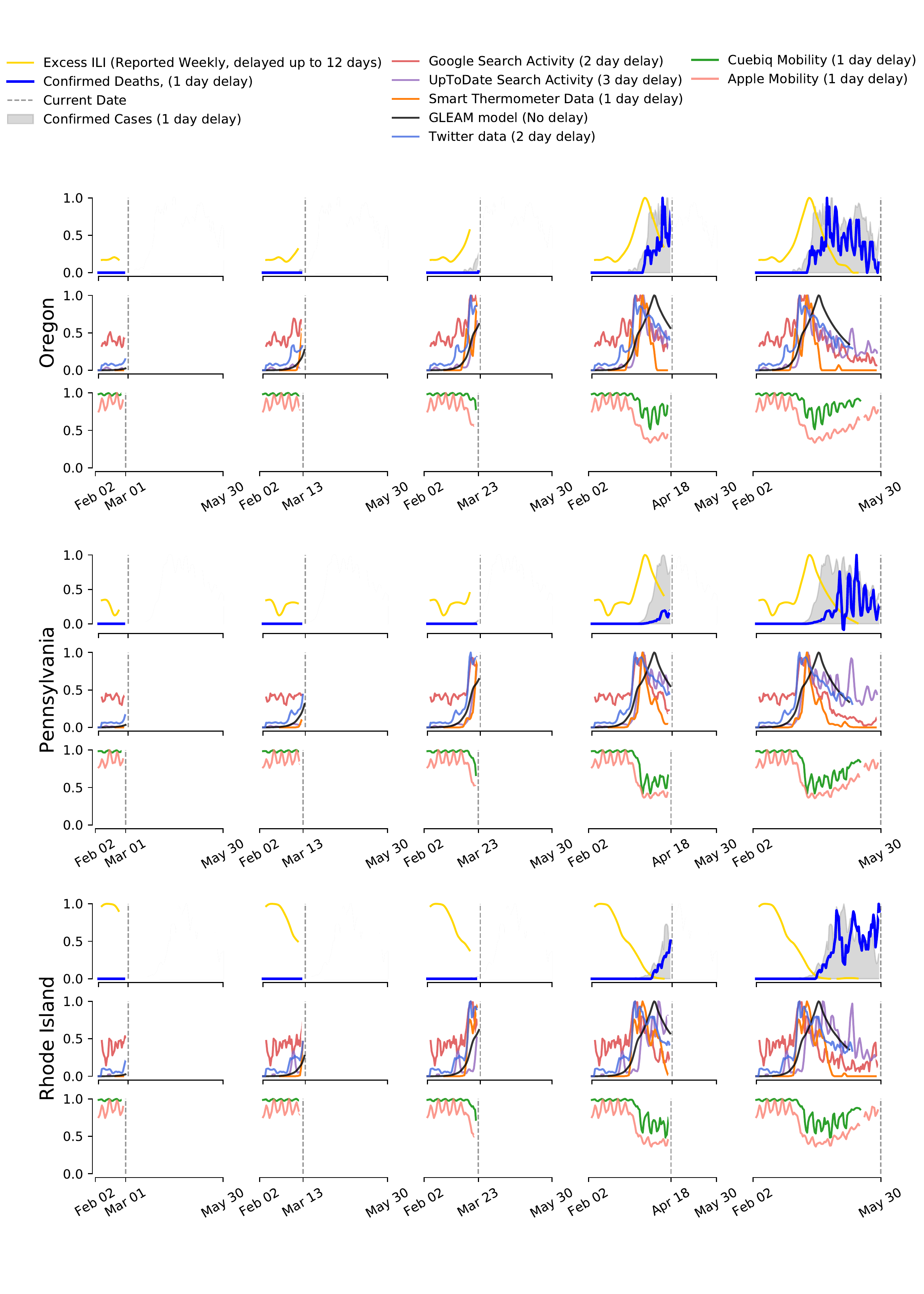}
    \label{fig:EventDetection} \caption{}
    \end{figure}
    
 \begin{figure}
    \centering
    \includegraphics[width=.75\textwidth]{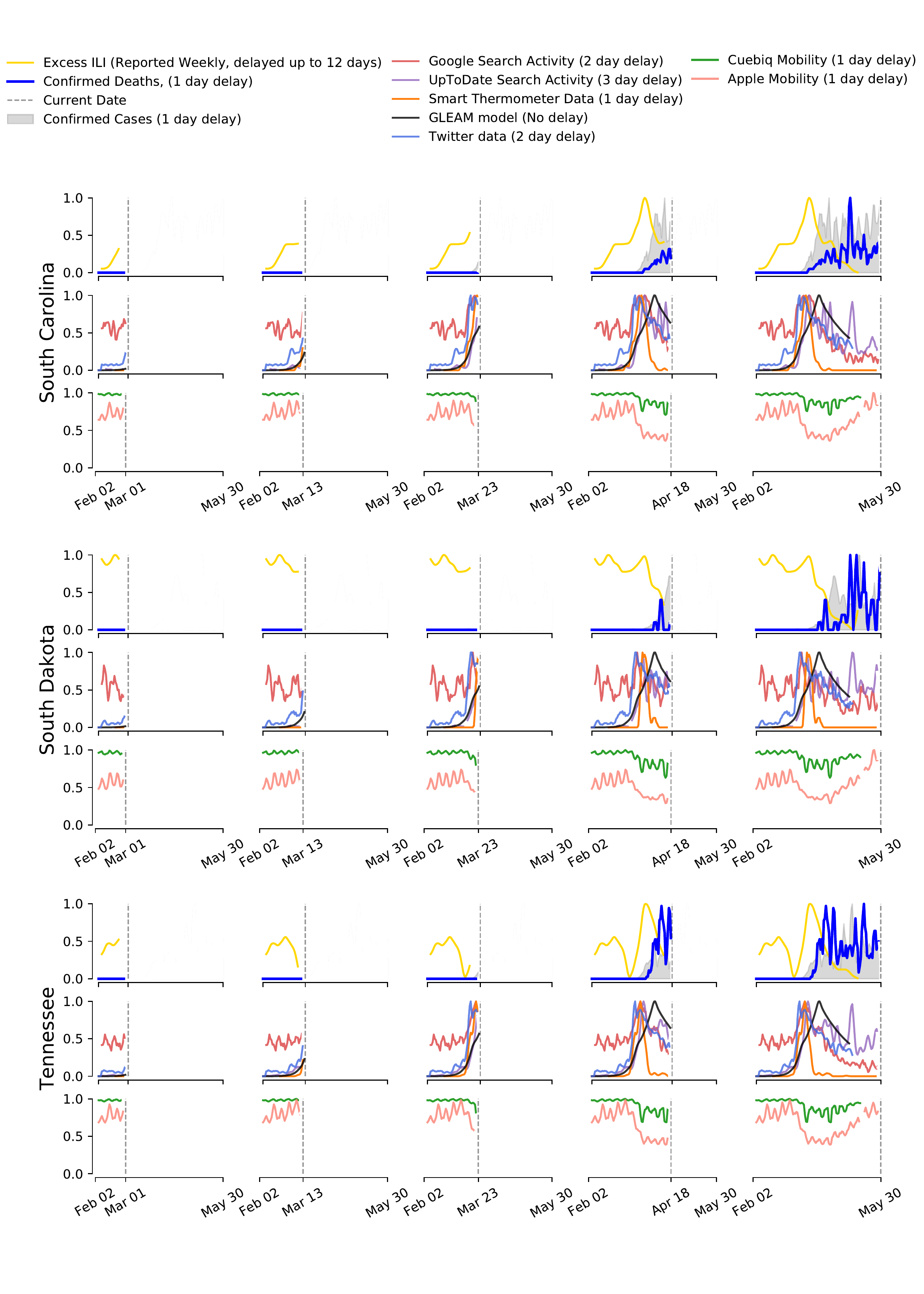}
    \label{fig:EventDetection} \caption{}
    \end{figure}
    
 \begin{figure}
    \centering
    \includegraphics[width=.75\textwidth]{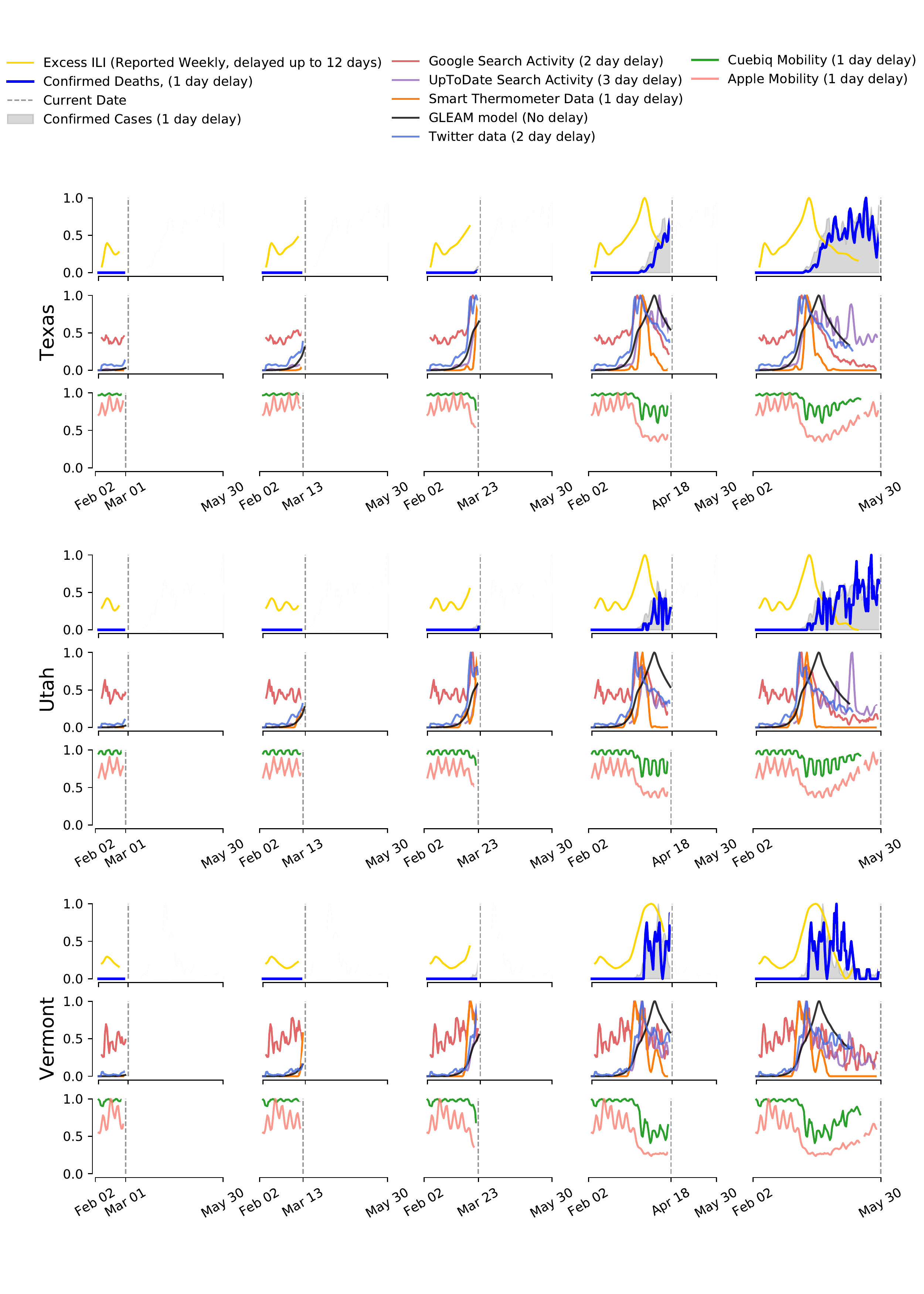}
    \label{fig:EventDetection} \caption{}
    \end{figure}
    
 \begin{figure}
    \centering
    \includegraphics[width=.75\textwidth]{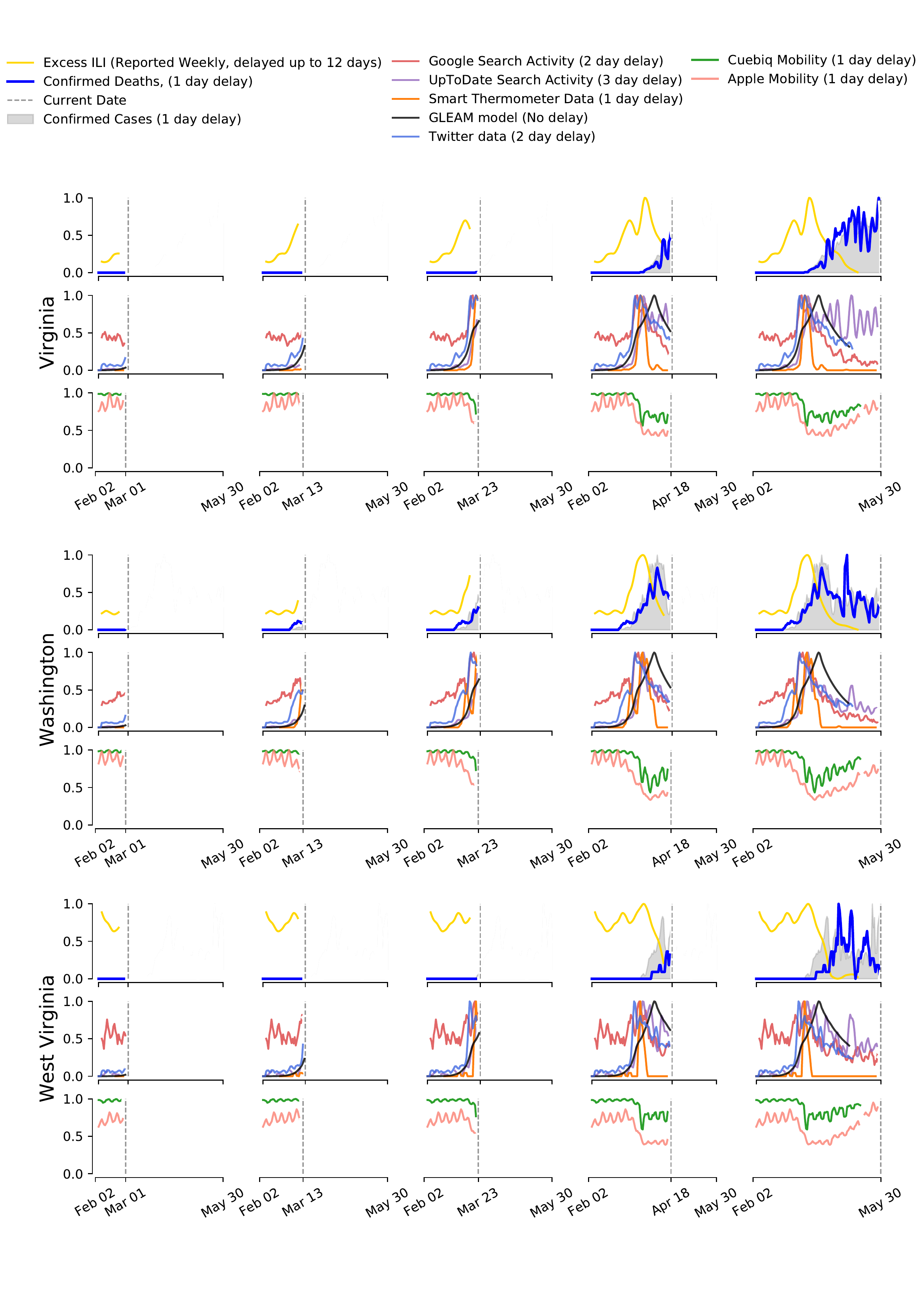}
    \label{fig:EventDetection} \caption{}
    \end{figure}
\FloatBarrier

\begin{center}
\textbf{Early warning system plots for every state}
\end{center}
 \begin{figure}
    \centering
    \includegraphics[width=.75\textwidth]{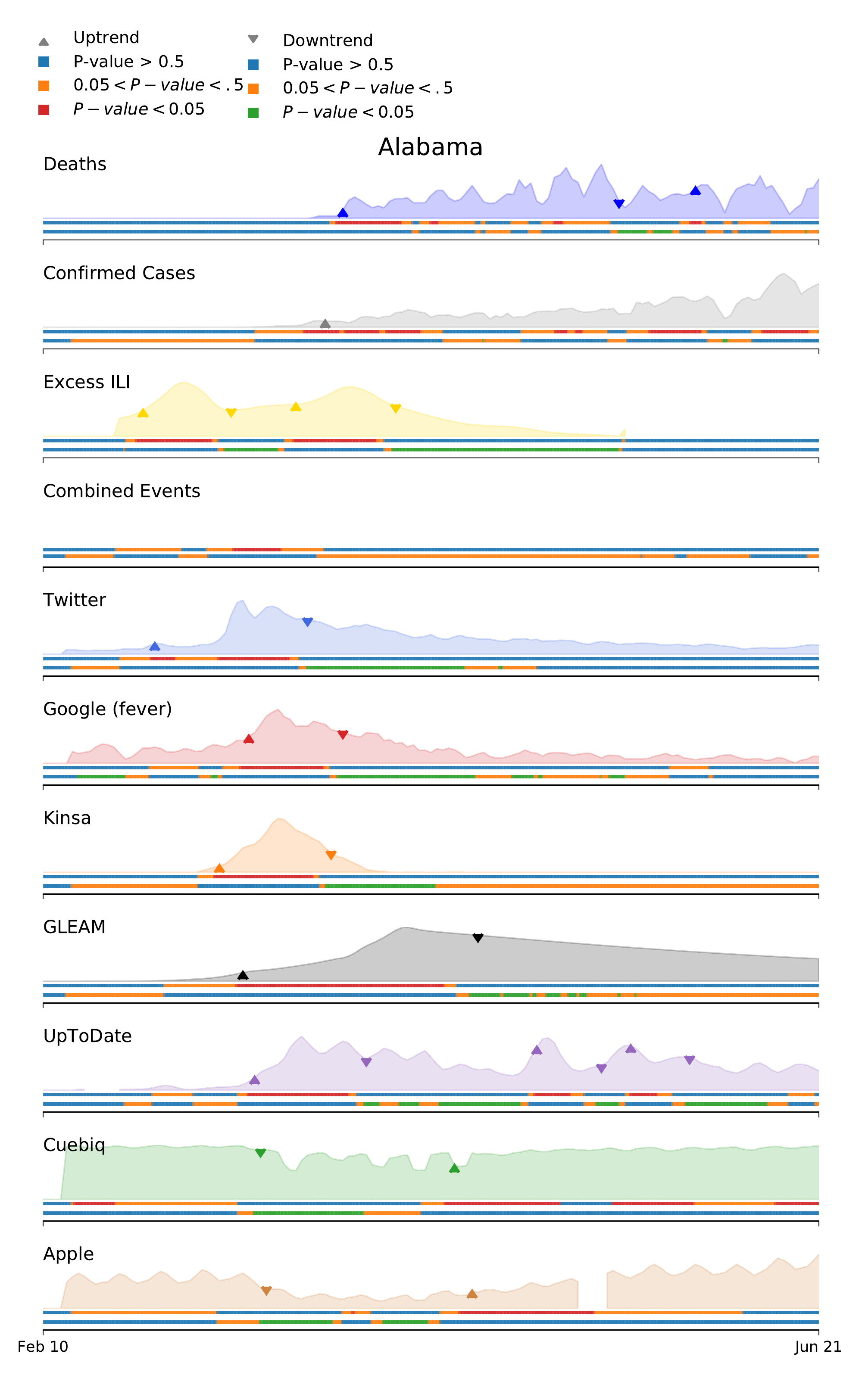}
    \label{fig:Alabama_event}
    \caption{}
\end{figure} 
 \begin{figure}
    \centering
    \includegraphics[width=.75\textwidth]{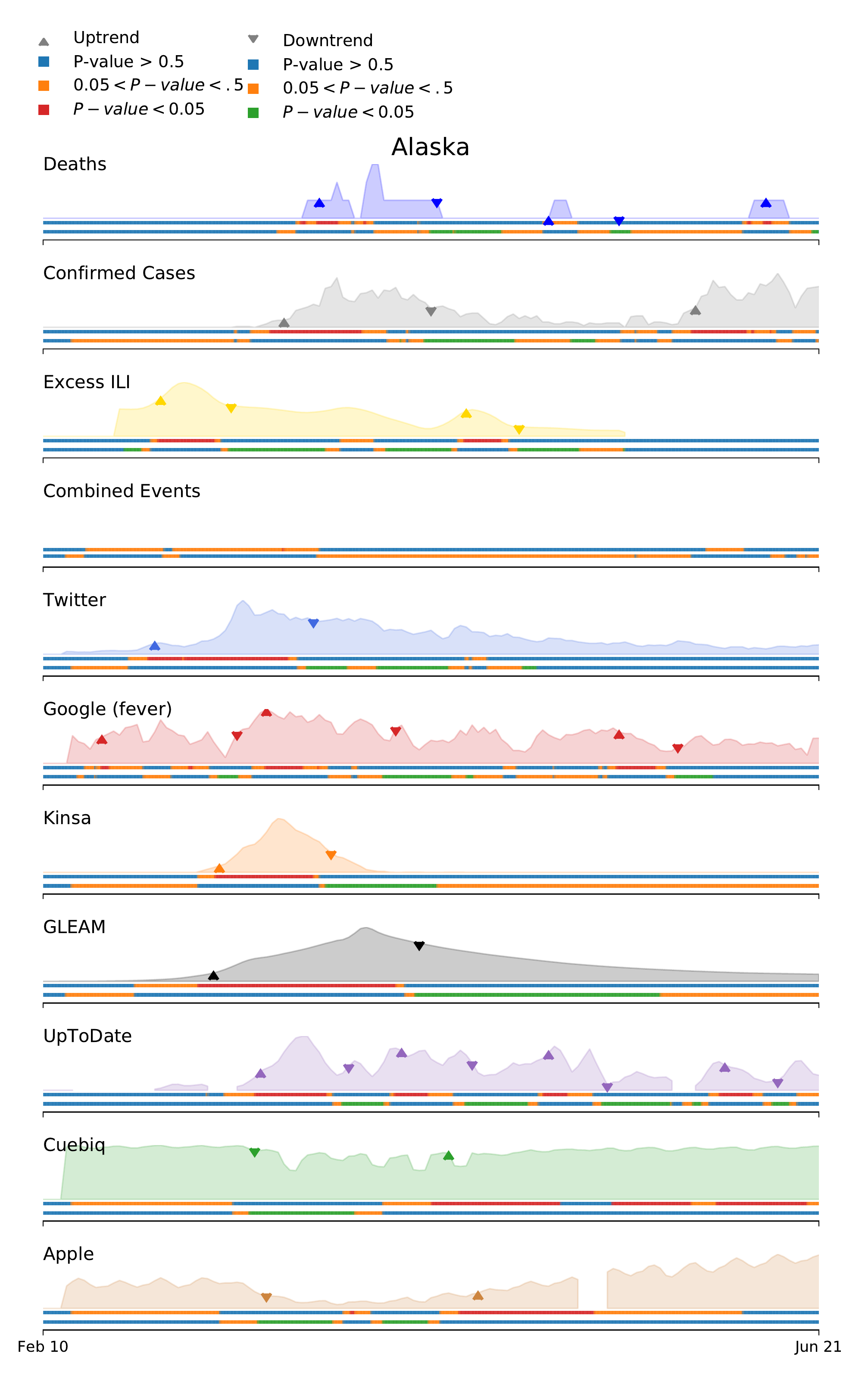}
    \label{fig:Alaska_event}
    \caption{}
\end{figure} 
 \begin{figure}
    \centering
    \includegraphics[width=.75\textwidth]{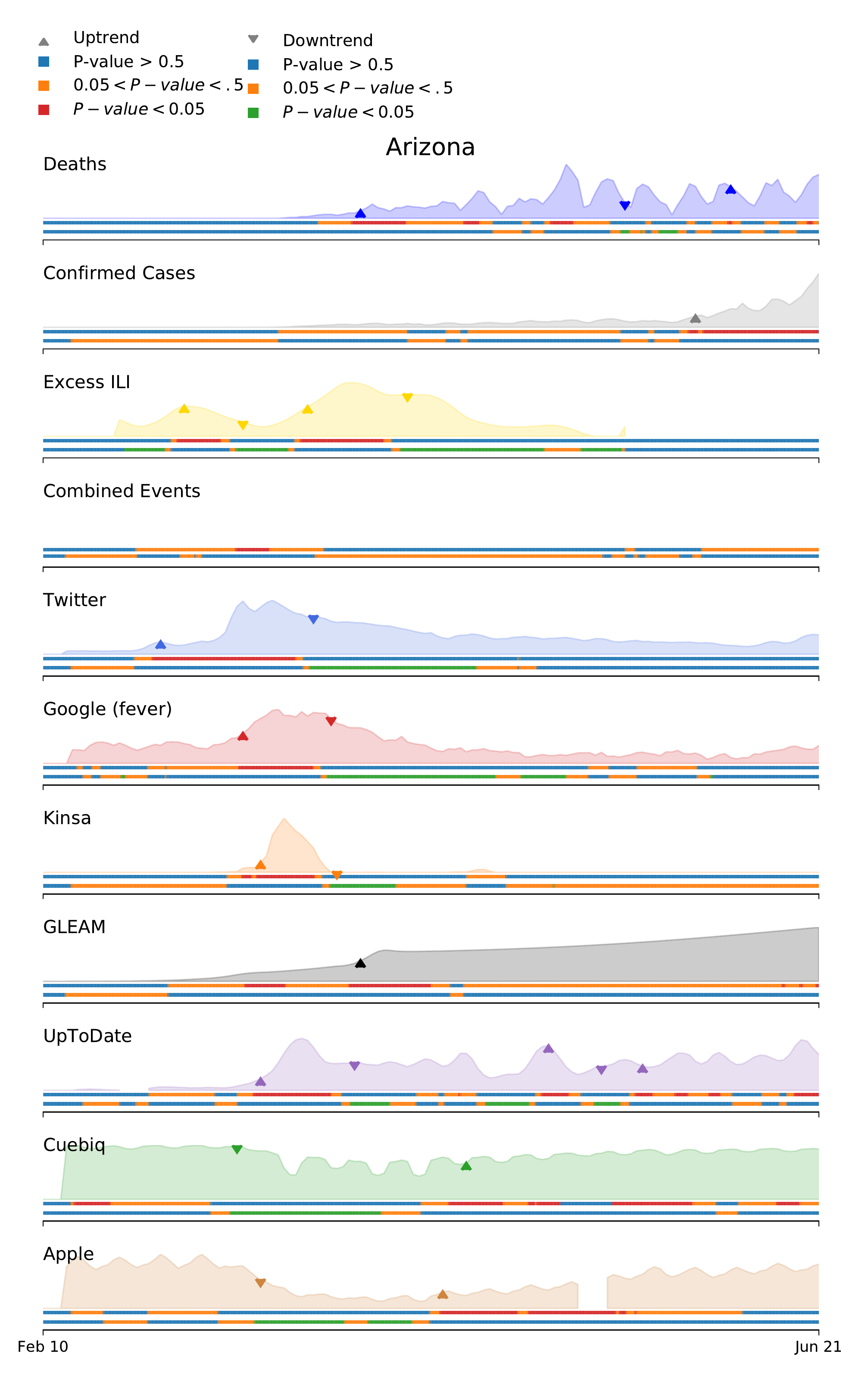}
    \label{fig:Arizona_event}
    \caption{}
\end{figure} 
 \begin{figure}
    \centering
    \includegraphics[width=.75\textwidth]{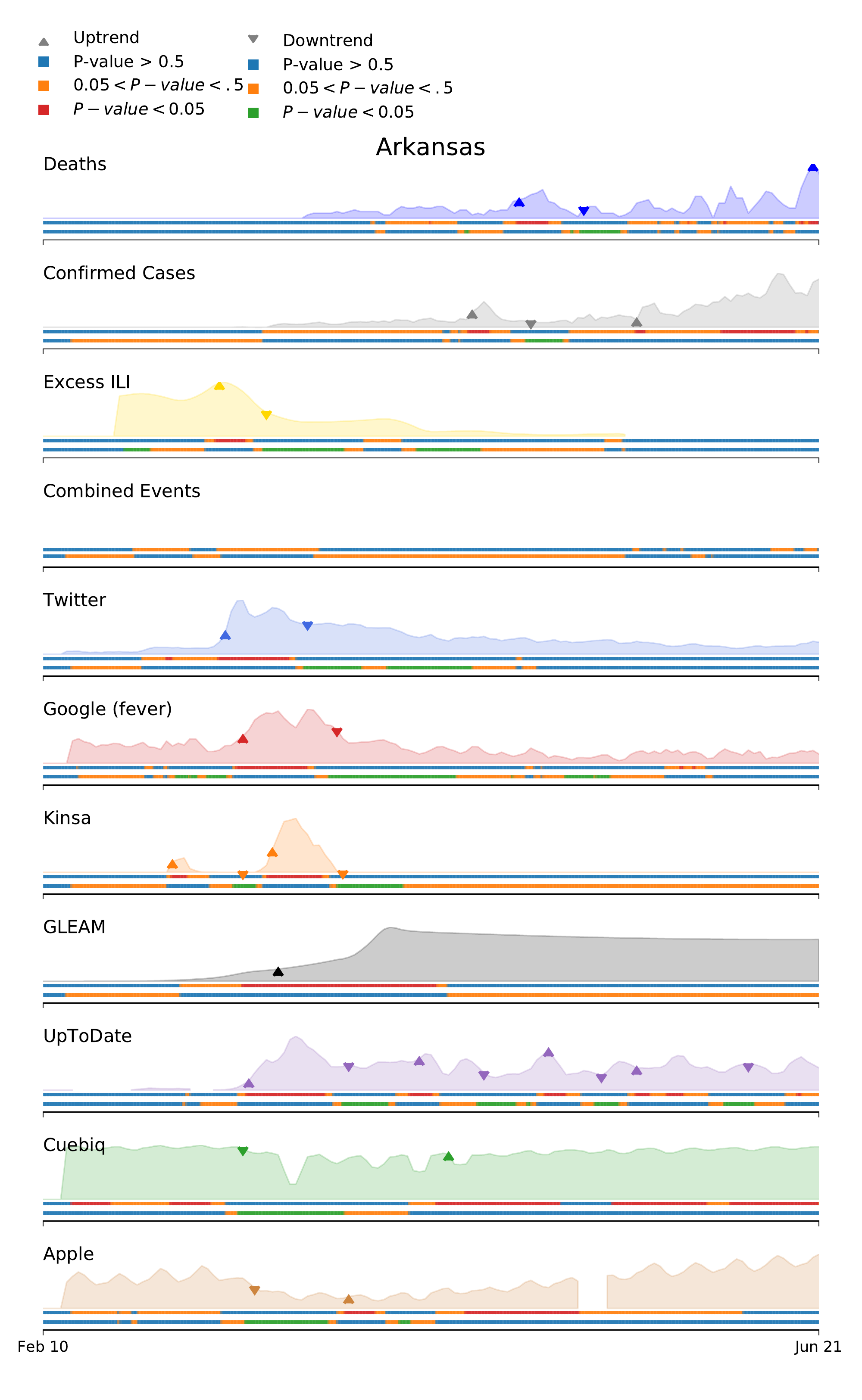}
    \label{fig:Arkansas_event}
    \caption{}
\end{figure} 
 \begin{figure}
    \centering
    \includegraphics[width=.75\textwidth]{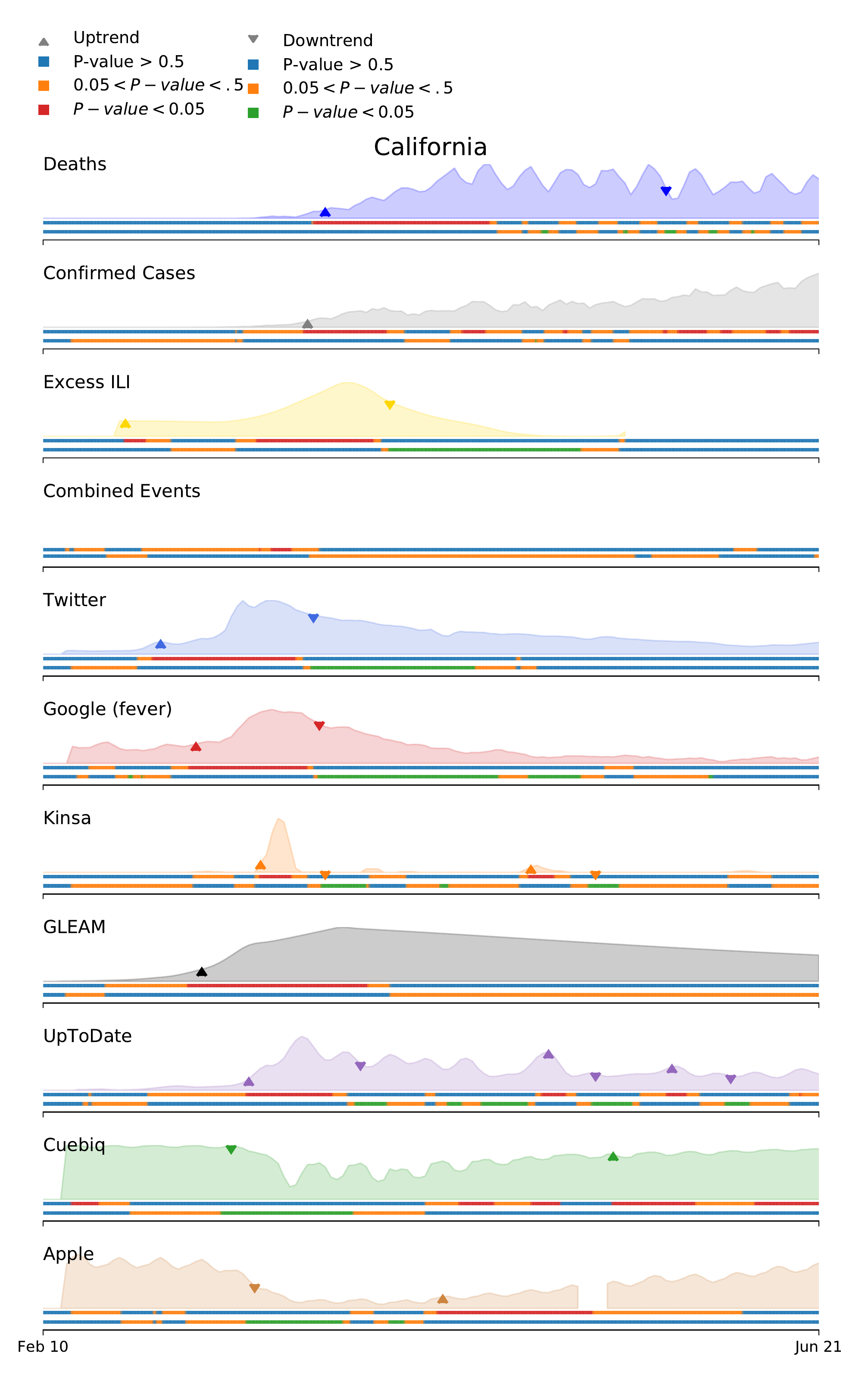}
    \label{fig:California_event}
    \caption{}
\end{figure} 
 \begin{figure}
    \centering
    \includegraphics[width=.75\textwidth]{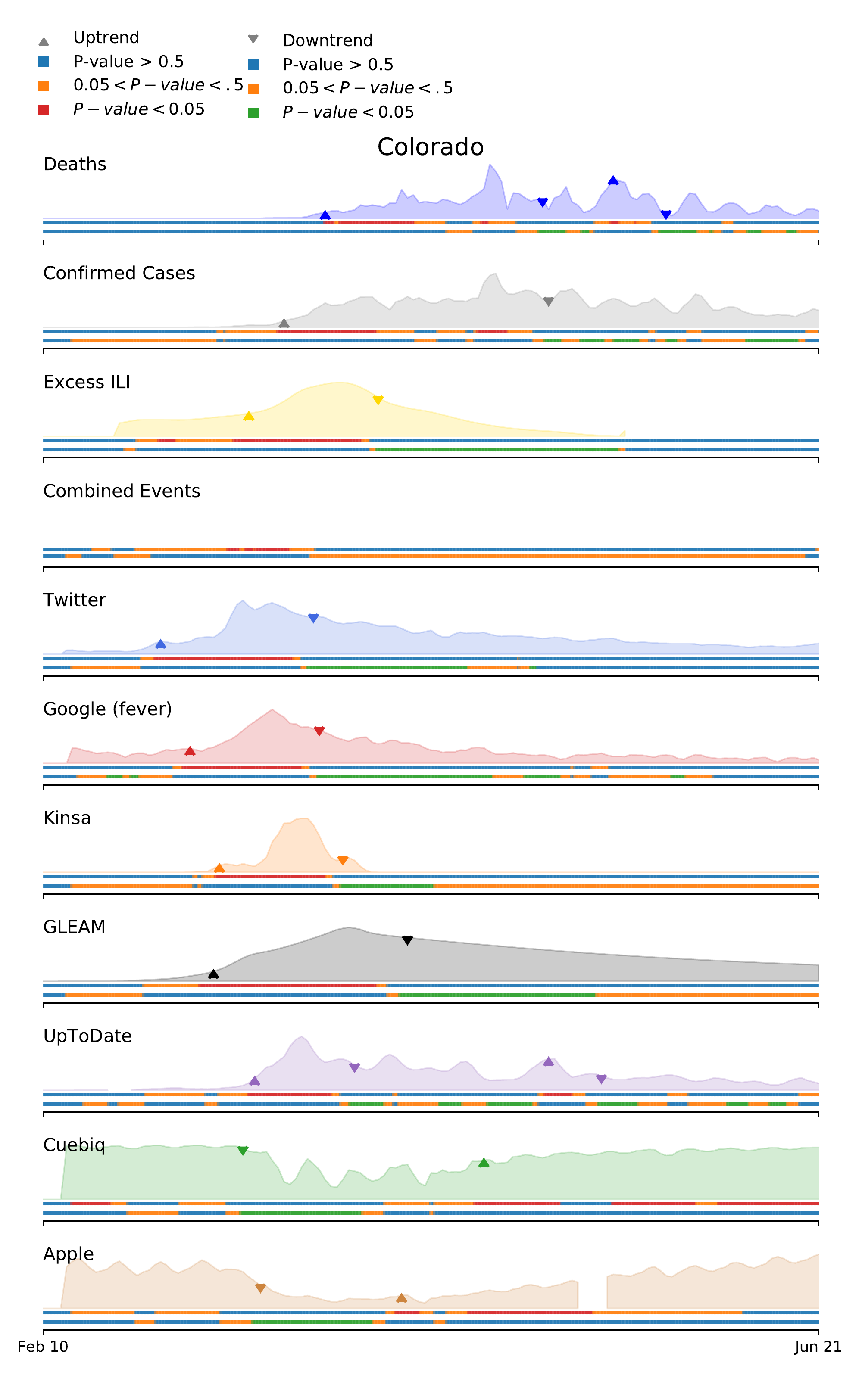}
    \label{fig:Colorado_event}
    \caption{}
\end{figure} 
 \begin{figure}
    \centering
    \includegraphics[width=.75\textwidth]{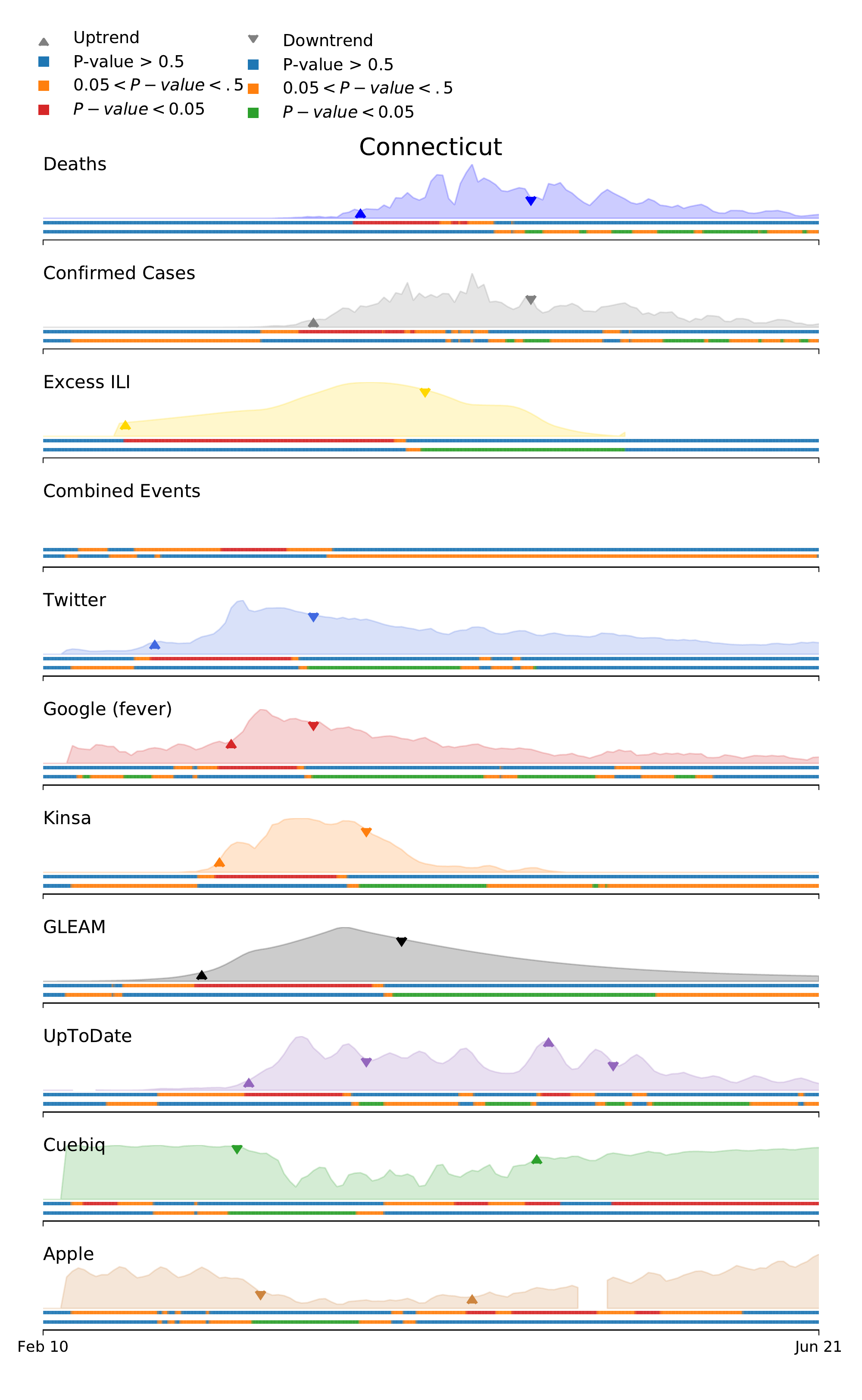}
    \label{fig:Connecticut_event}
    \caption{}
\end{figure} 
 \begin{figure}
    \centering
    \includegraphics[width=.75\textwidth]{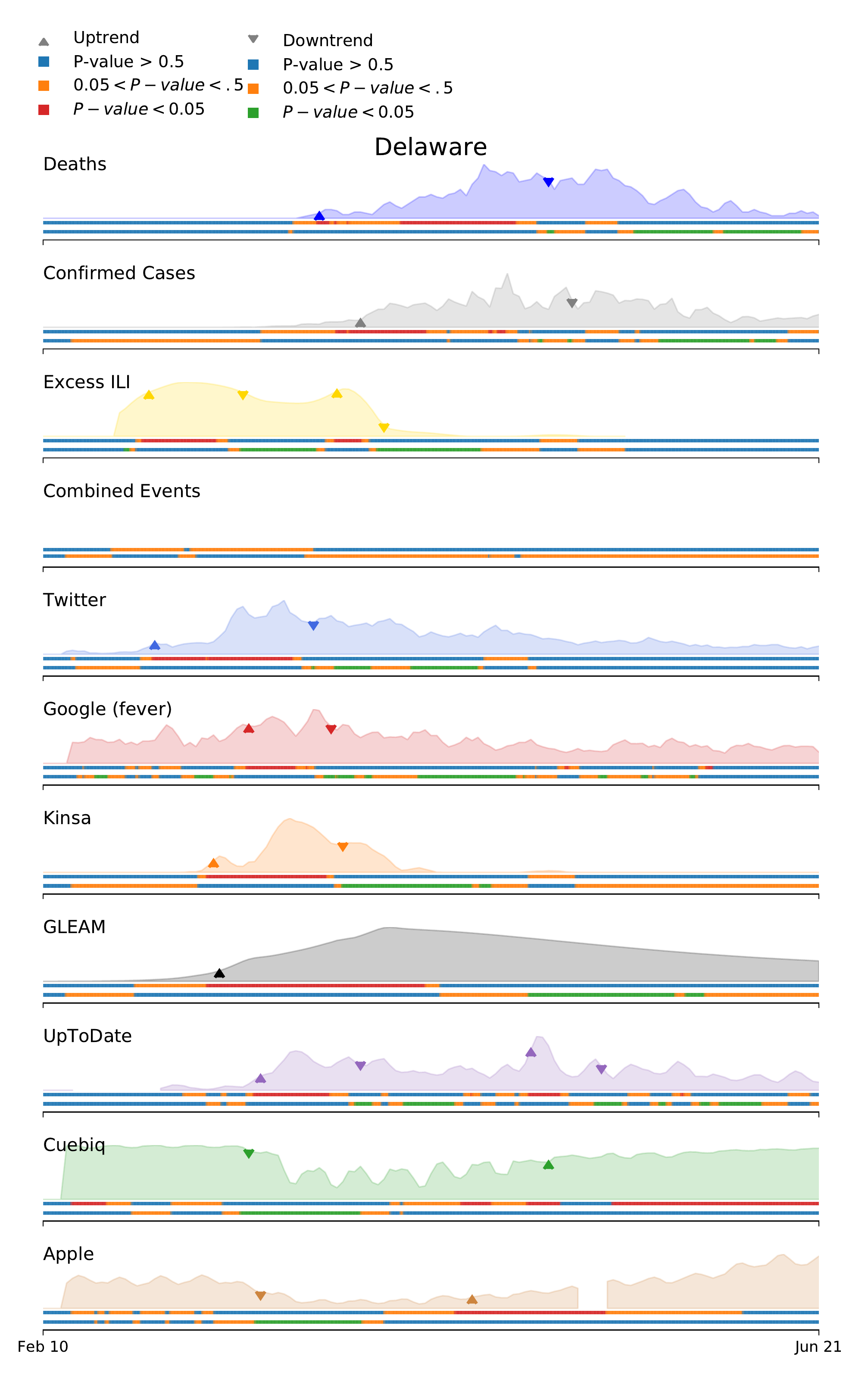}
    \label{fig:Delaware_event}
    \caption{}
\end{figure} 
 \begin{figure}
    \centering
    \includegraphics[width=.75\textwidth]{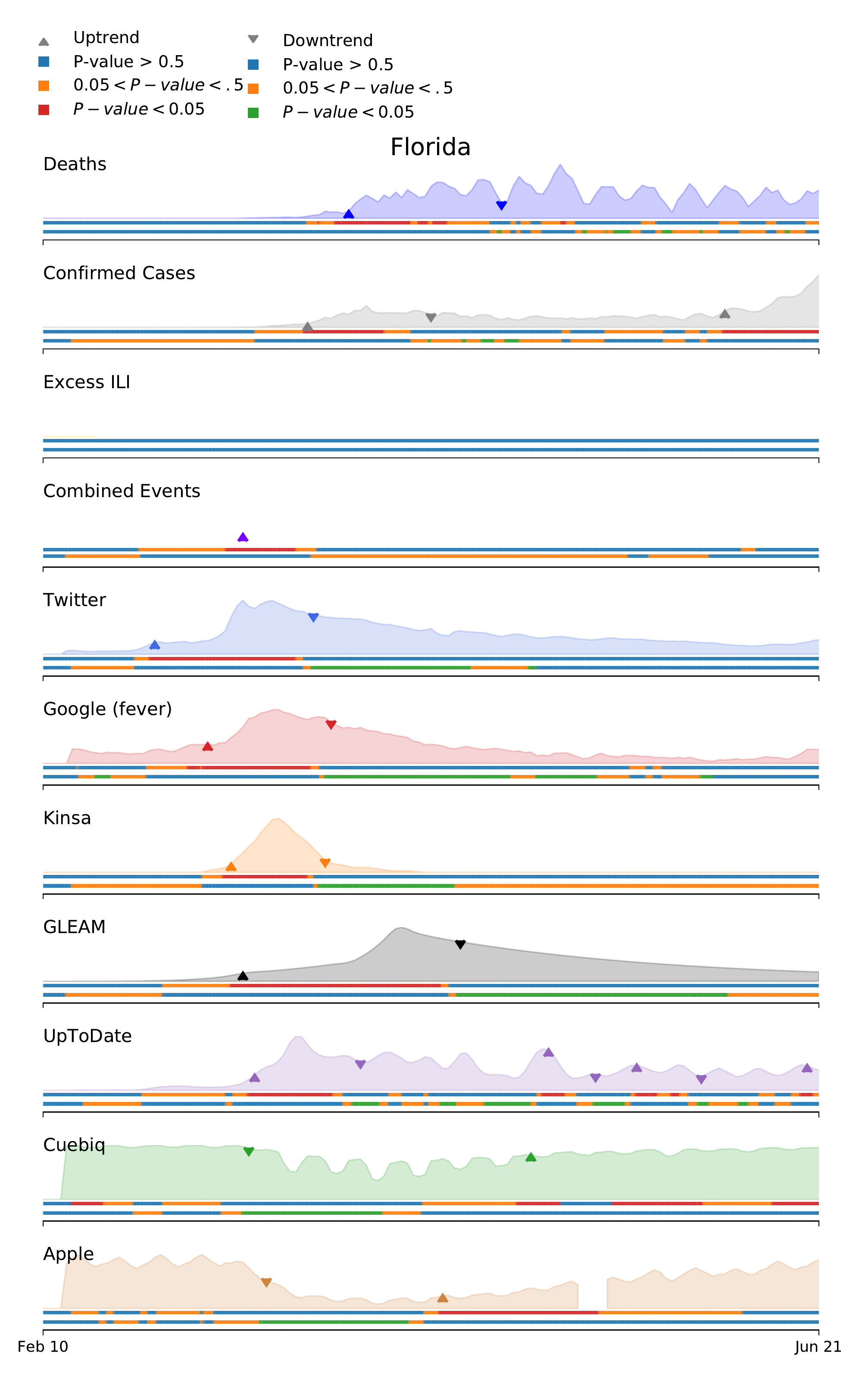}
    \label{fig:Florida_event}
    \caption{}
\end{figure} 
 \begin{figure}
    \centering
    \includegraphics[width=.75\textwidth]{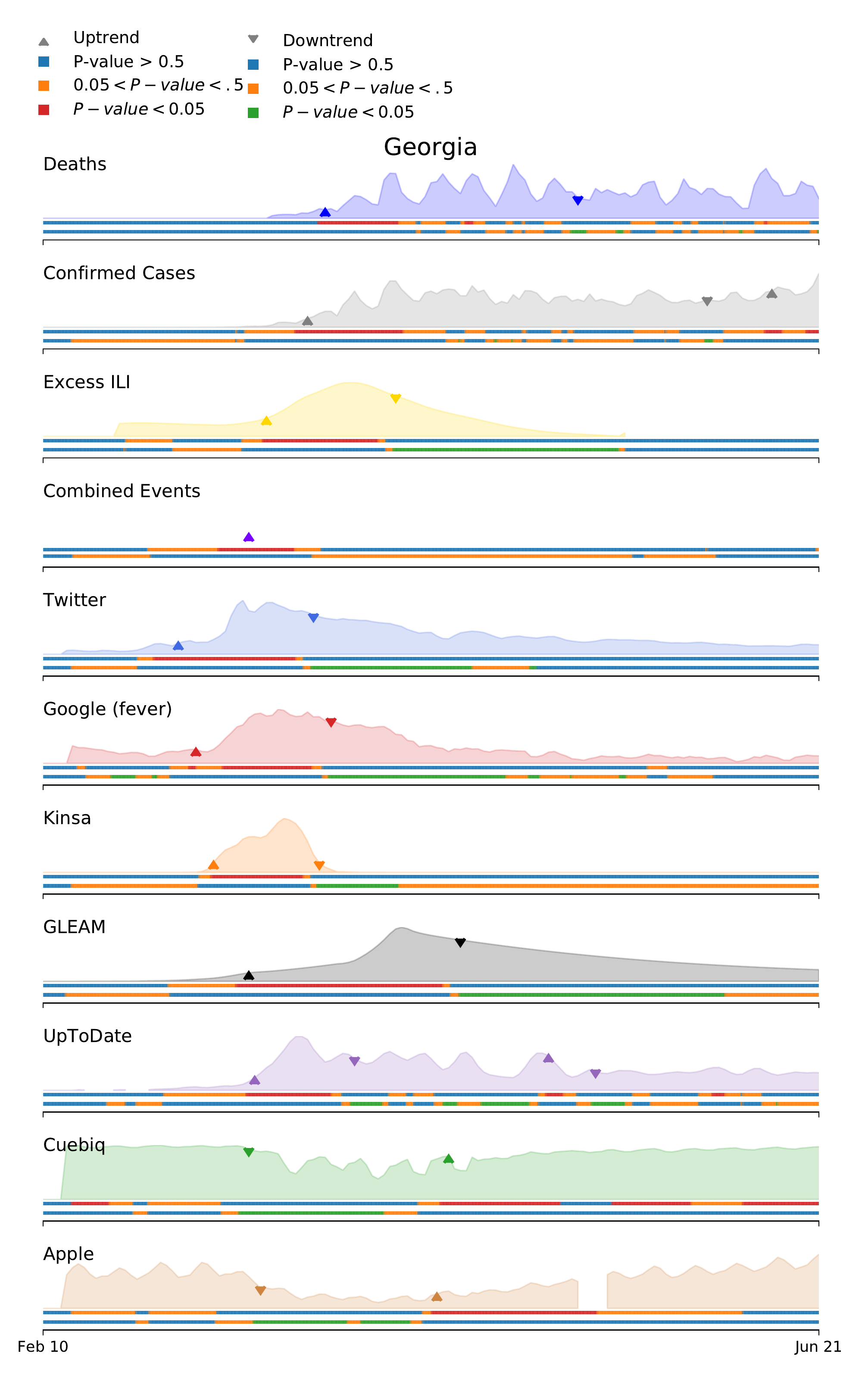}
    \label{fig:Georgia_event}
    \caption{}
\end{figure} 
 \begin{figure}
    \centering
    \includegraphics[width=.75\textwidth]{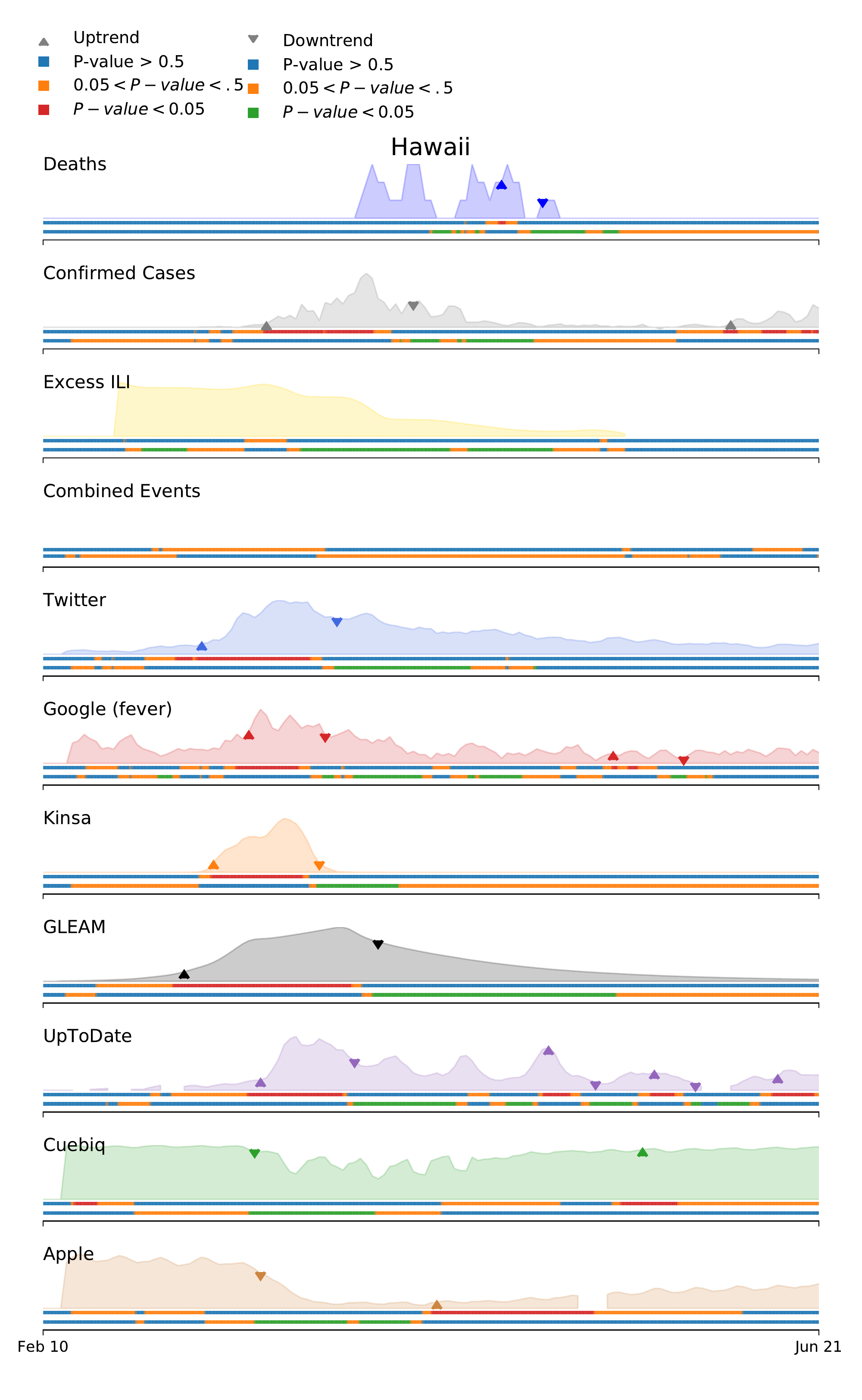}
    \label{fig:Hawaii_event}
    \caption{}
\end{figure} 
 \begin{figure}
    \centering
    \includegraphics[width=.75\textwidth]{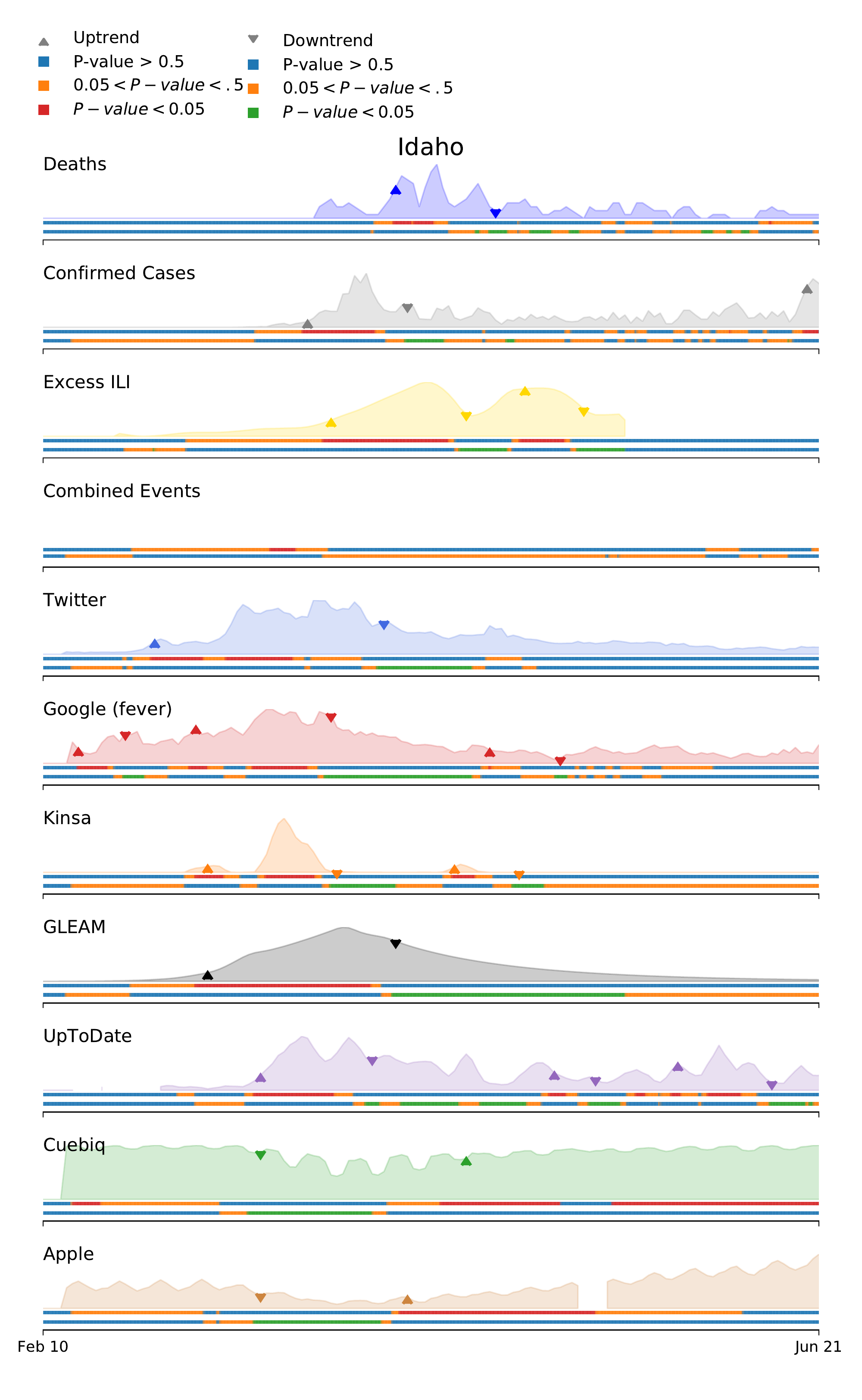}
    \label{fig:Idaho_event}
    \caption{}
\end{figure} 
 \begin{figure}
    \centering
    \includegraphics[width=.75\textwidth]{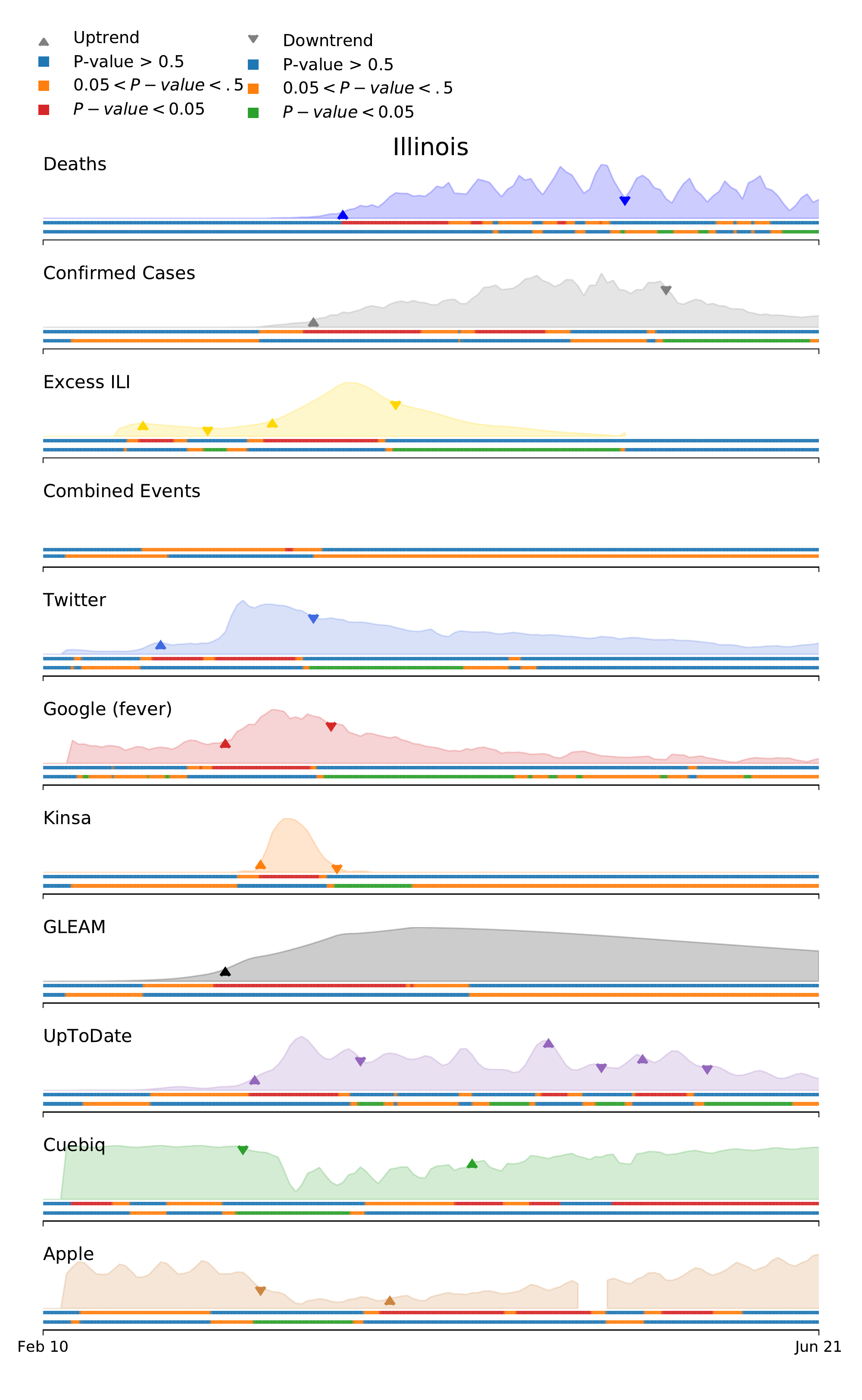}
    \label{fig:Illinois_event}
    \caption{}
\end{figure} 
 \begin{figure}
    \centering
    \includegraphics[width=.75\textwidth]{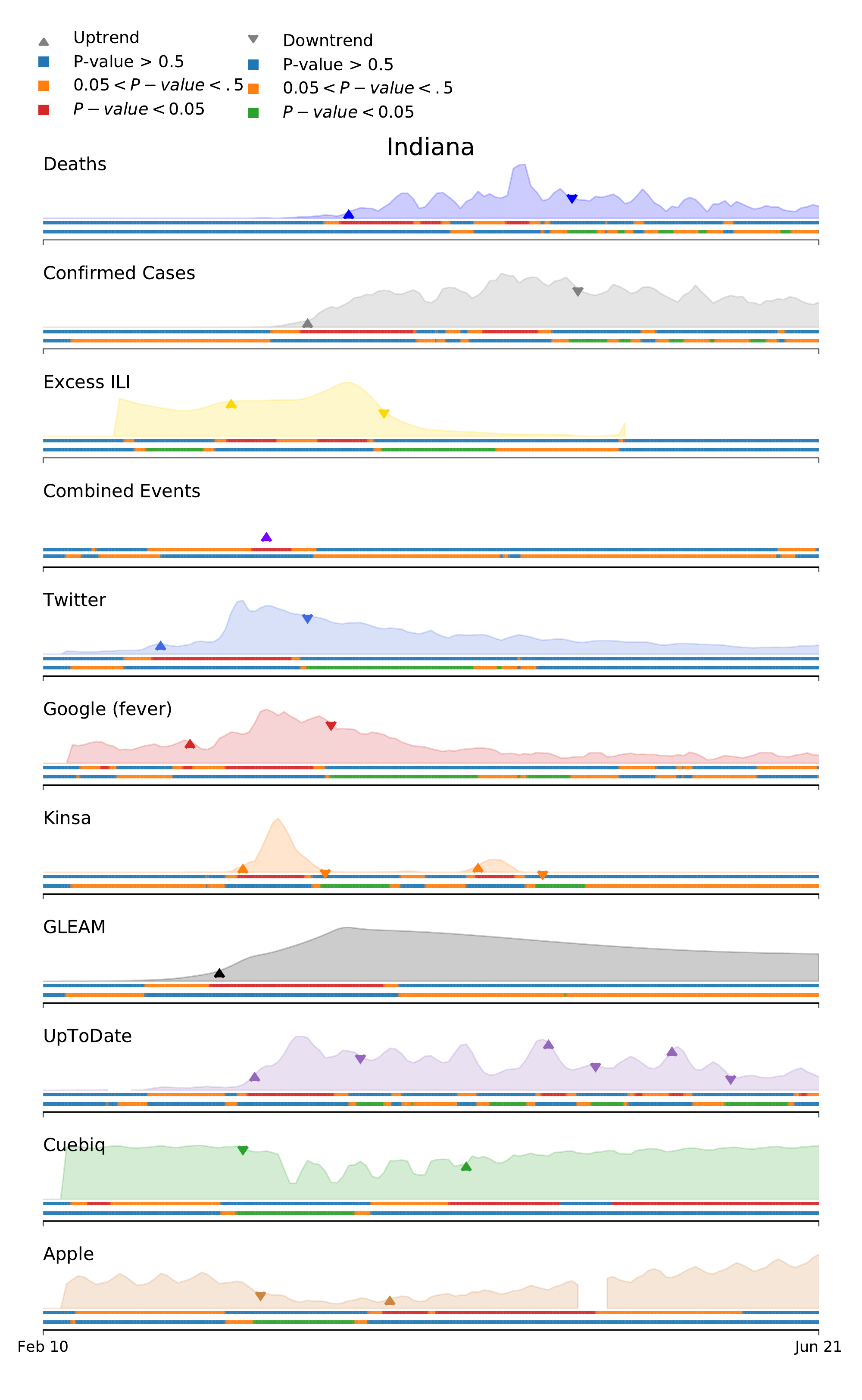}
    \label{fig:Indiana_event}
    \caption{}
\end{figure} 
 \begin{figure}
    \centering
    \includegraphics[width=.75\textwidth]{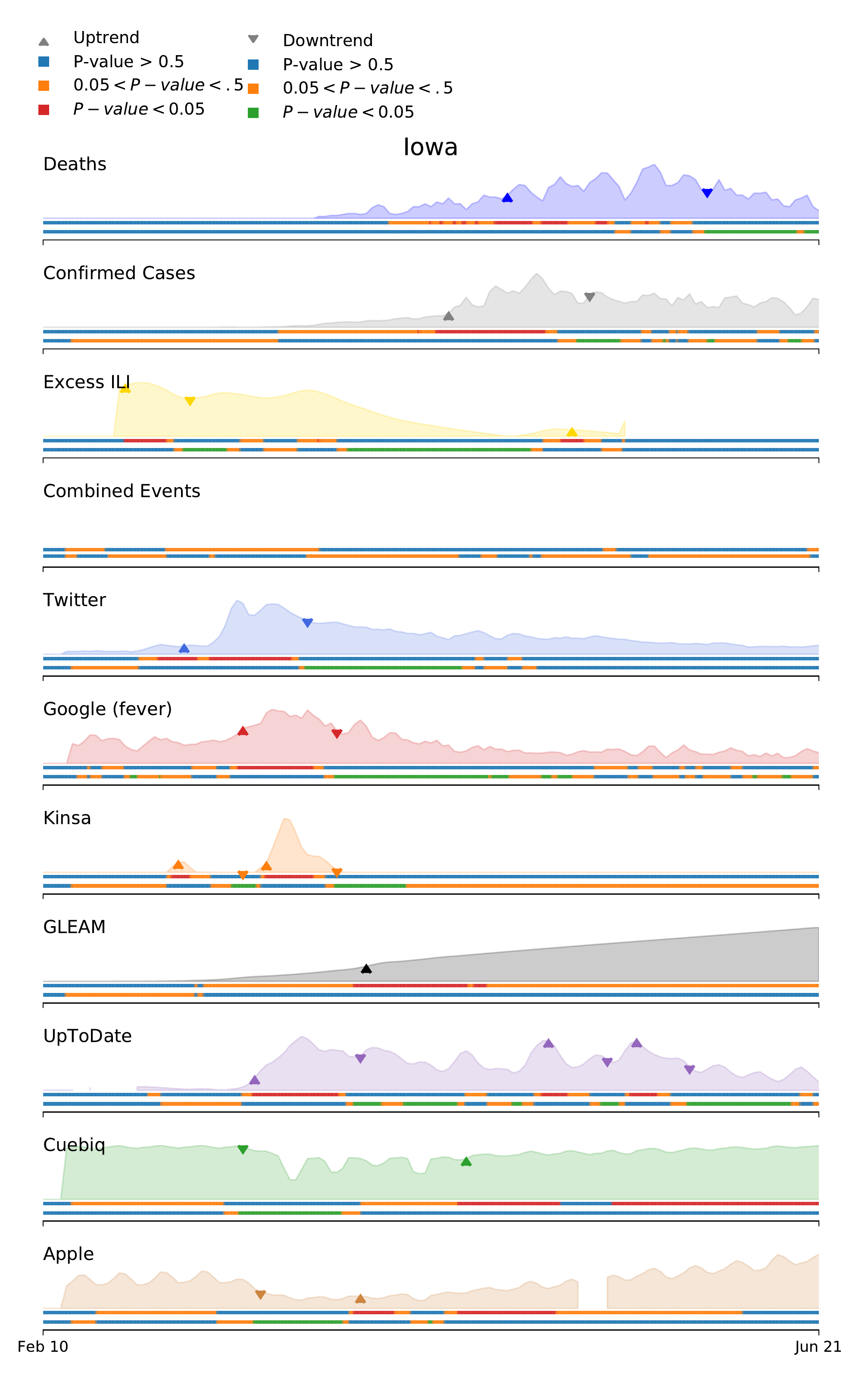}
    \label{fig:Iowa_event}
    \caption{}
\end{figure} 
 \begin{figure}
    \centering
    \includegraphics[width=.75\textwidth]{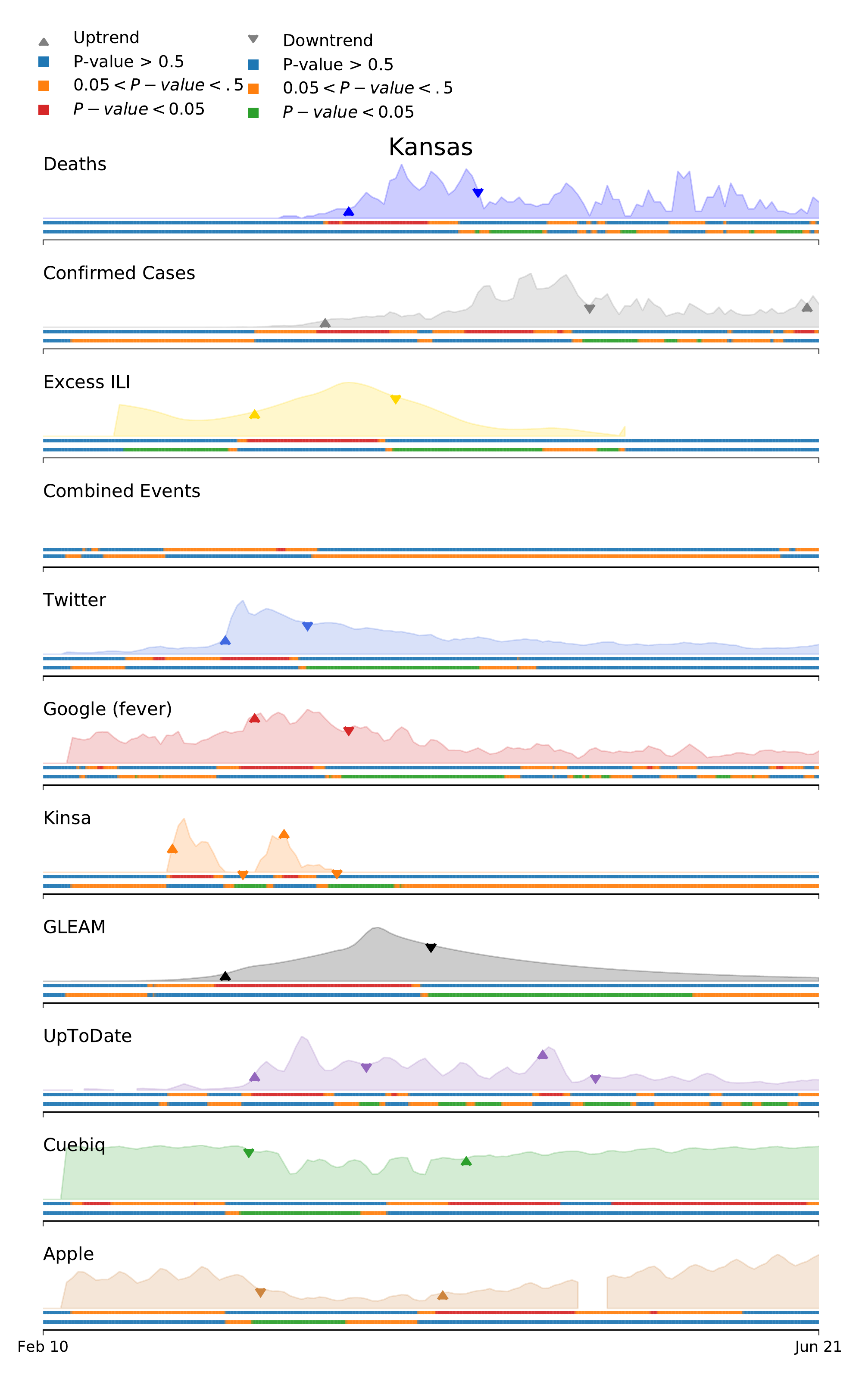}
    \label{fig:Kansas_event}
    \caption{}
\end{figure} 
 \begin{figure}
    \centering
    \includegraphics[width=.75\textwidth]{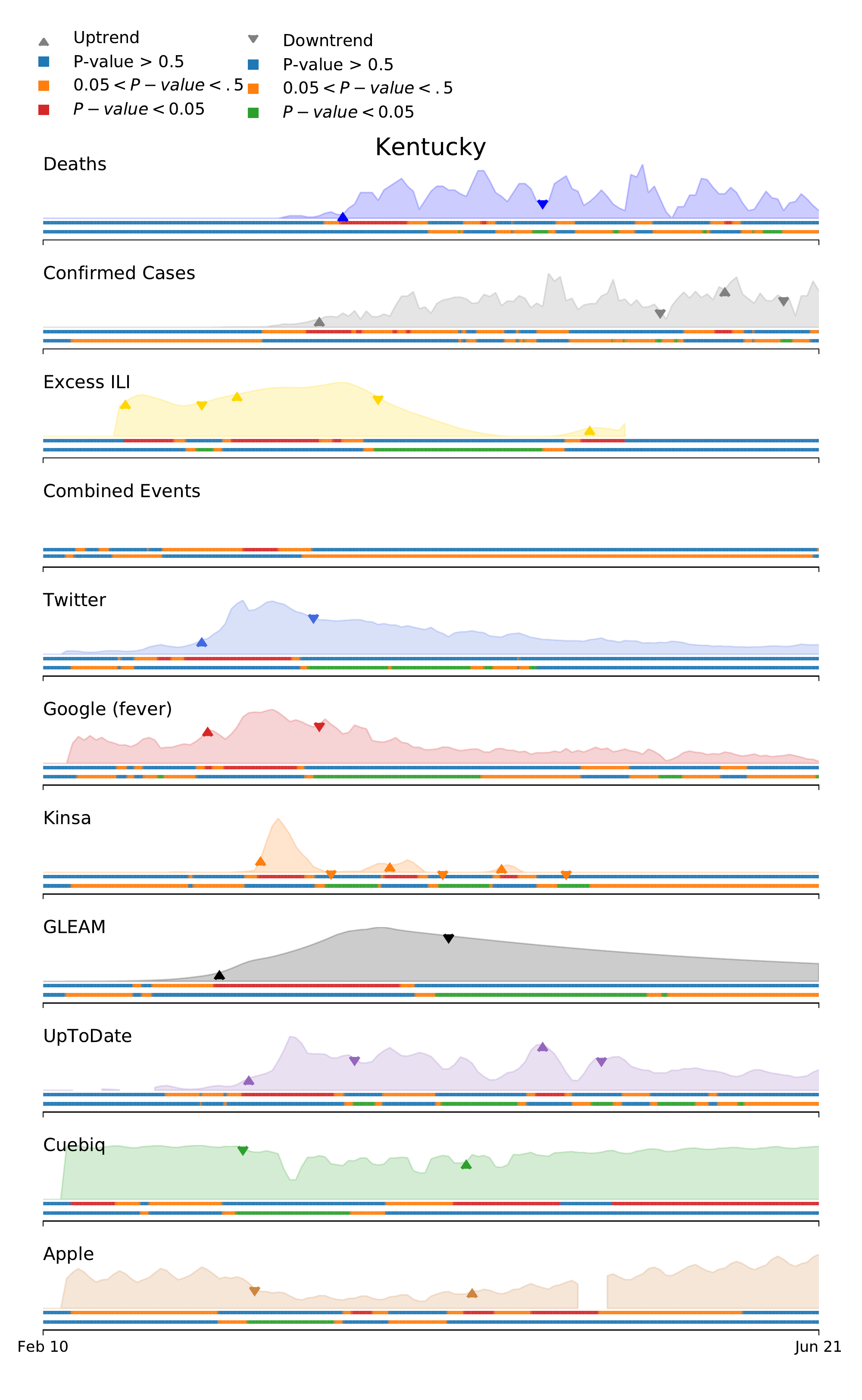}
    \label{fig:Kentucky_event}
    \caption{}
\end{figure} 
 \begin{figure}
    \centering
    \includegraphics[width=.75\textwidth]{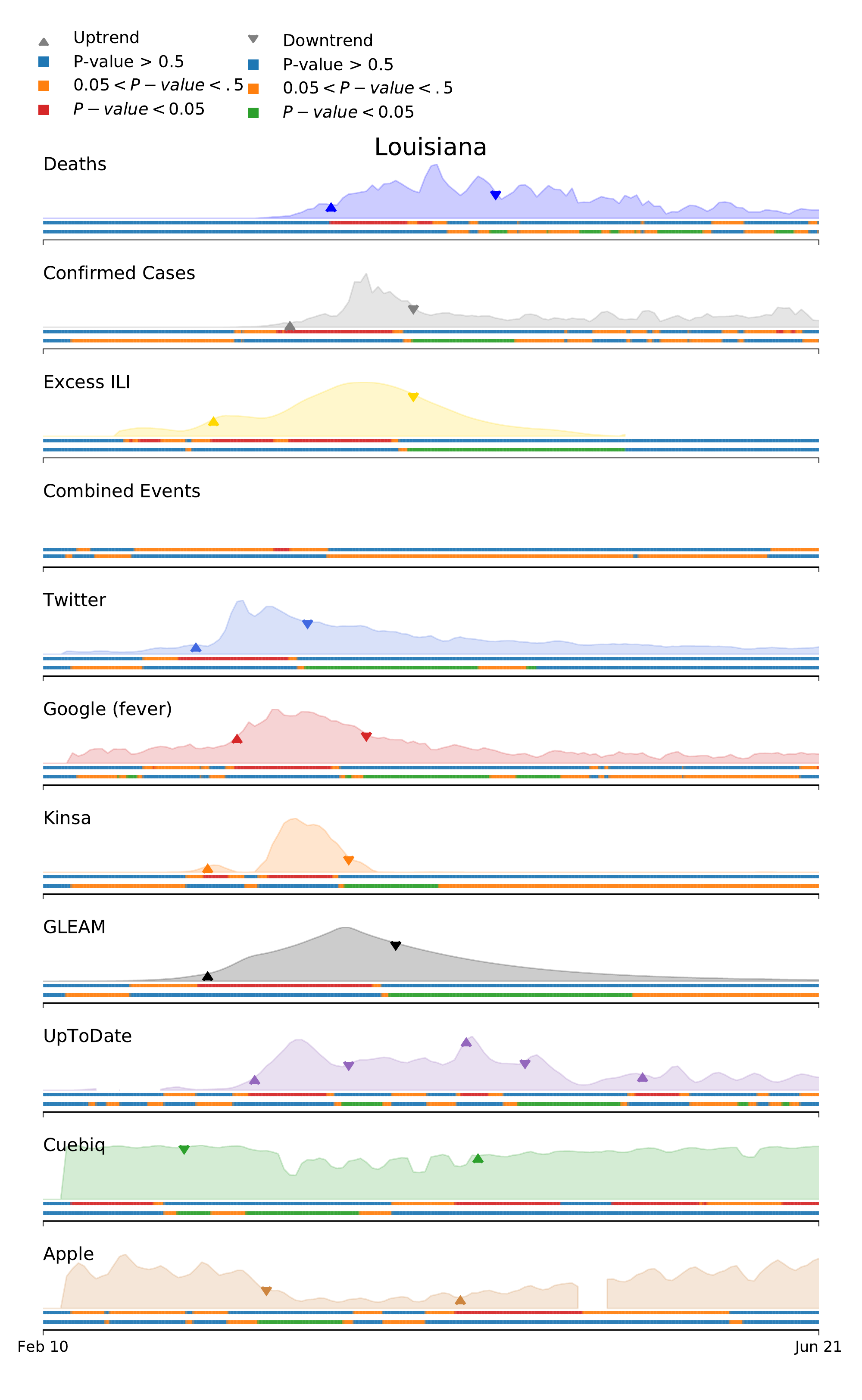}
    \label{fig:Louisiana_event}
    \caption{}
\end{figure} 
 \begin{figure}
    \centering
    \includegraphics[width=.75\textwidth]{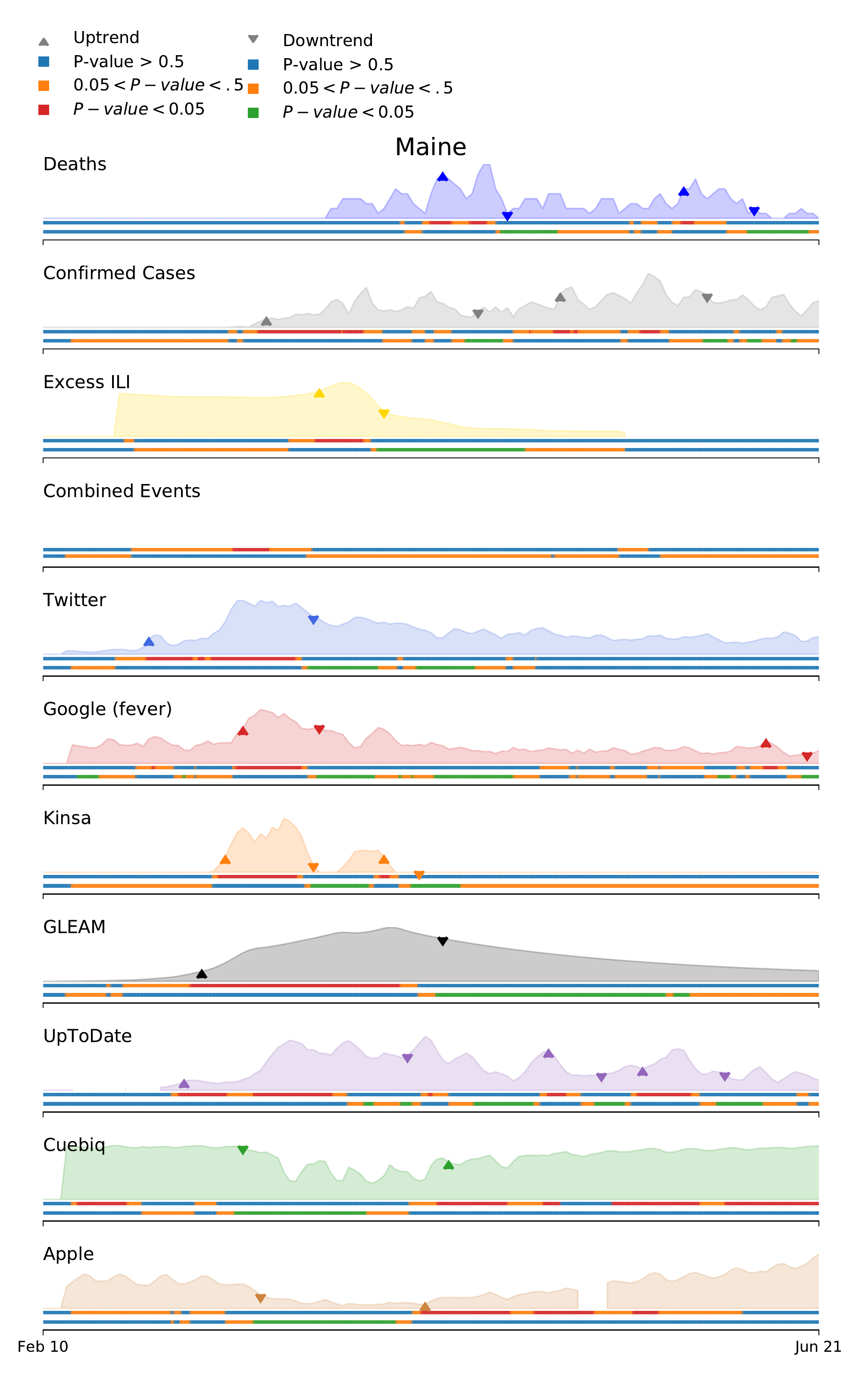}
    \label{fig:Maine_event}
    \caption{}
\end{figure} 
 \begin{figure}
    \centering
    \includegraphics[width=.75\textwidth]{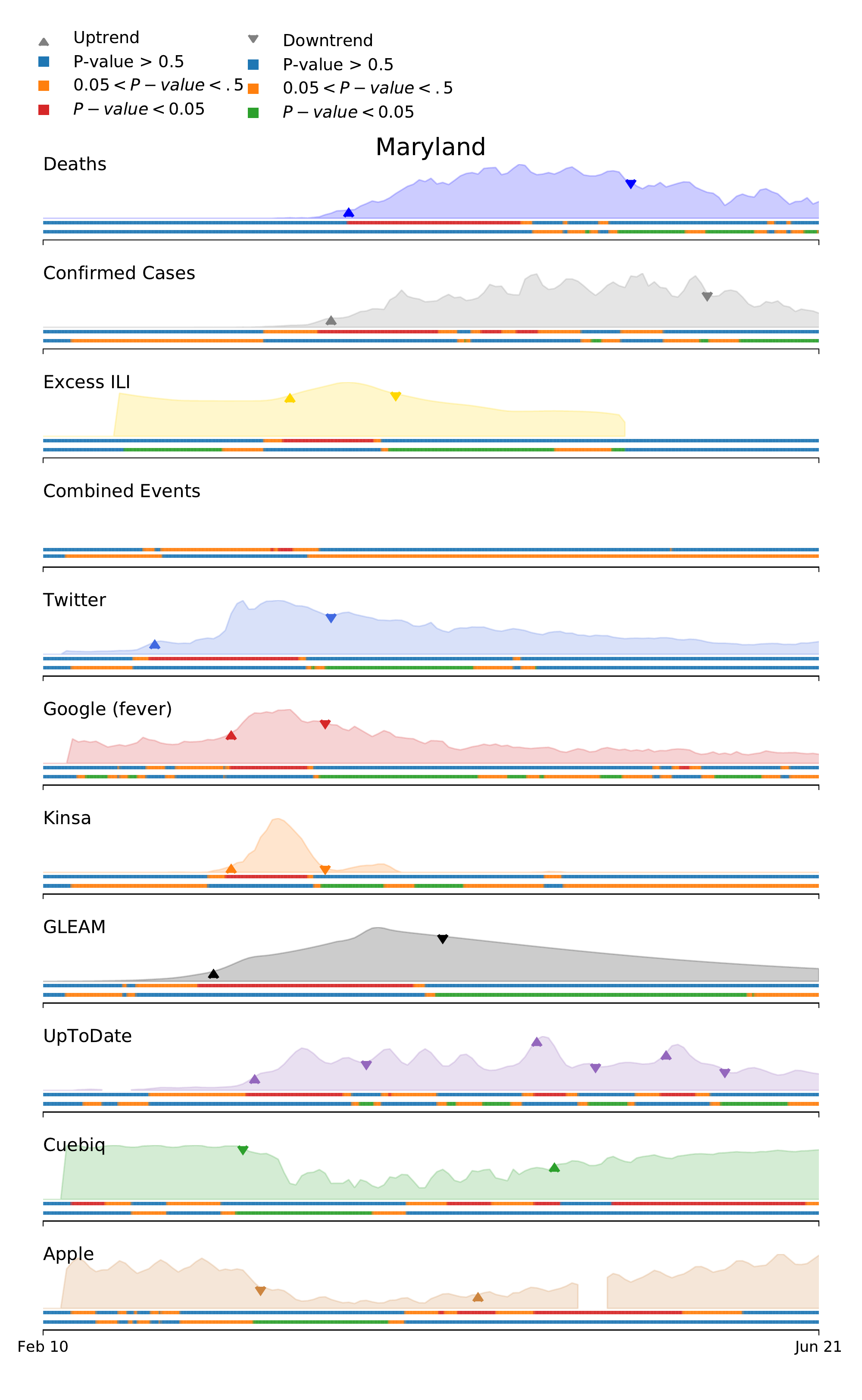}
    \label{fig:Maryland_event}
    \caption{}
\end{figure} 
 \begin{figure}
    \centering
    \includegraphics[width=.75\textwidth]{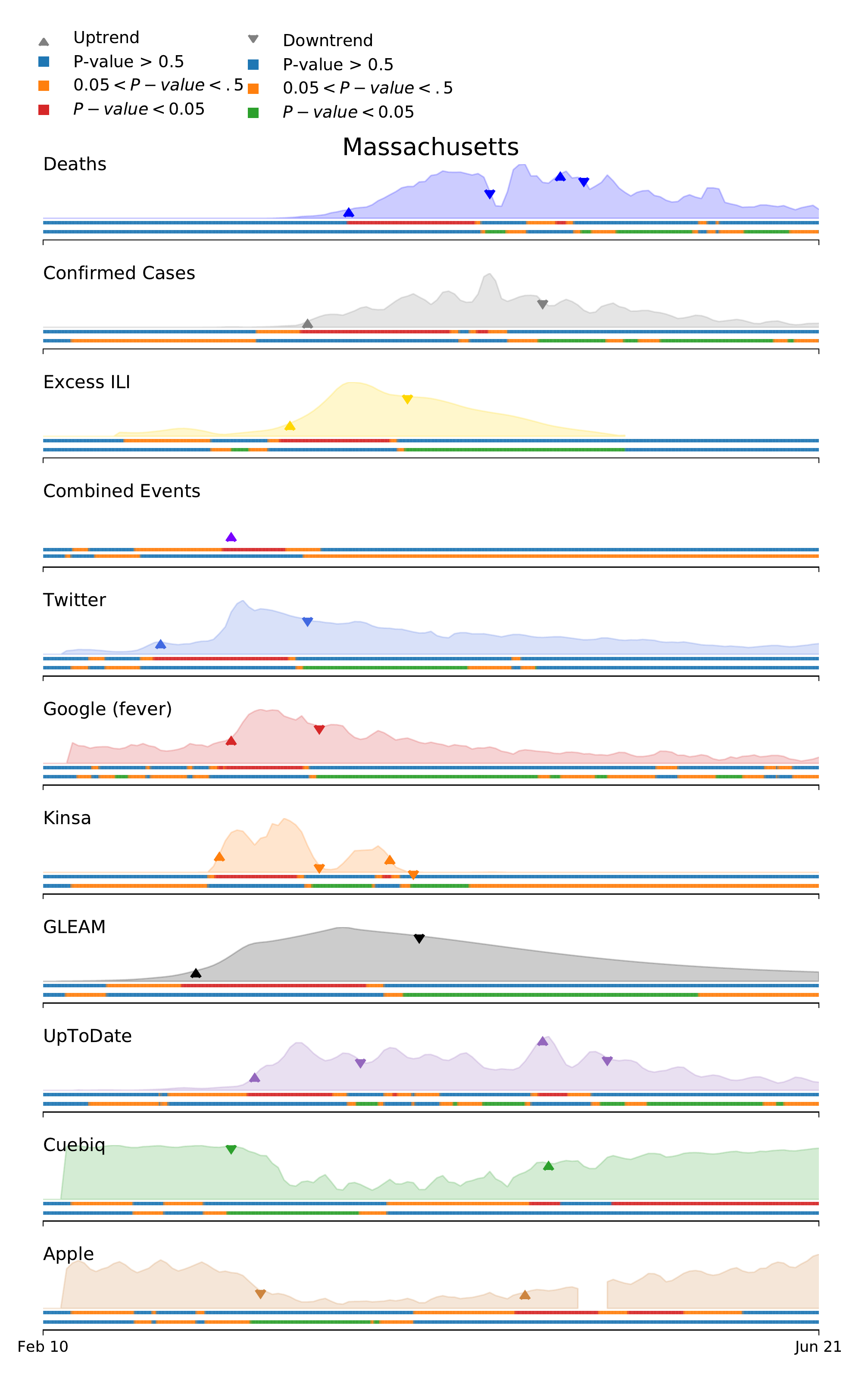}
    \label{fig:Massachusetts_event}
    \caption{}
\end{figure} 
 \begin{figure}
    \centering
    \includegraphics[width=.75\textwidth]{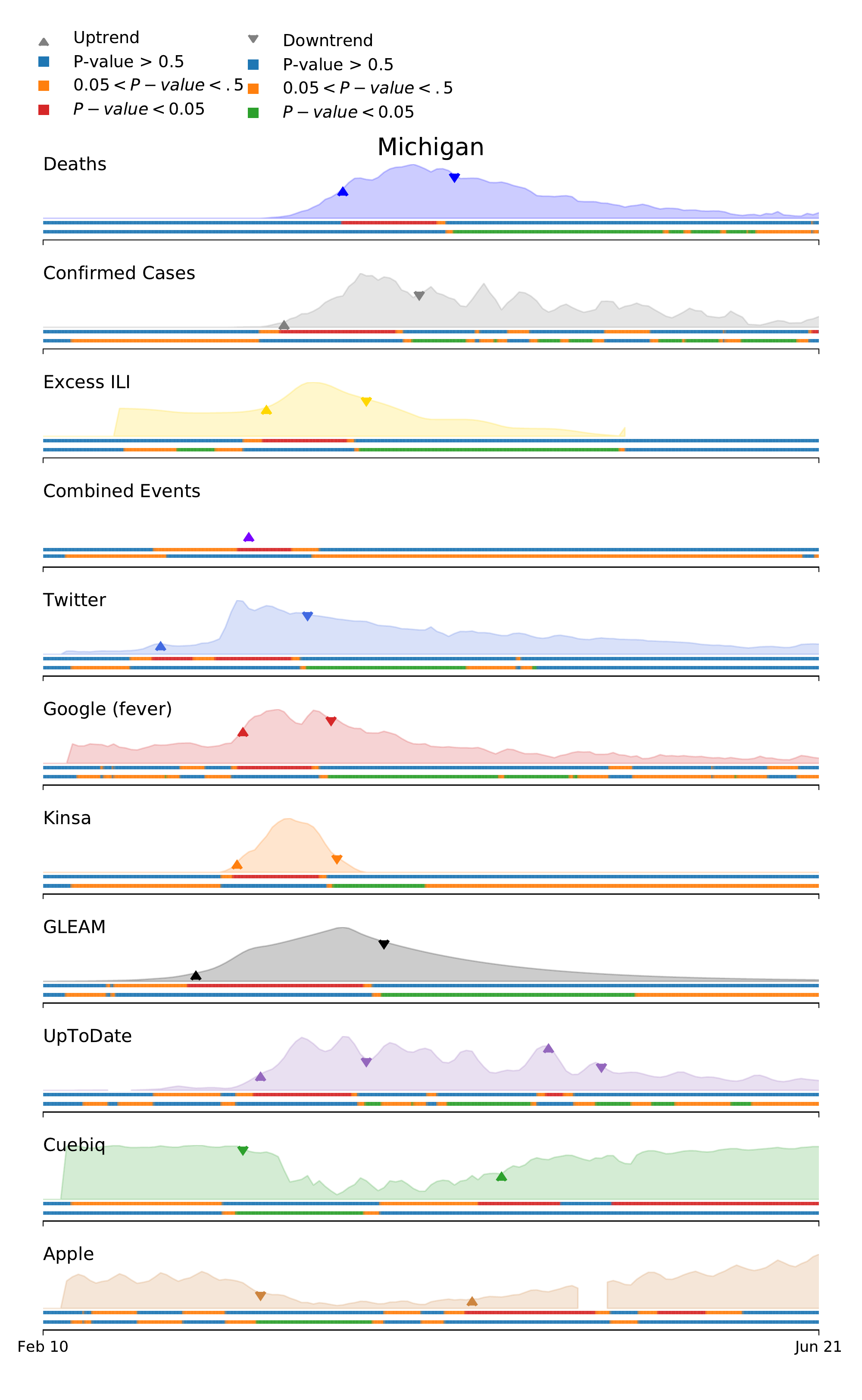}
    \label{fig:Michigan_event}
    \caption{}
\end{figure} 
 \begin{figure}
    \centering
    \includegraphics[width=.75\textwidth]{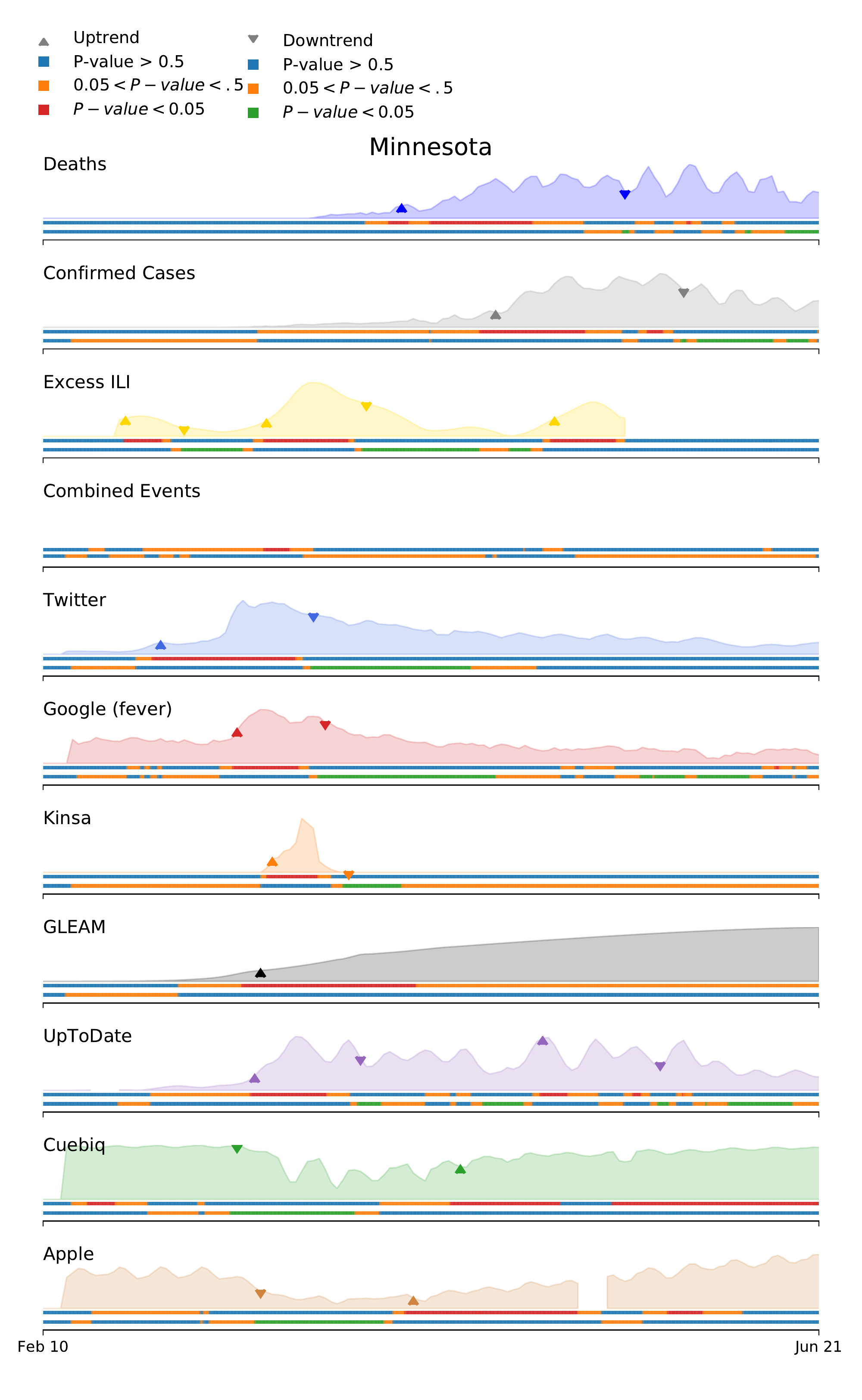}
    \label{fig:Minnesota_event}
    \caption{}
\end{figure} 
 \begin{figure}
    \centering
    \includegraphics[width=.75\textwidth]{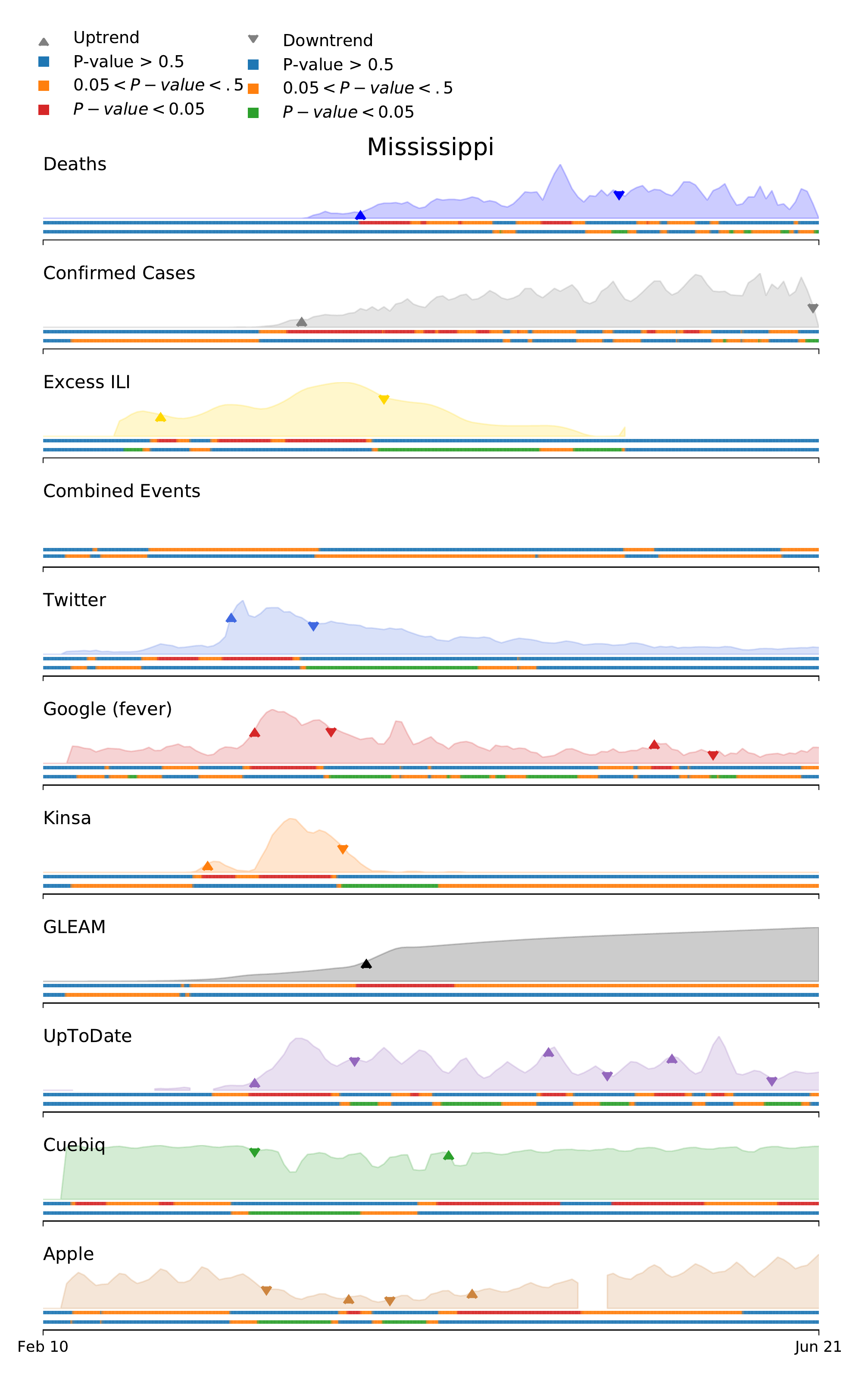}
    \label{fig:Mississippi_event}
    \caption{}
\end{figure} 
 \begin{figure}
    \centering
    \includegraphics[width=.75\textwidth]{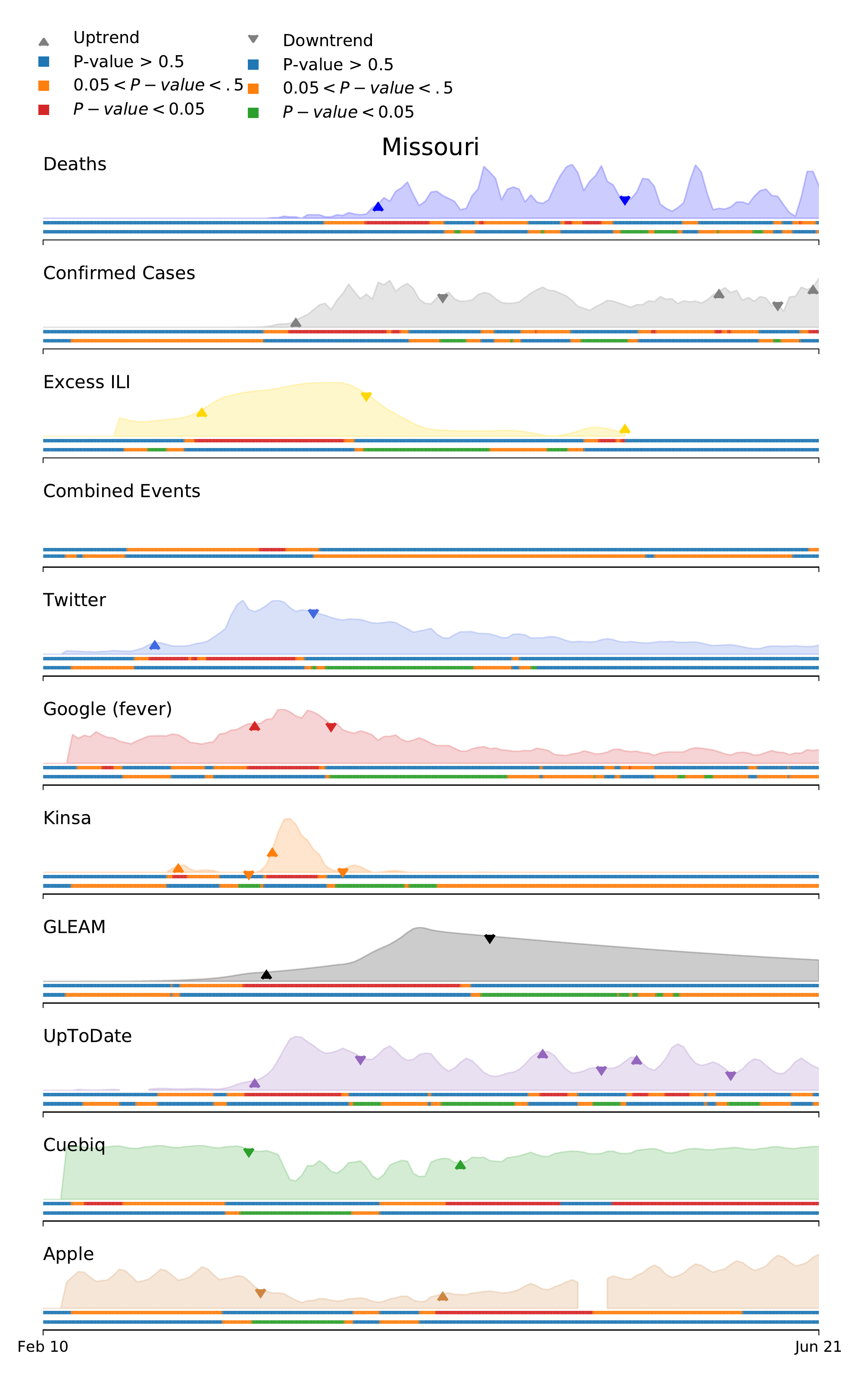}
    \label{fig:Missouri_event}
    \caption{}
\end{figure} 
 \begin{figure}
    \centering
    \includegraphics[width=.75\textwidth]{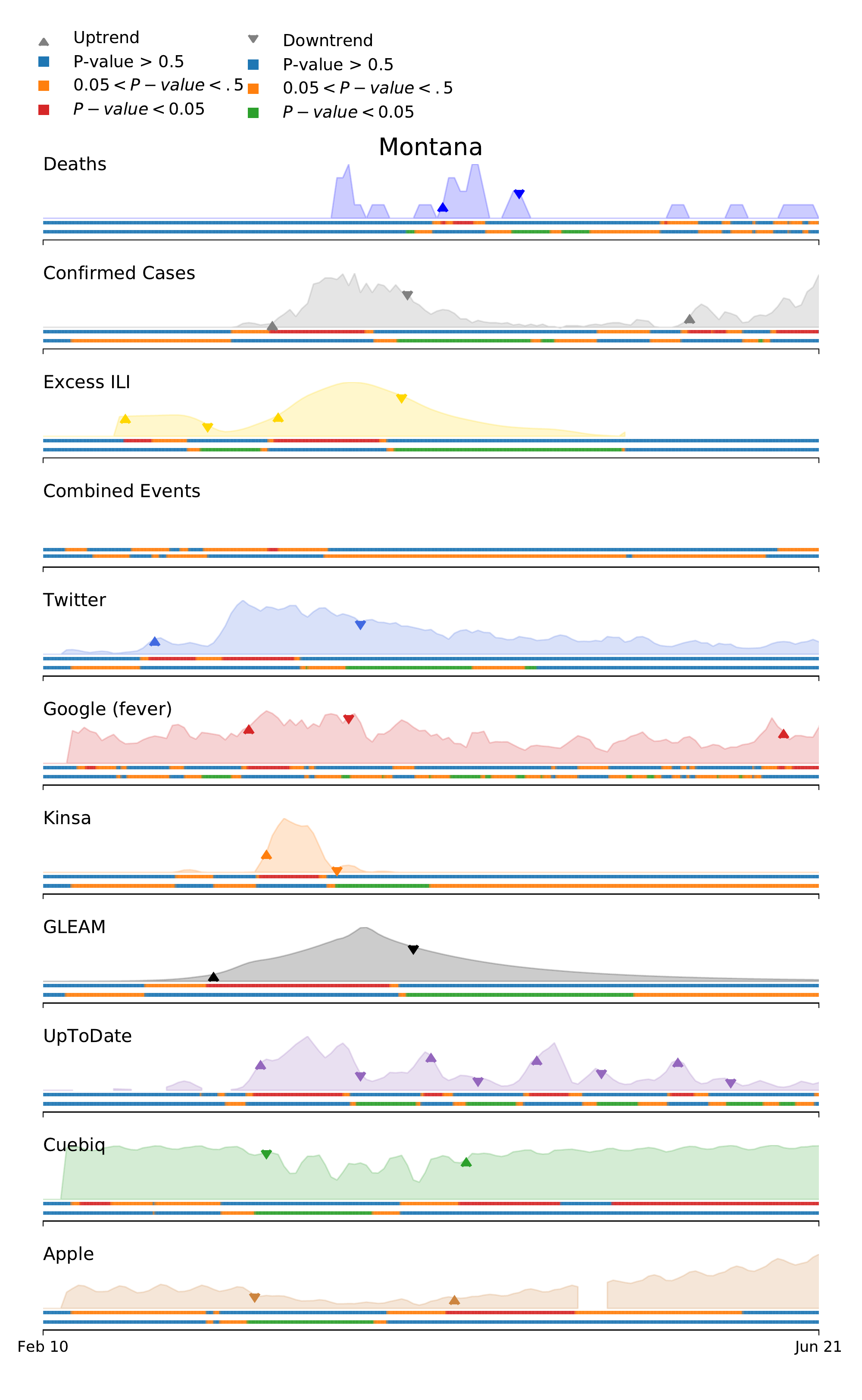}
    \label{fig:Montana_event}
    \caption{}
\end{figure} 
 \begin{figure}
    \centering
    \includegraphics[width=.75\textwidth]{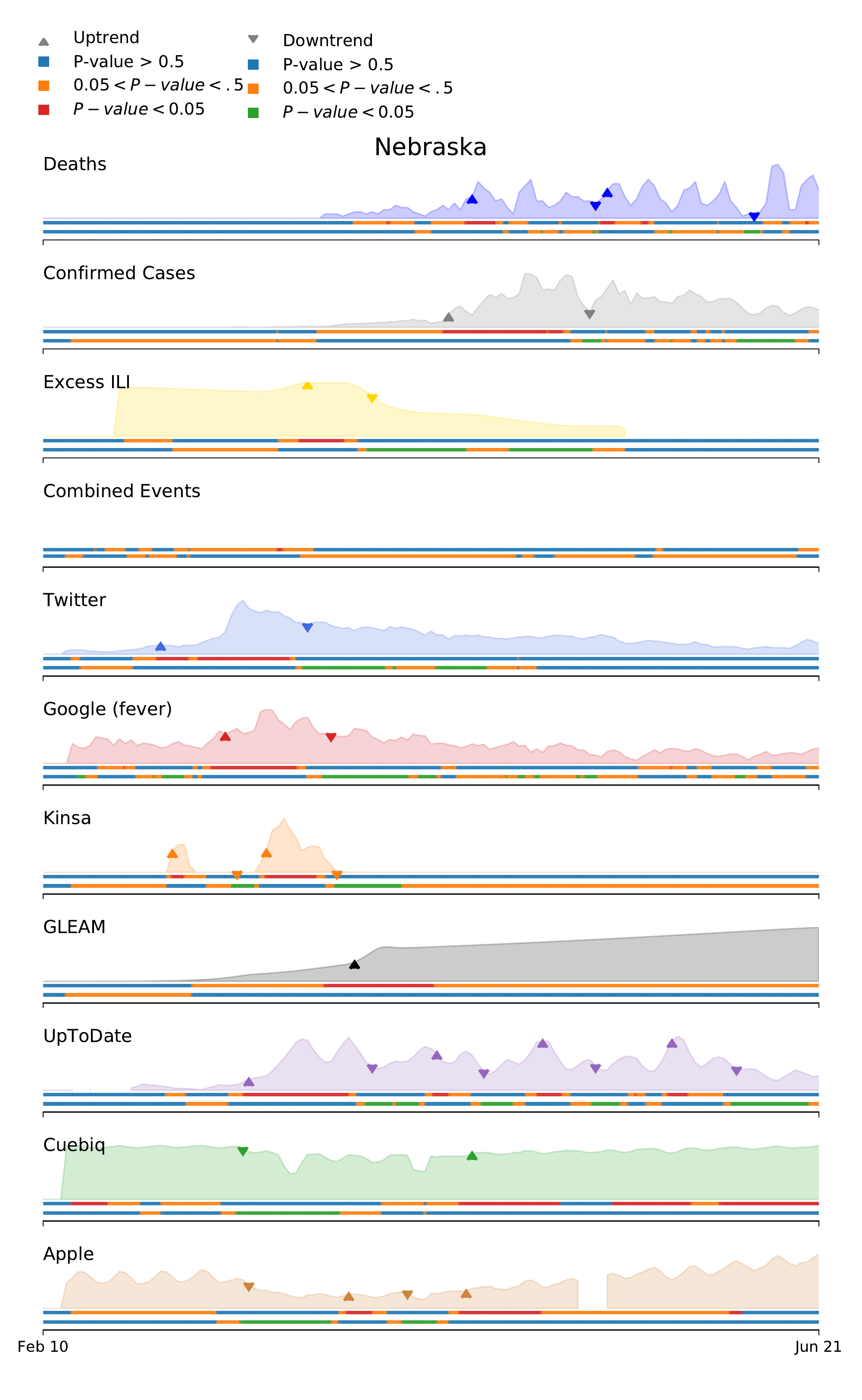}
    \label{fig:Nebraska_event}
    \caption{}
\end{figure} 
 \begin{figure}
    \centering
    \includegraphics[width=.75\textwidth]{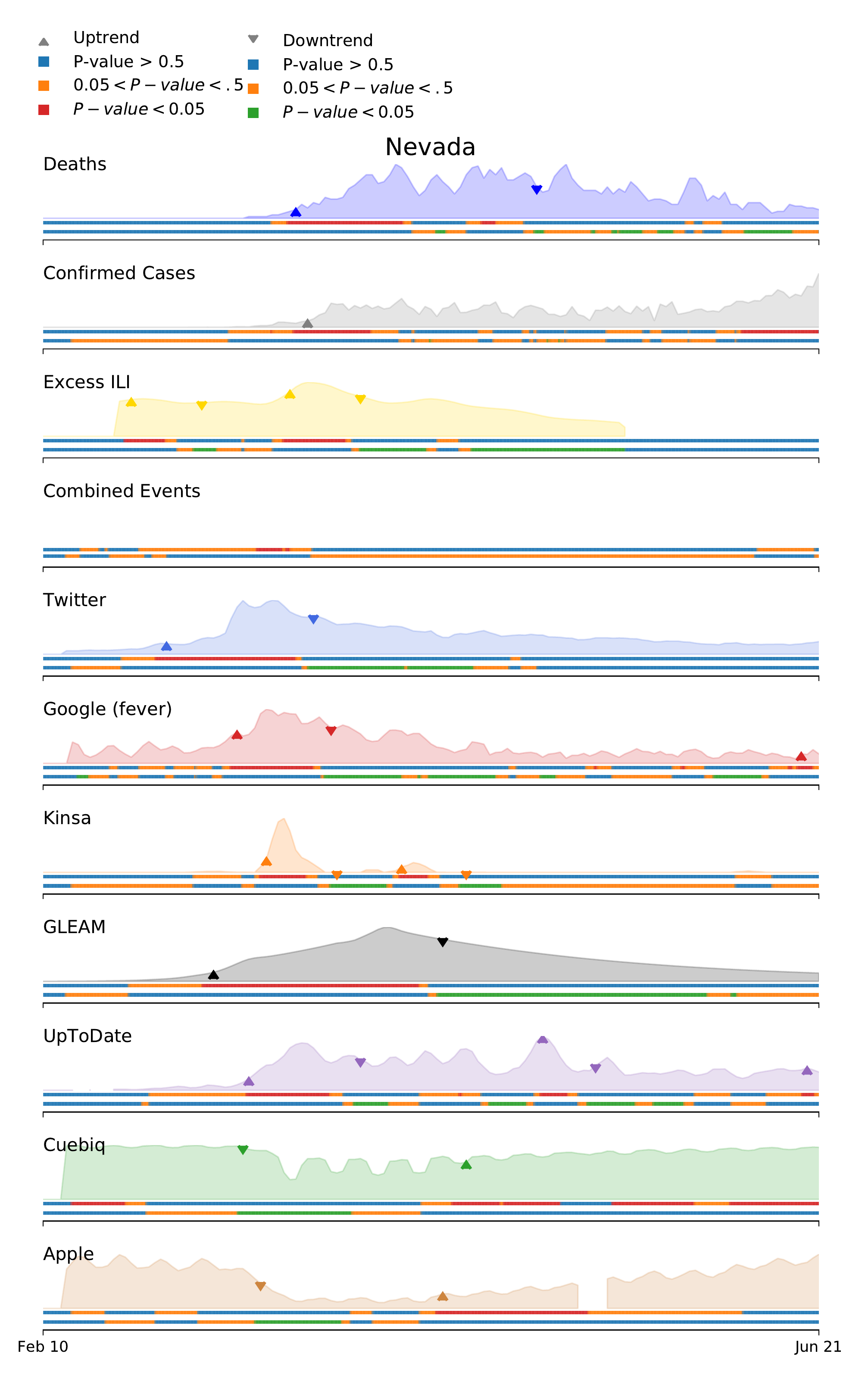}
    \label{fig:Nevada_event}
    \caption{}
\end{figure} 
 \begin{figure}
    \centering
    \includegraphics[width=.75\textwidth]{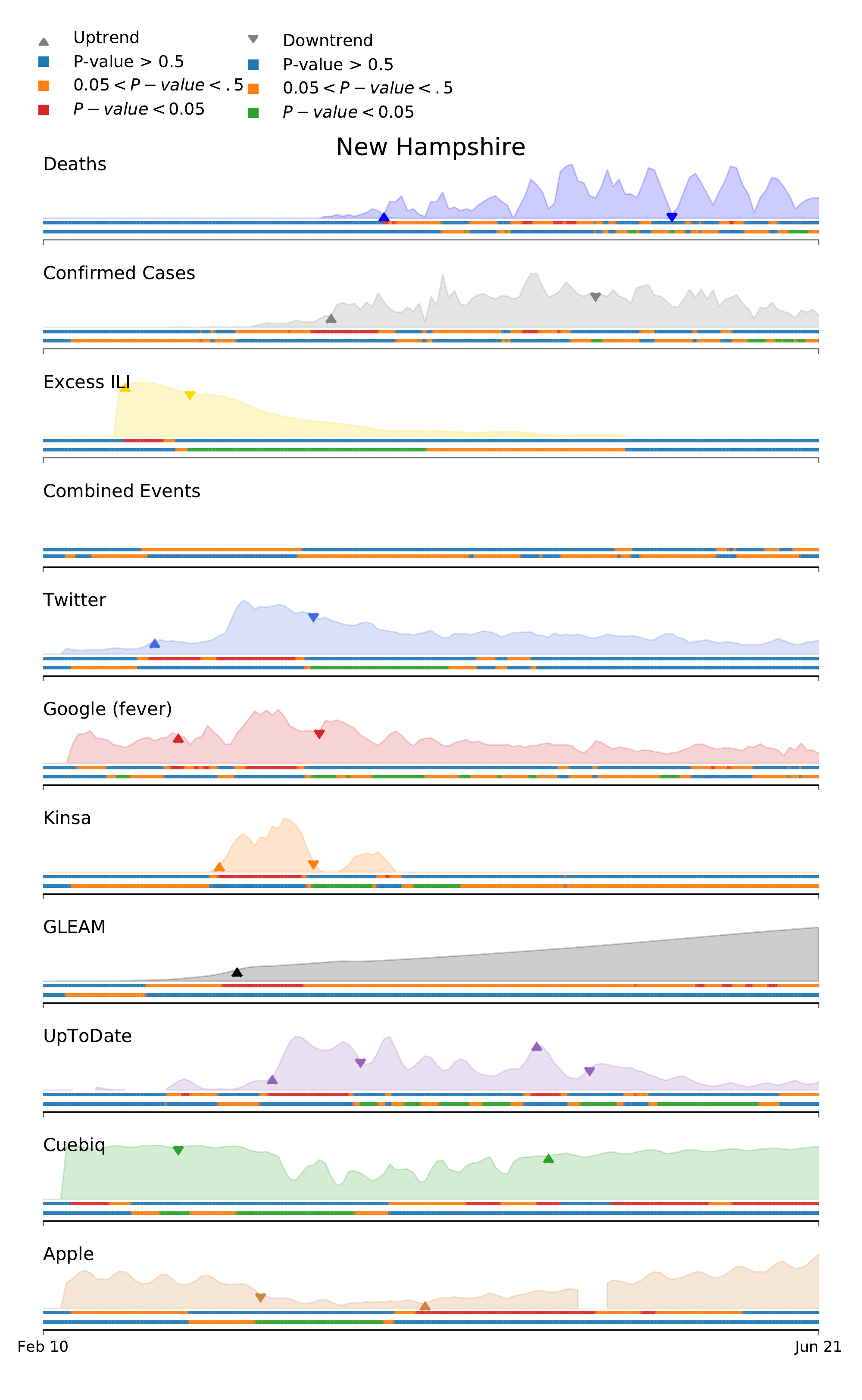}
    \label{fig:New Hampshire_event}
    \caption{}
\end{figure} 
 \begin{figure}
    \centering
    \includegraphics[width=.75\textwidth]{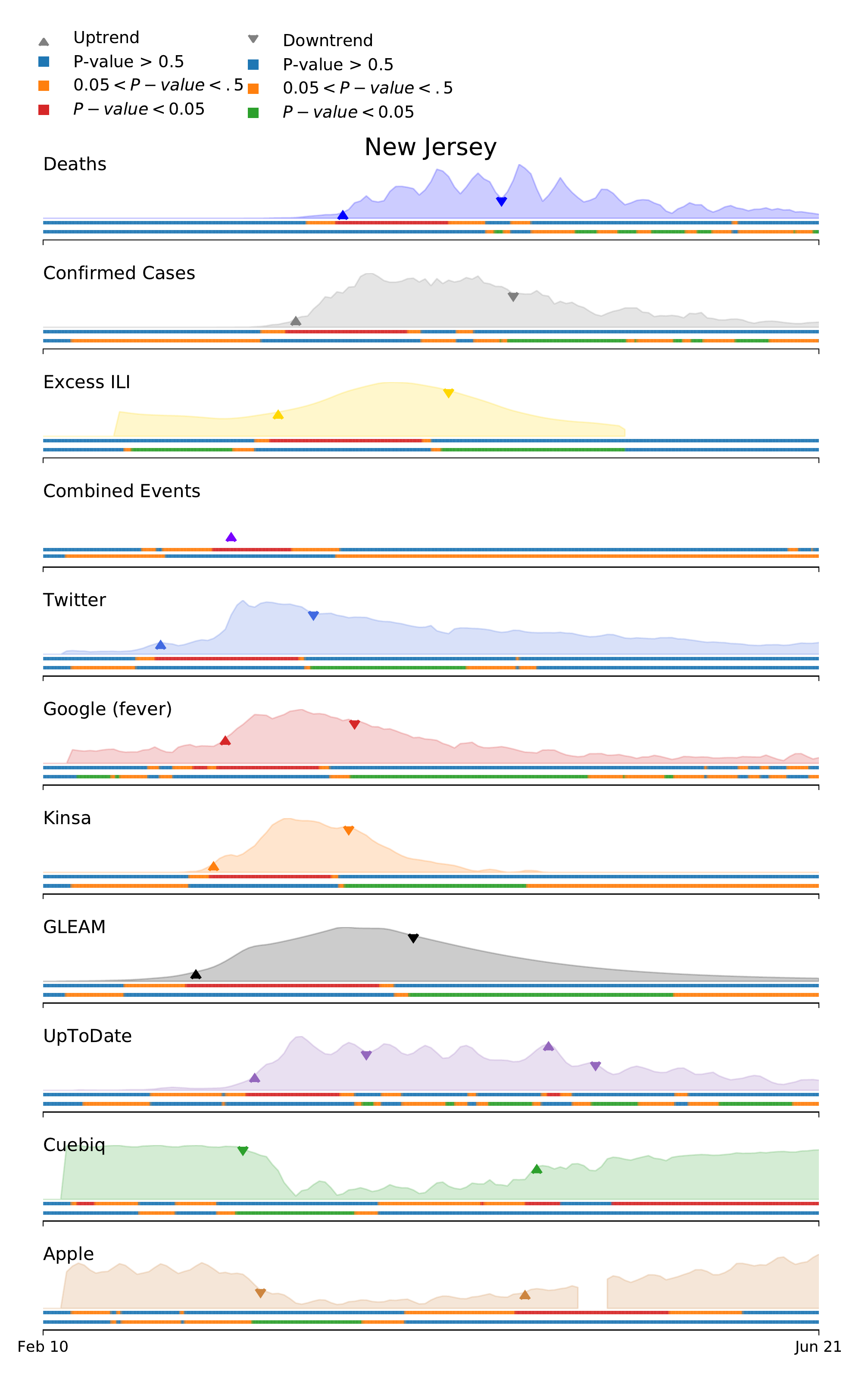}
    \label{fig:New Jersey_event}
    \caption{}
\end{figure} 
 \begin{figure}
    \centering
    \includegraphics[width=.75\textwidth]{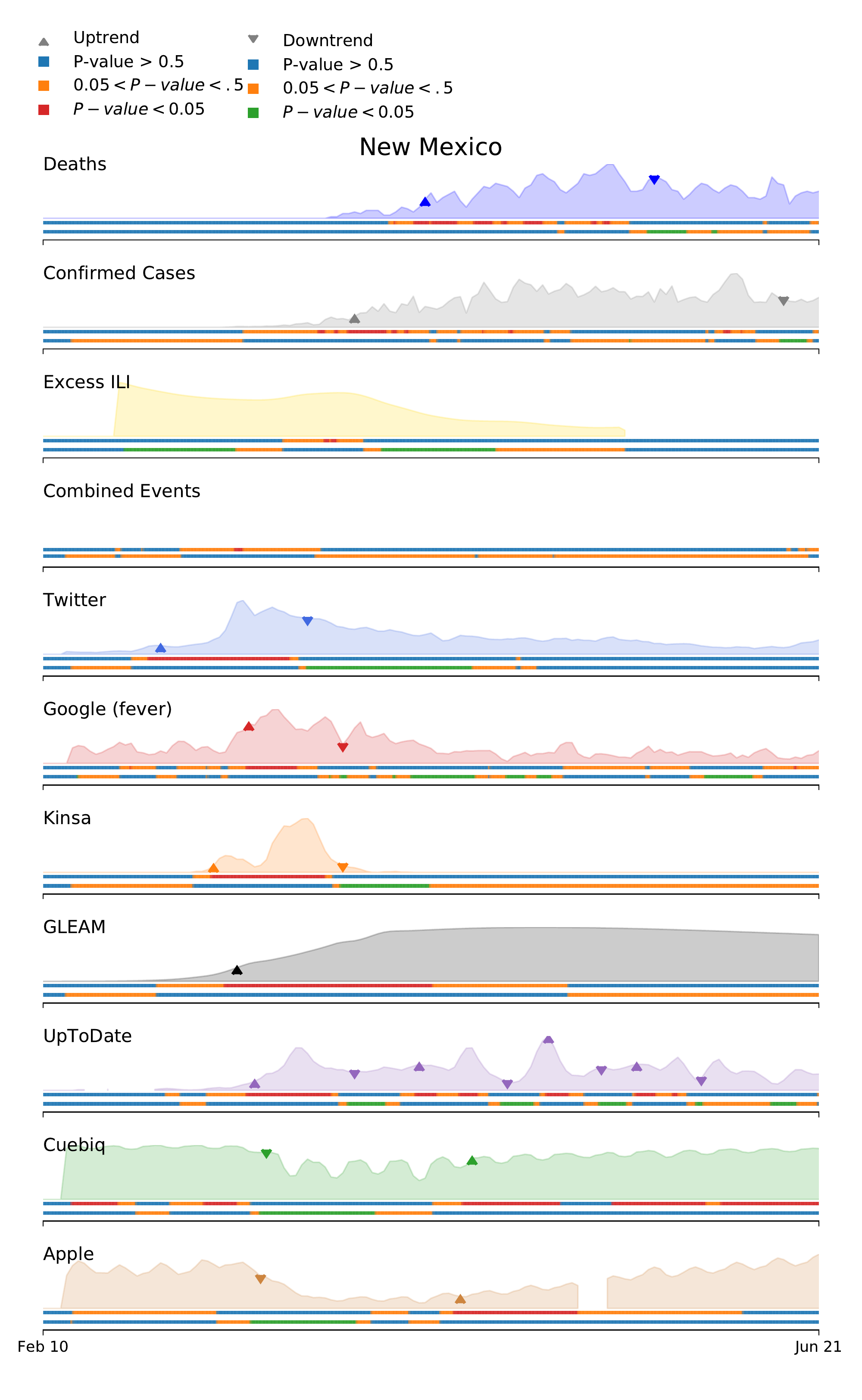}
    \label{fig:New Mexico_event}
    \caption{}
\end{figure} 
 \begin{figure}
    \centering
    \includegraphics[width=.75\textwidth]{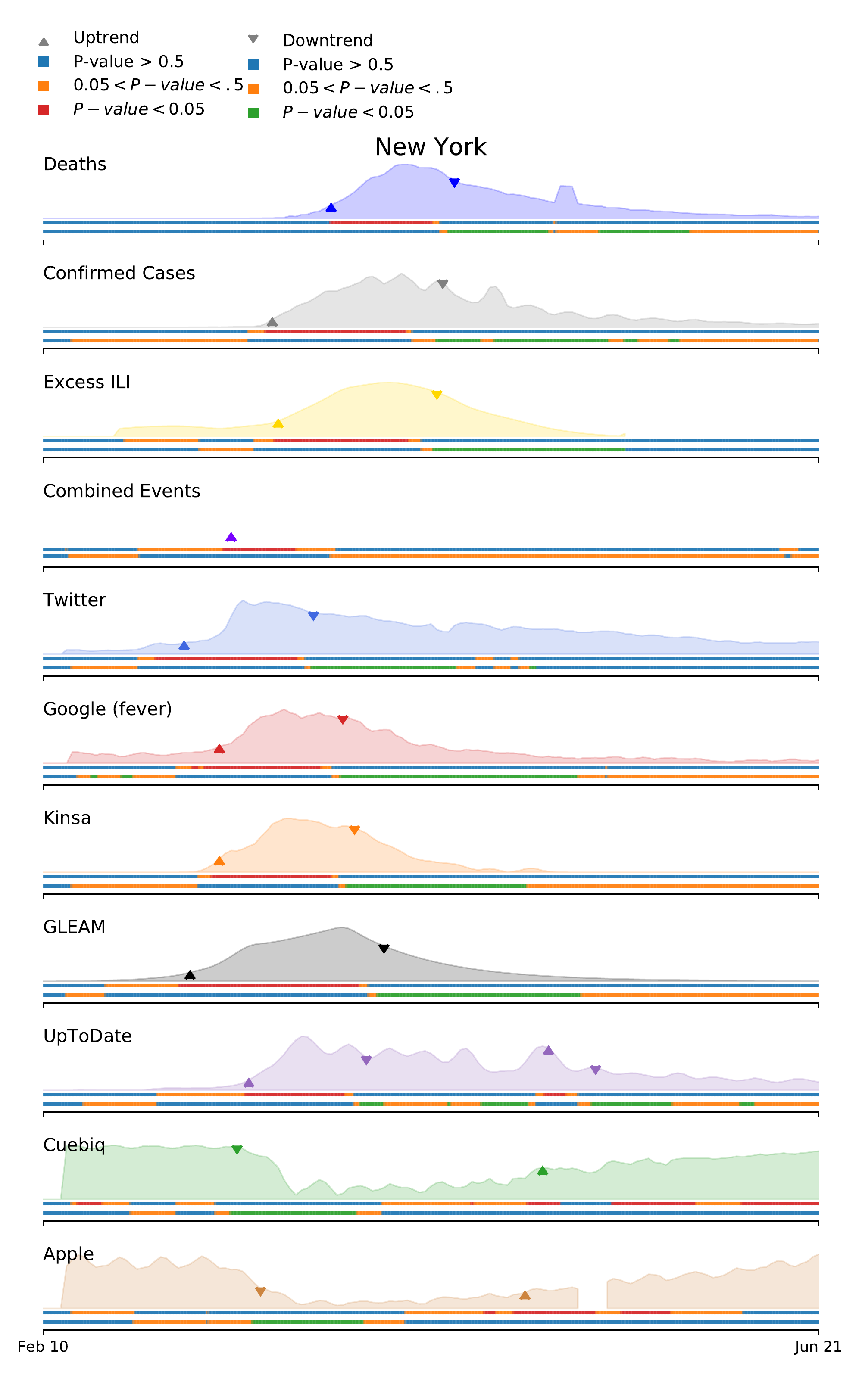}
    \label{fig:New York_event}
    \caption{}
\end{figure} 
 \begin{figure}
    \centering
    \includegraphics[width=.75\textwidth]{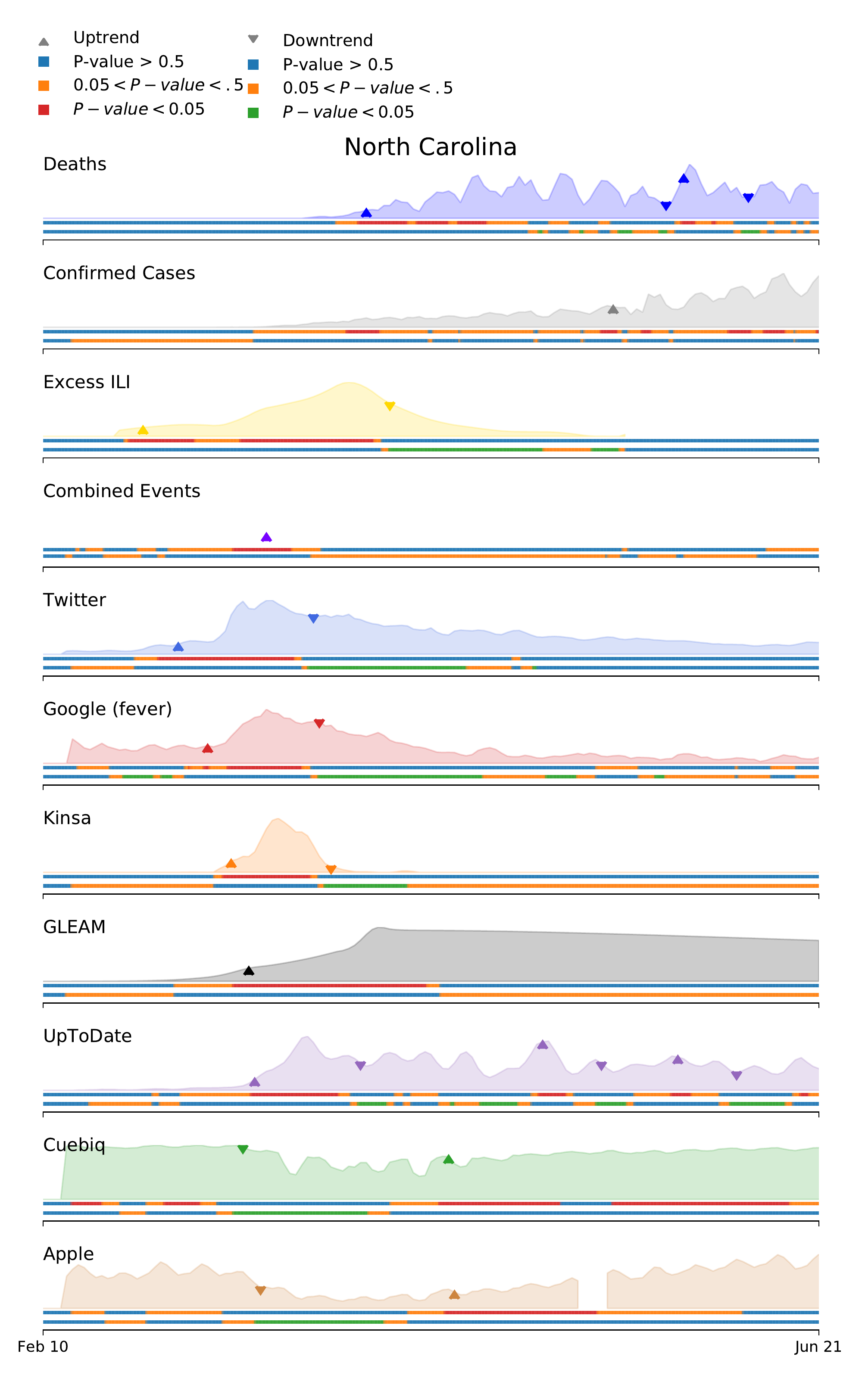}
    \label{fig:North Carolina_event}
    \caption{}
\end{figure} 
 \begin{figure}
    \centering
    \includegraphics[width=.75\textwidth]{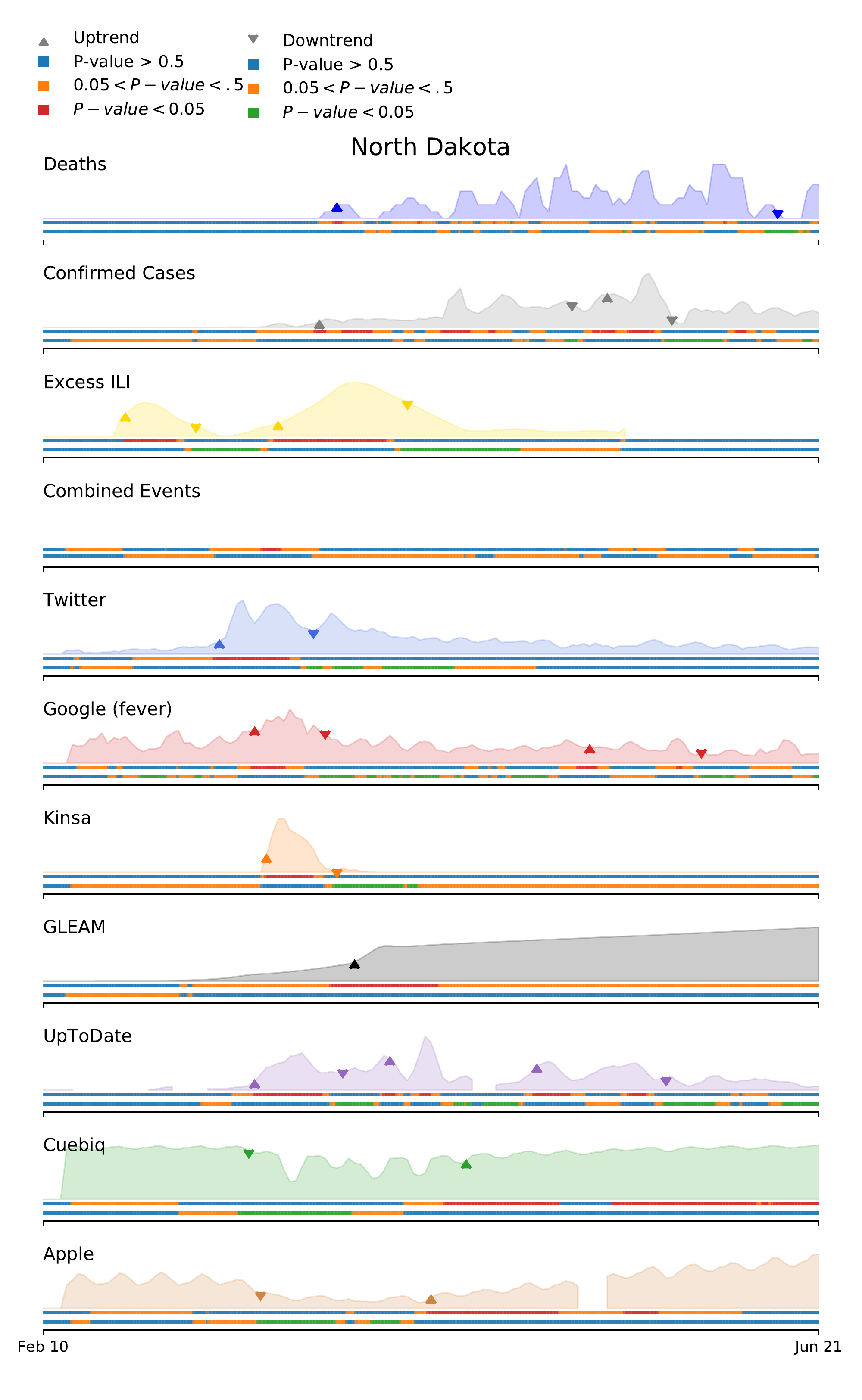}
    \label{fig:North Dakota_event}
    \caption{}
\end{figure} 
 \begin{figure}
    \centering
    \includegraphics[width=.75\textwidth]{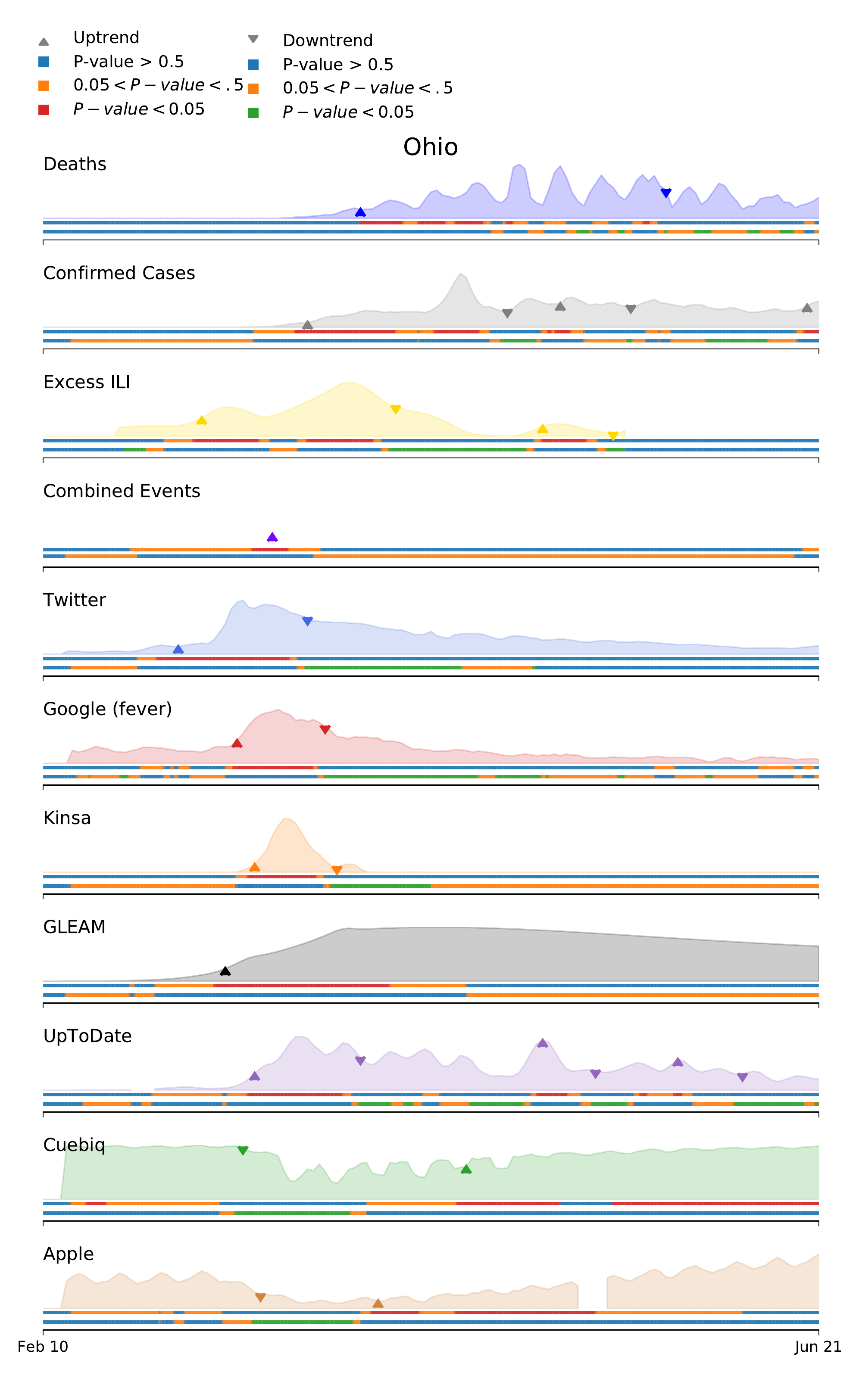}
    \label{fig:Ohio_event}
    \caption{}
\end{figure} 
 \begin{figure}
    \centering
    \includegraphics[width=.75\textwidth]{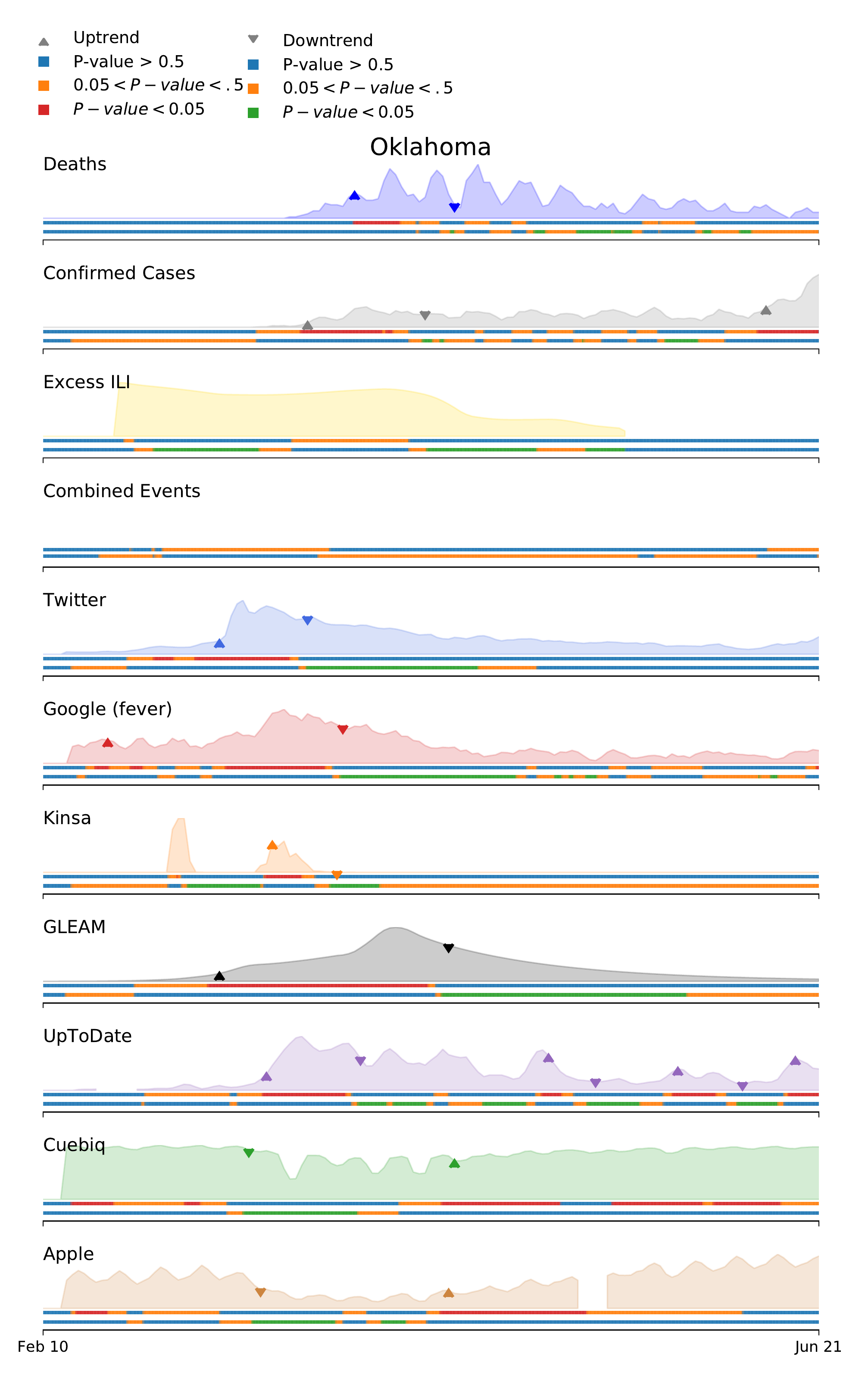}
    \label{fig:Oklahoma_event}
    \caption{}
\end{figure} 
 \begin{figure}
    \centering
    \includegraphics[width=.75\textwidth]{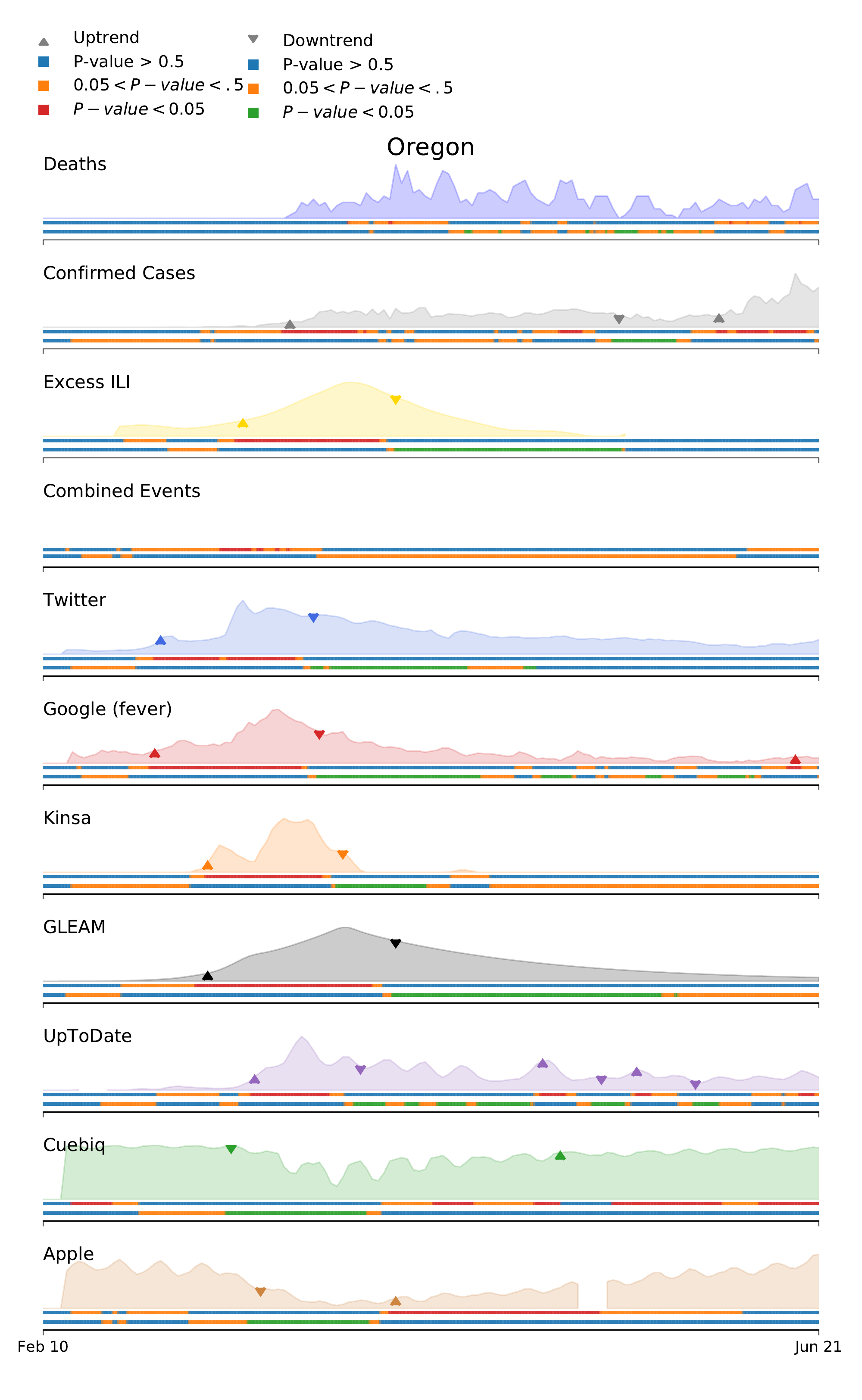}
    \label{fig:Oregon_event}
    \caption{}
\end{figure} 
 \begin{figure}
    \centering
    \includegraphics[width=.75\textwidth]{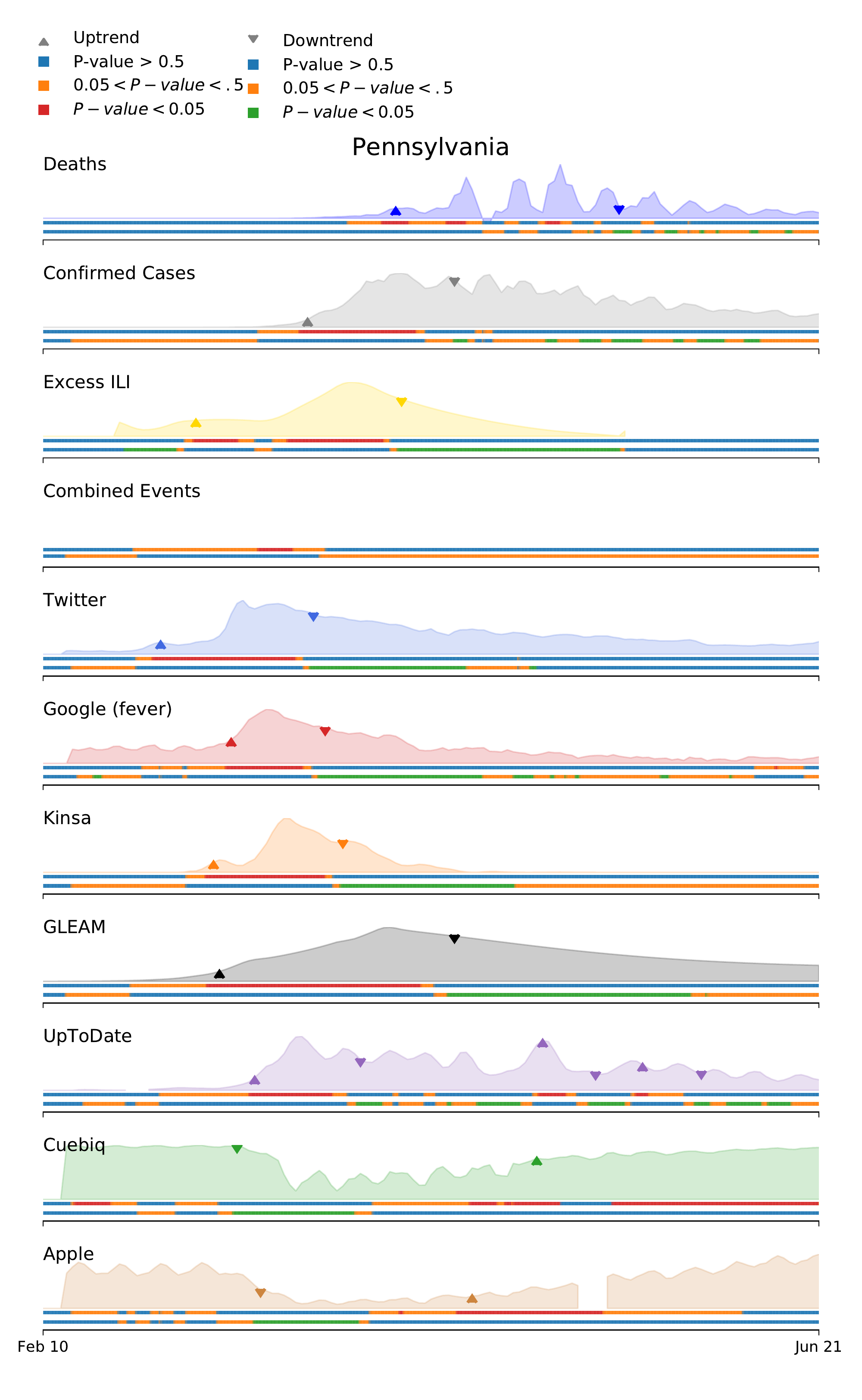}
    \label{fig:Pennsylvania_event}
    \caption{}
\end{figure} 
 \begin{figure}
    \centering
    \includegraphics[width=.75\textwidth]{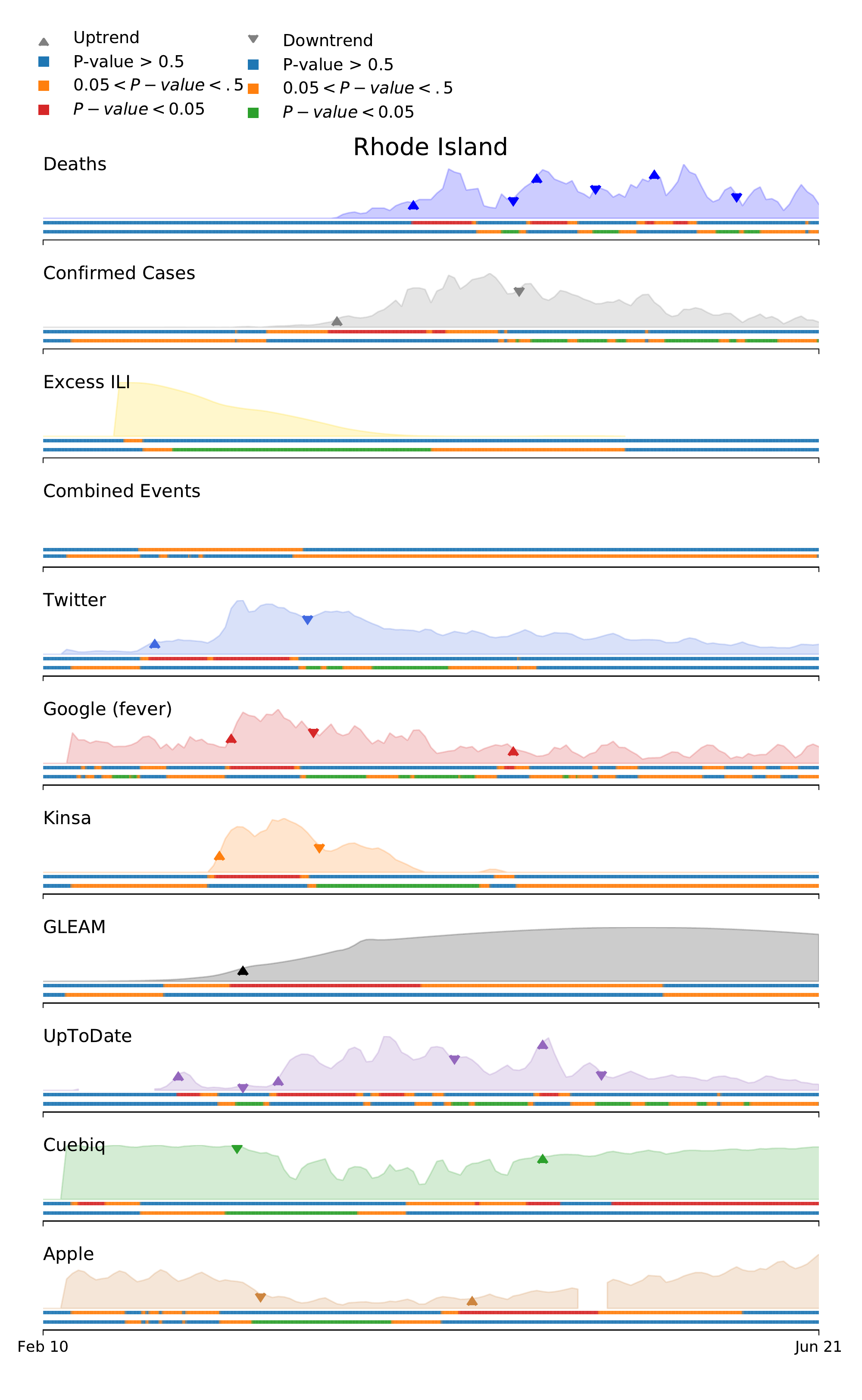}
    \label{fig:Rhode Island_event}
    \caption{}
\end{figure} 
 \begin{figure}
    \centering
    \includegraphics[width=.75\textwidth]{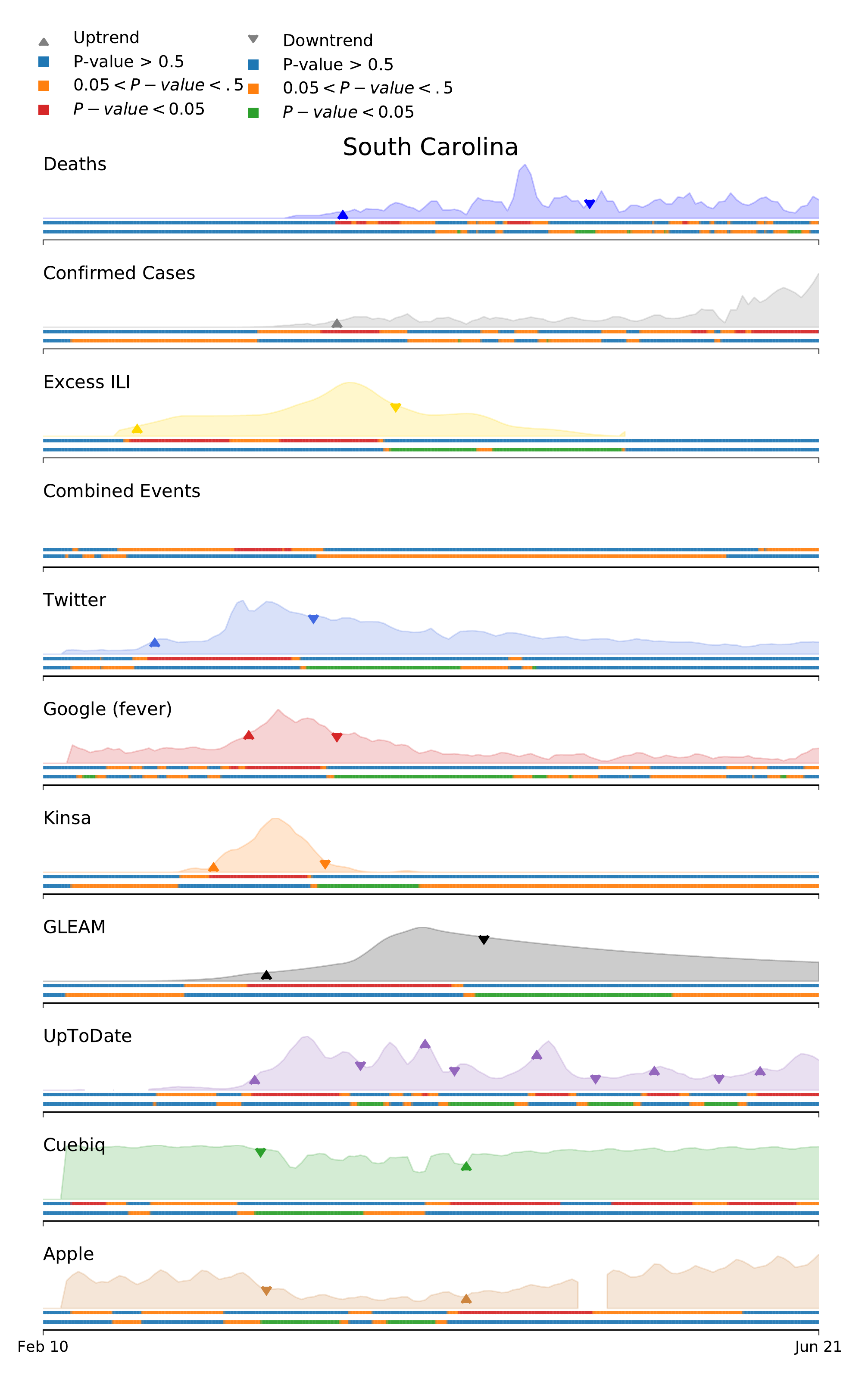}
    \label{fig:South Carolina_event}
    \caption{}
\end{figure} 
 \begin{figure}
    \centering
    \includegraphics[width=.75\textwidth]{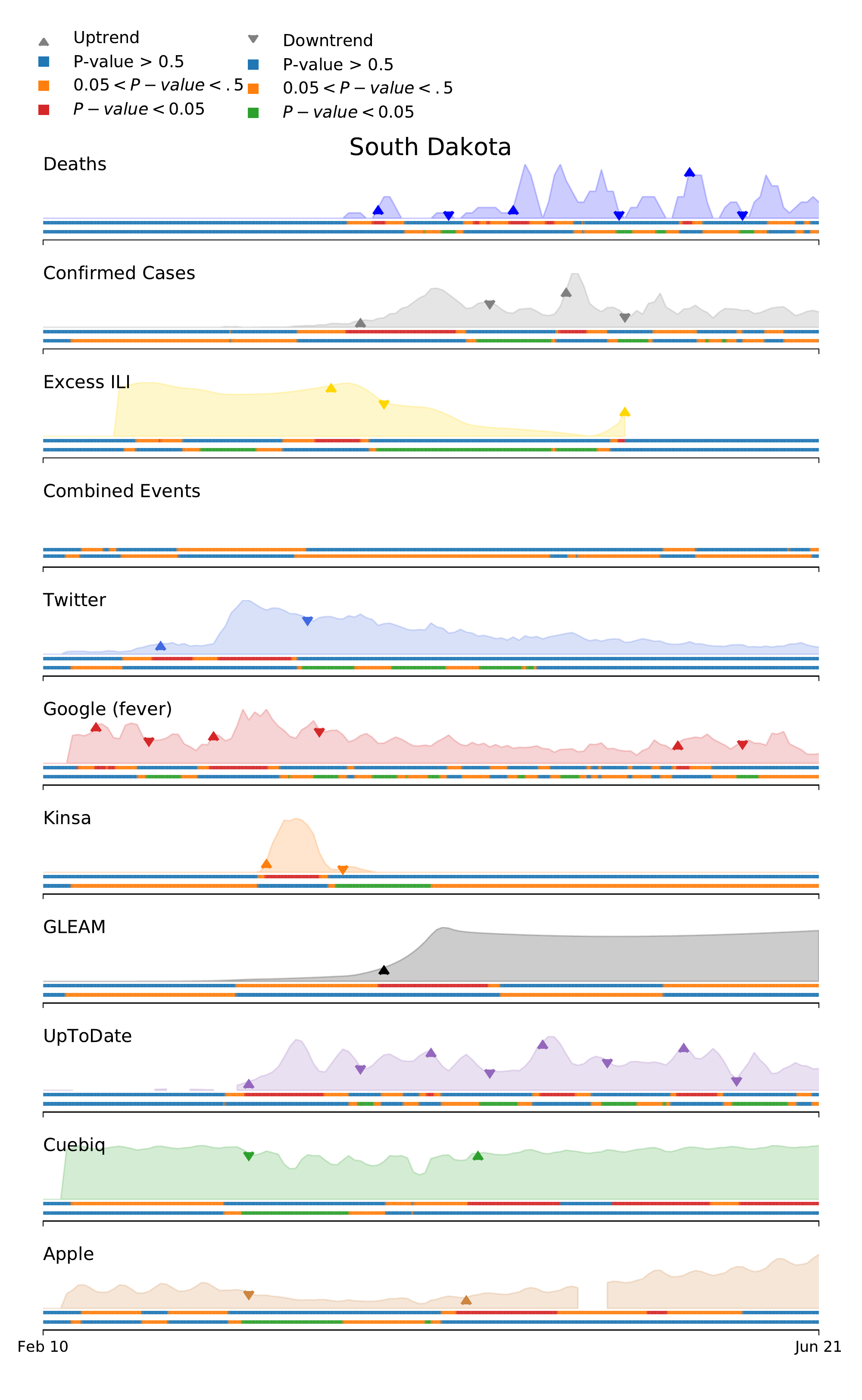}
    \label{fig:South Dakota_event}
    \caption{}
\end{figure} 
 \begin{figure}
    \centering
    \includegraphics[width=.75\textwidth]{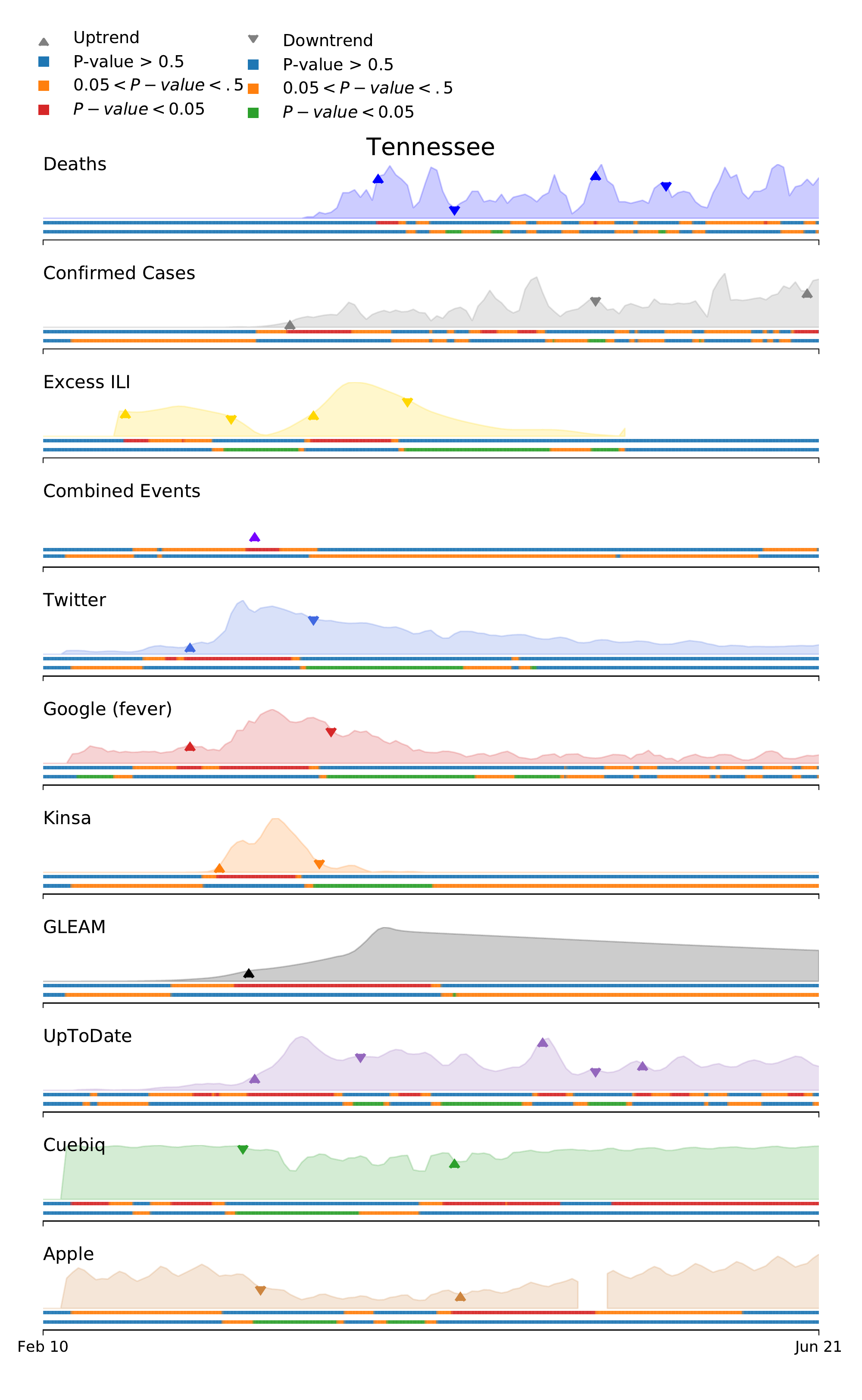}
    \label{fig:Tennessee_event}
    \caption{}
\end{figure} 
 \begin{figure}
    \centering
    \includegraphics[width=.75\textwidth]{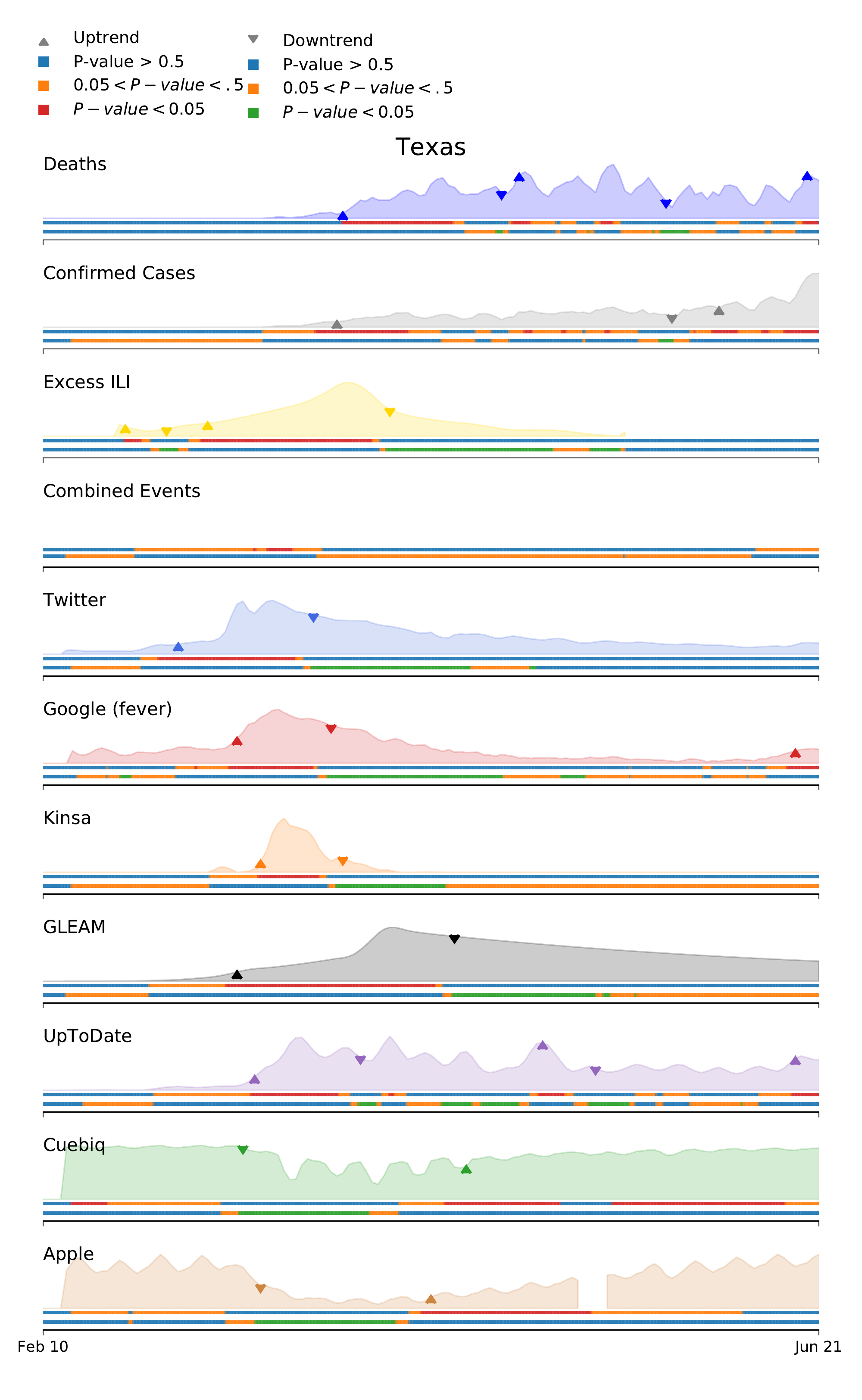}
    \label{fig:Texas_event}
    \caption{}
\end{figure} 
 \begin{figure}
    \centering
    \includegraphics[width=.75\textwidth]{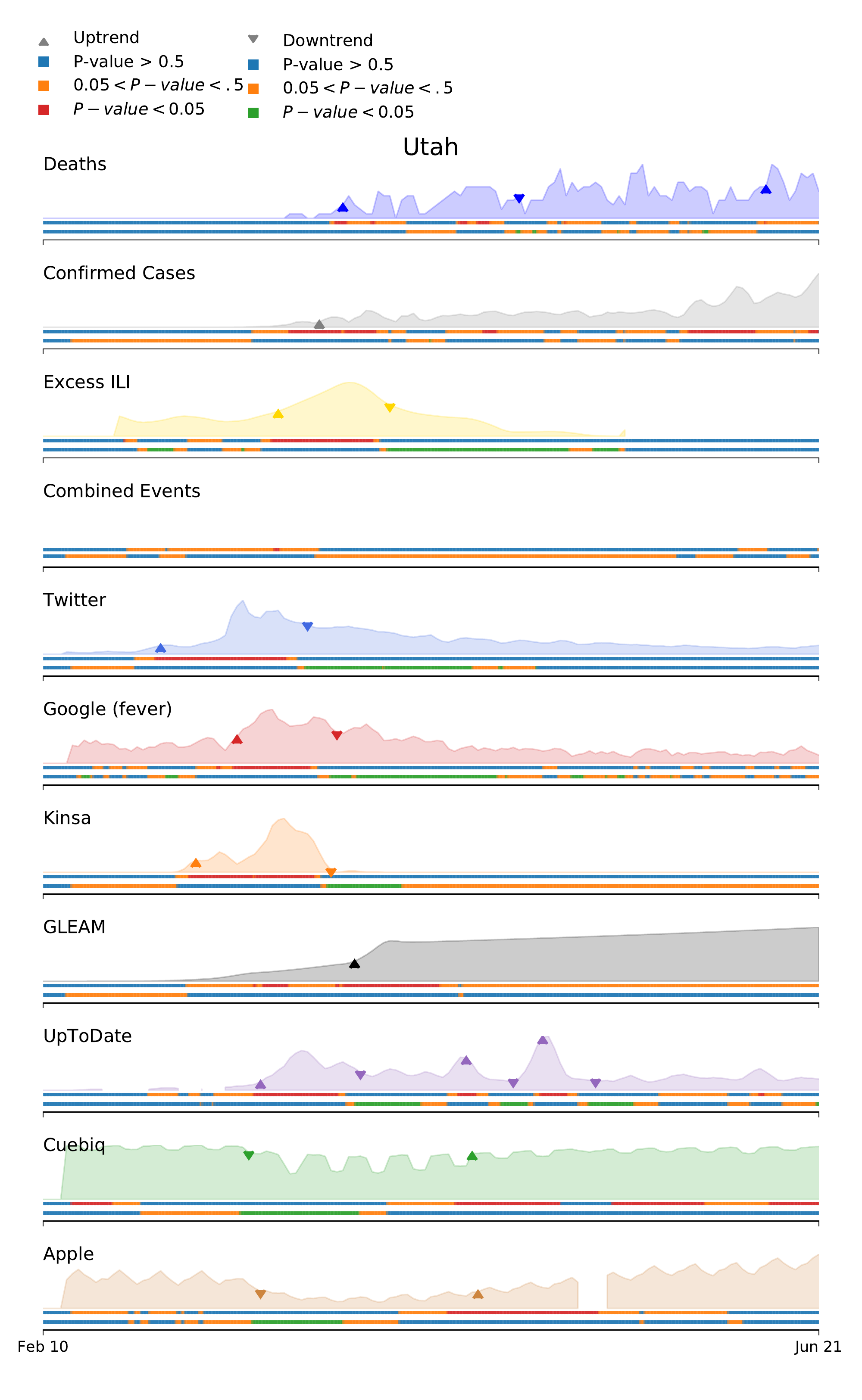}
    \label{fig:Utah_event}
    \caption{}
\end{figure} 
 \begin{figure}
    \centering
    \includegraphics[width=.75\textwidth]{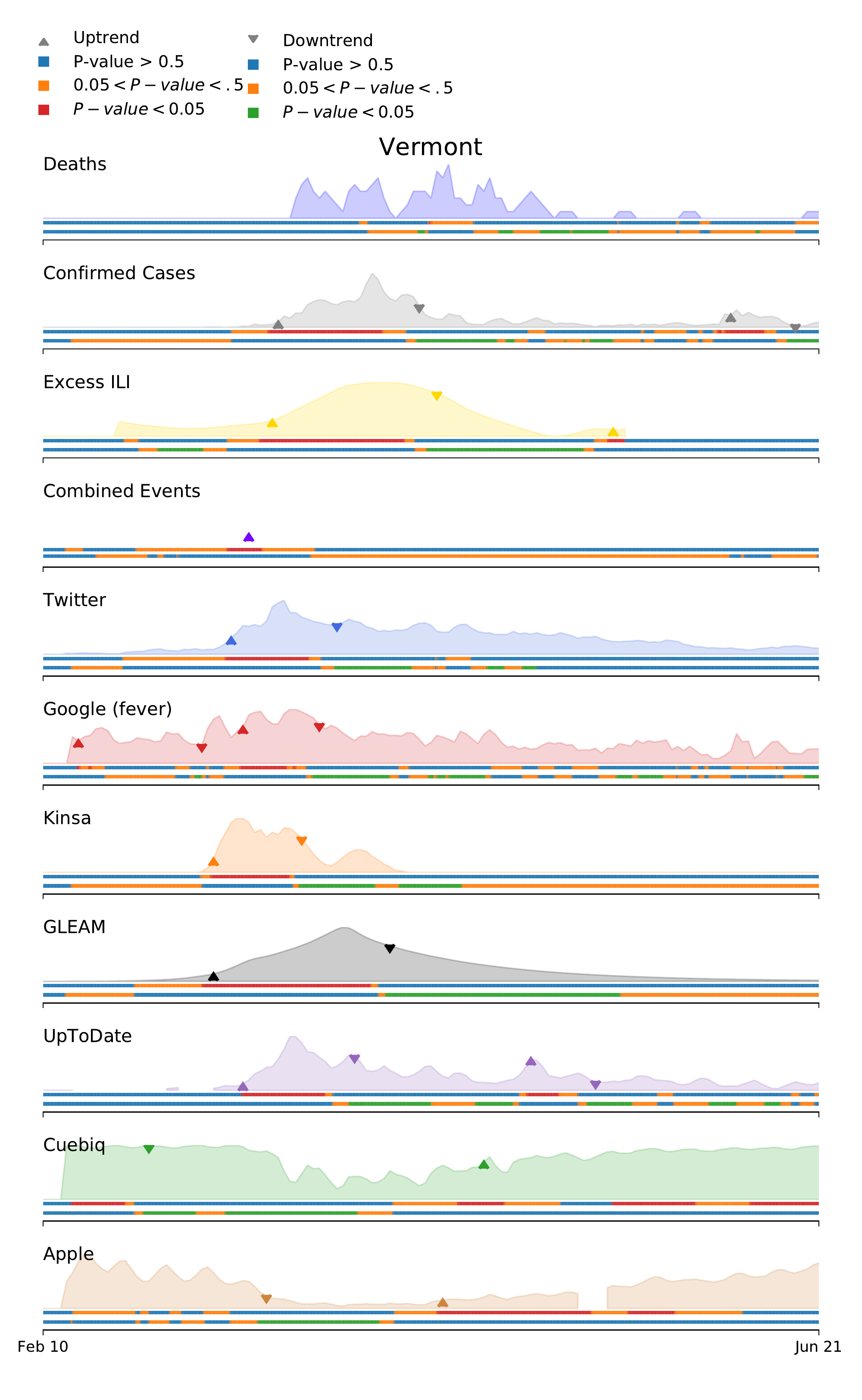}
    \label{fig:Vermont_event}
    \caption{}
\end{figure} 
 \begin{figure}
    \centering
    \includegraphics[width=.75\textwidth]{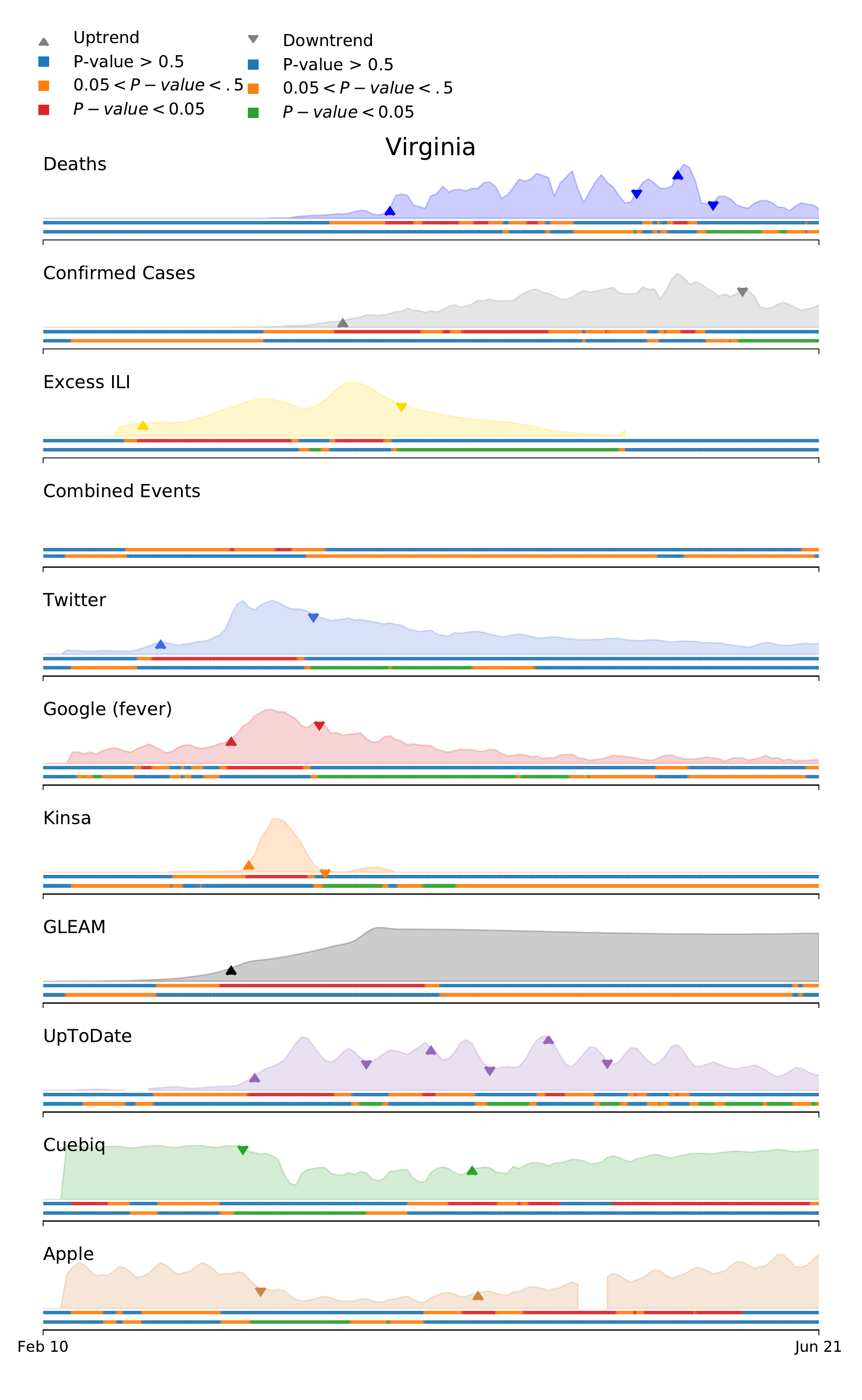}
    \label{fig:Virginia_event}
    \caption{}
\end{figure} 
 \begin{figure}
    \centering
    \includegraphics[width=.75\textwidth]{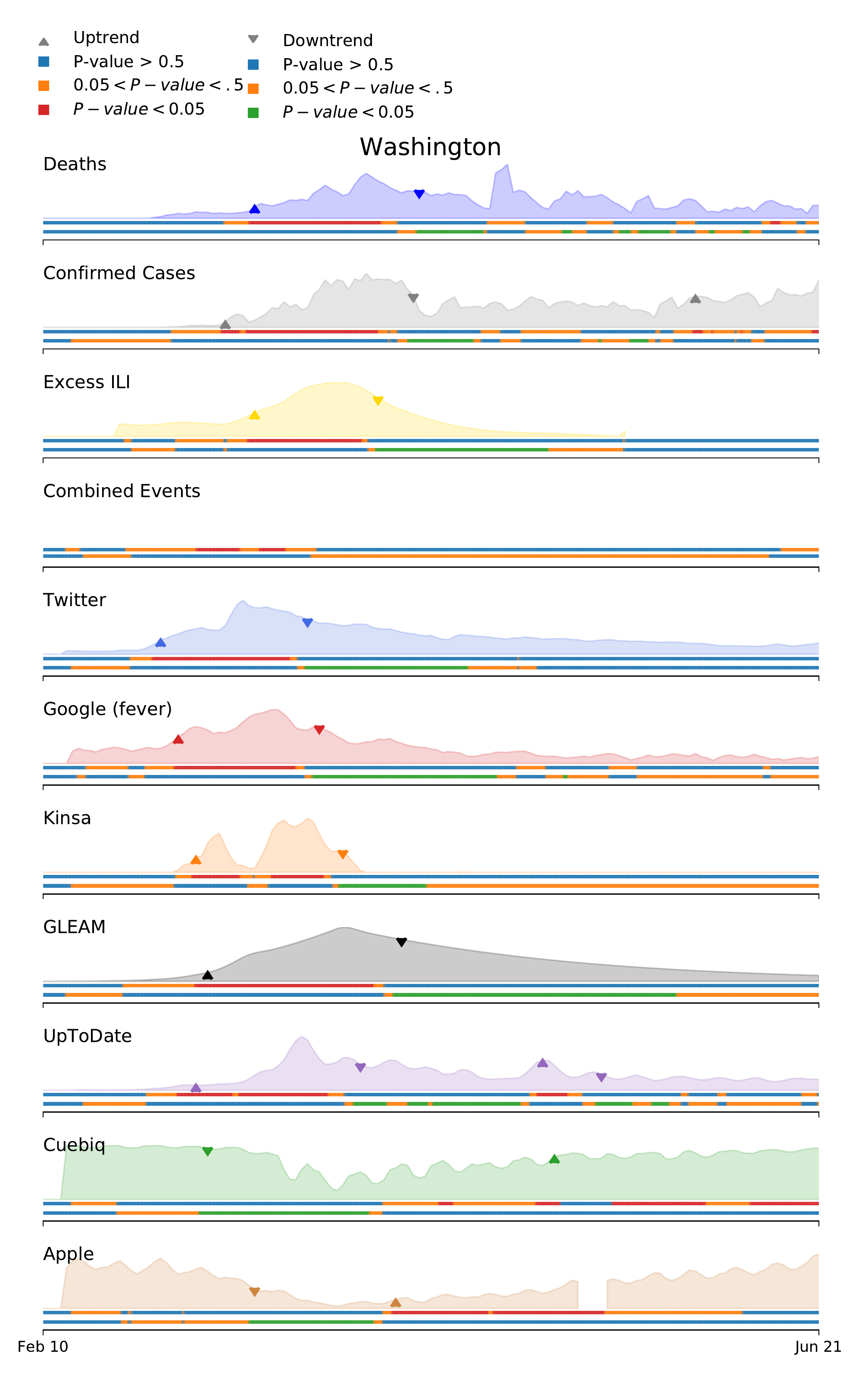}
    \label{fig:Washington_event}
    \caption{}
\end{figure} 
 \begin{figure}
    \centering
    \includegraphics[width=.75\textwidth]{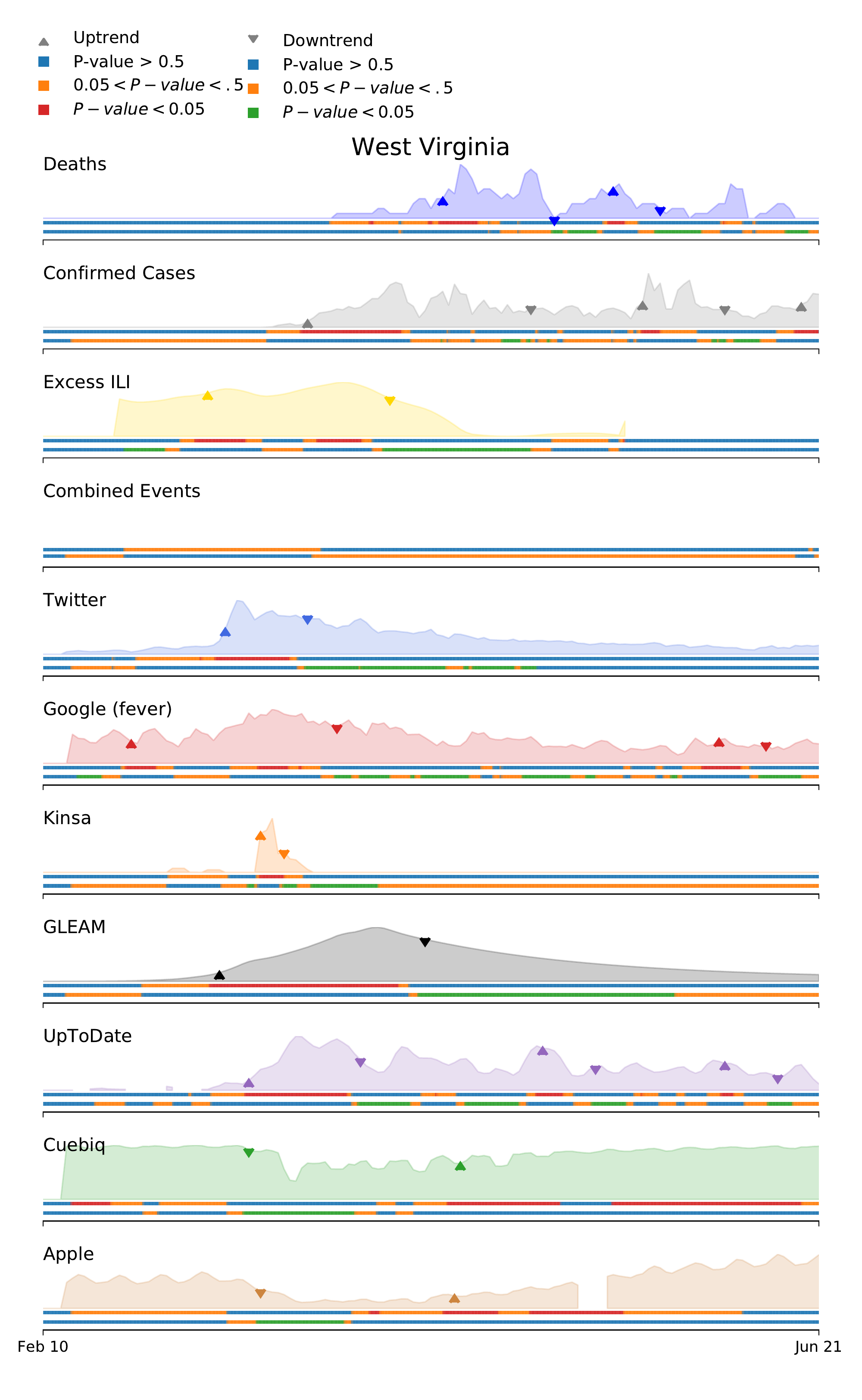}
    \label{fig:West Virginia_event}
    \caption{}
\end{figure}

 \begin{figure}
    \centering
    \includegraphics[width=.75\textwidth]{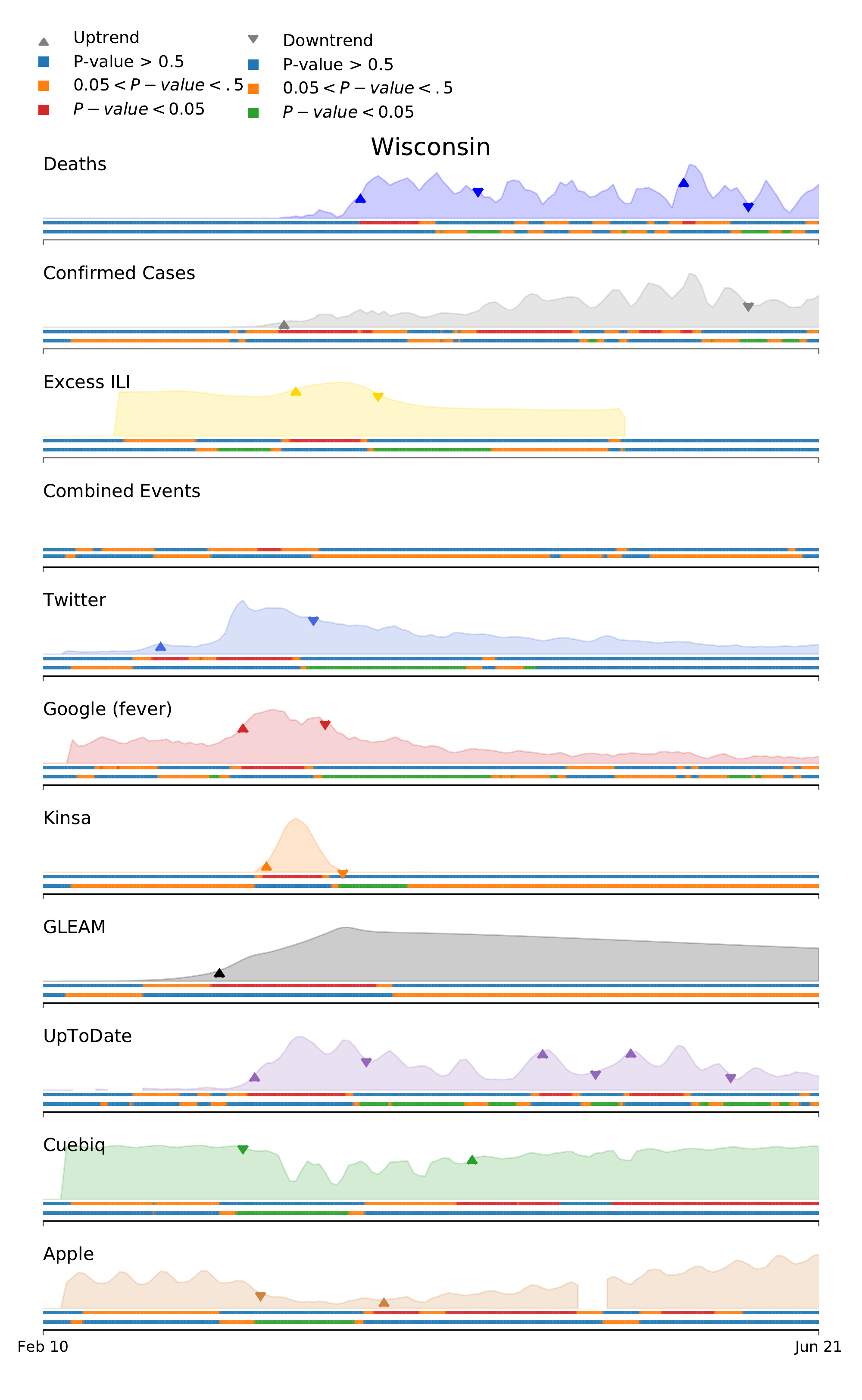}
    \label{fig:Wisconsin_event}
    \caption{}
\end{figure}

 \begin{figure}
    \centering
    \includegraphics[width=.75\textwidth]{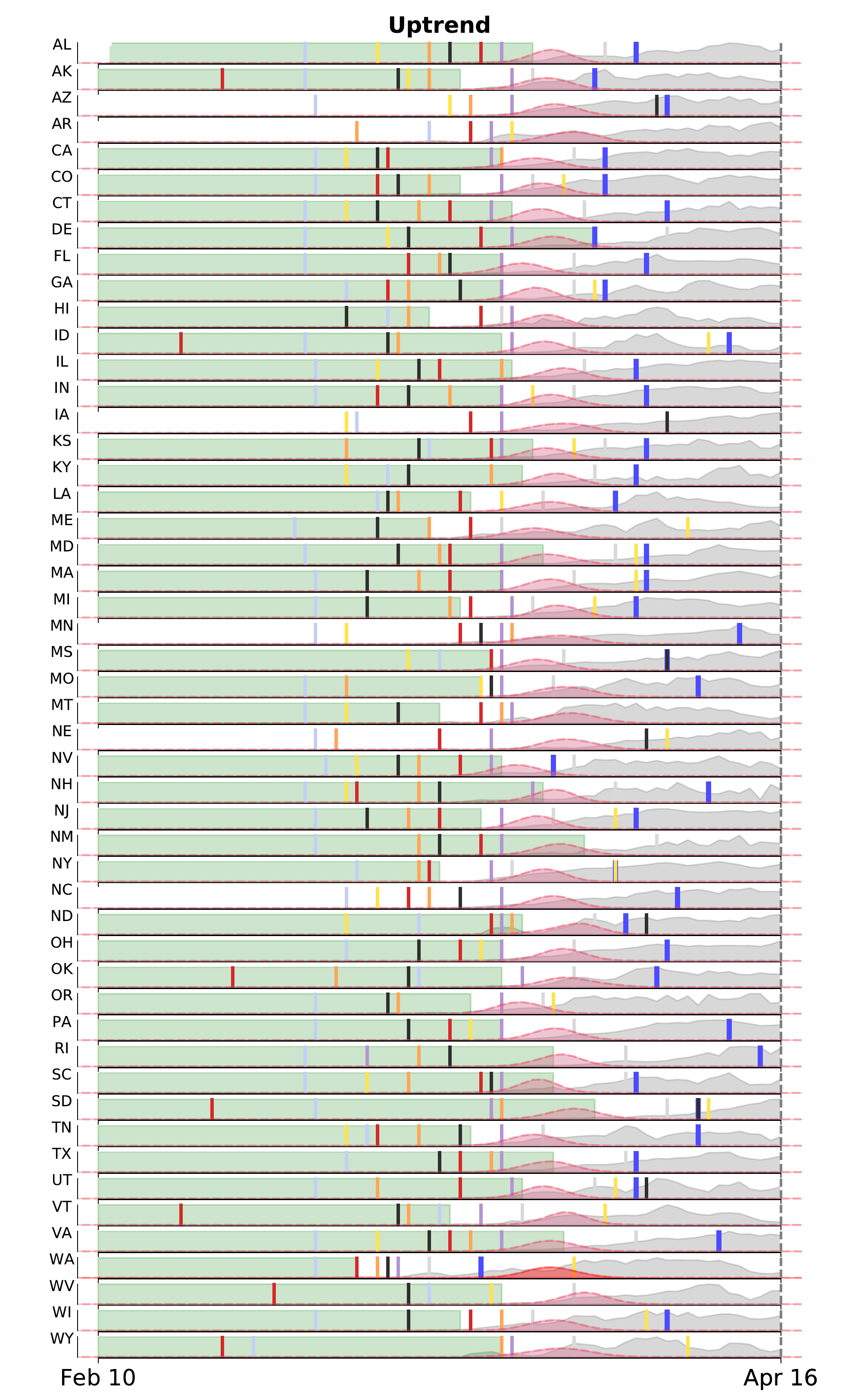}
    \label{fig:uptrend_probability}
    \caption{}
\end{figure} 

 \begin{figure}
    \centering
    \includegraphics[width=.75\textwidth]{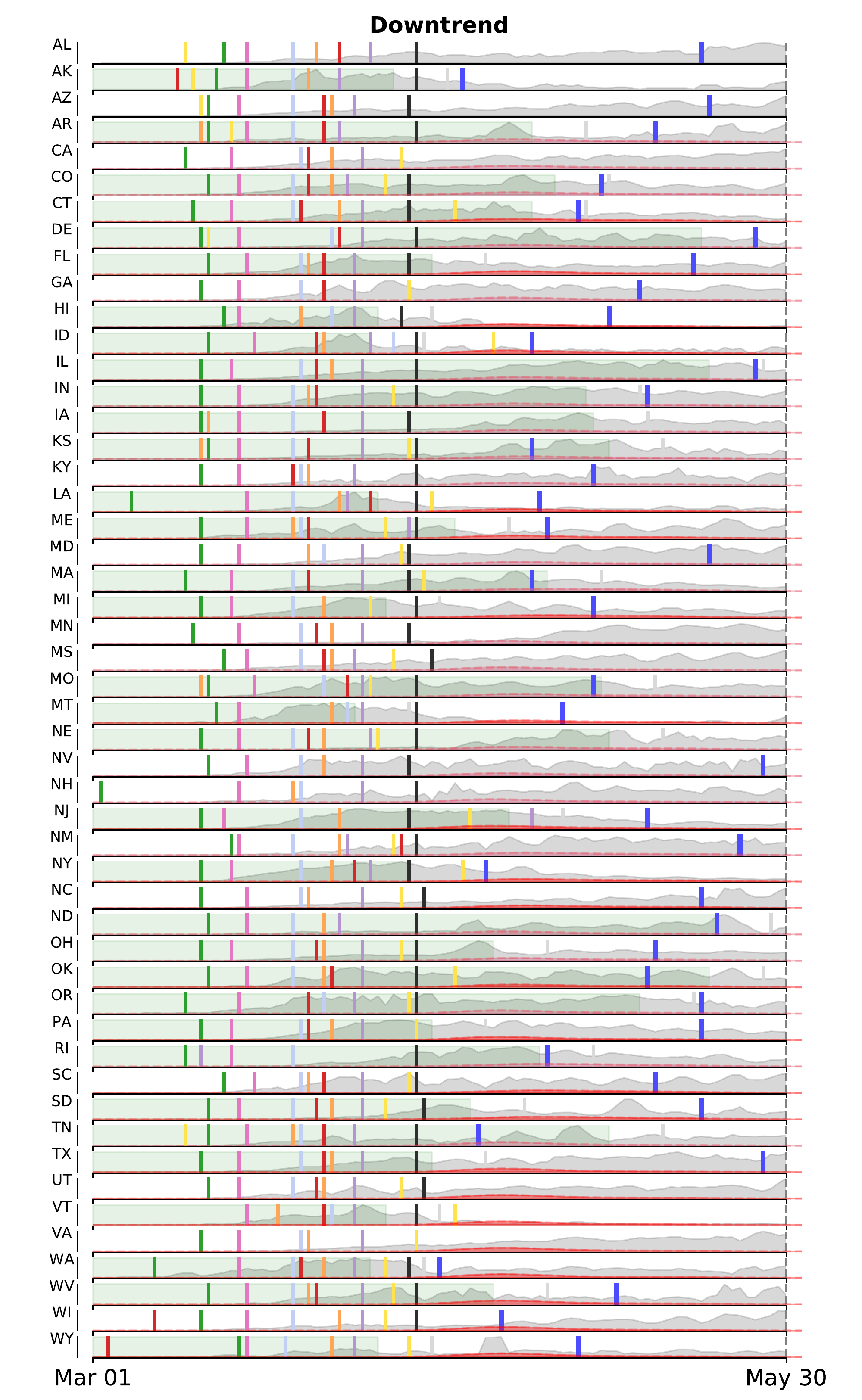}
    \label{fig:uptrend_probability}
    \caption{}
\end{figure} 


\FloatBarrier

\begin{figure}
    \centering
    \vspace*{-0.8in}
    \includegraphics[width=1\textwidth]{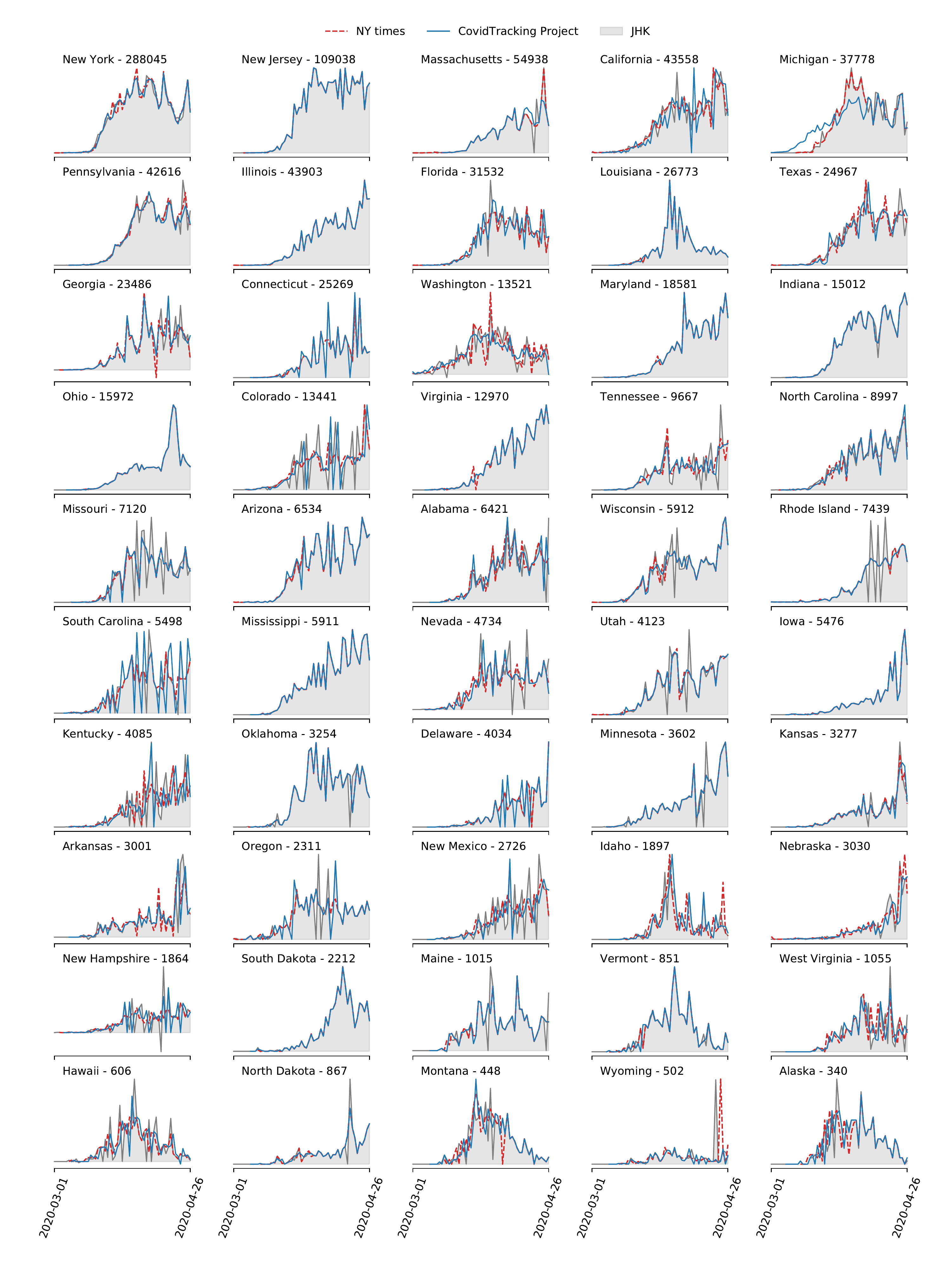}
    \caption{Visualization of the daily number of confirmed cases, as reported by the New York Times (red), the CovidTracking Project (blue), and the John Hopkins University COVID-19 Resource Center (gray). Most states follow a similar case trajectory, with exceptions arising from reporting inconsistencies between data sources.}
    \label{fig:confirmed_cases_comparison}
\end{figure}

\textbf{Comparing COVID-19 activity datasets}: We visualized COVID-19 confirmed case activity as a way to detect any inconsistencies between datasets prepared by  different organizations (New York Times, CovidTracking Project, and Johns Hopkins University). Figure S\ref{fig:confirmed_cases_comparison} shows a state-by-state plot of the number of confirmed cases (positive tests) for COVID-19, as collected from the New York Times (red), the CovidTracking Project (blue), and John Hopkins University COVID-19 resource center (gray). Overall, most of the time series follow a similar trend, except for Michigan (row 1), where the CovidTracking Project reported an increase in cases earlier than the other organizations. For most states, there also exist days for which an organization may have not reported any new confirmed cases, followed by a sudden increase in confirmed cases (see Kentucky). Additionally, there are instances where confirmed cases seems to decrease (see Georgia and New Hampshire) or where the same reported number of cases is delayed for a few days (see Wyoming).

\FloatBarrier

\textbf{Lagged Correlation Analysis}: We implemented a lagged correlation analysis as a retrospective counterpart to the event detection analysis described in the manuscript. This post-hoc method calls for calculating pairwise Pearson correlations of COVID-19 outcome time series (confirmed cases, attributed deaths, and Excess ILI) with COVID-19 proxy time series at different temporal offsets.\vspace{12pt}

\begin{figure}
    \centering
    \includegraphics[width=0.65\textwidth]{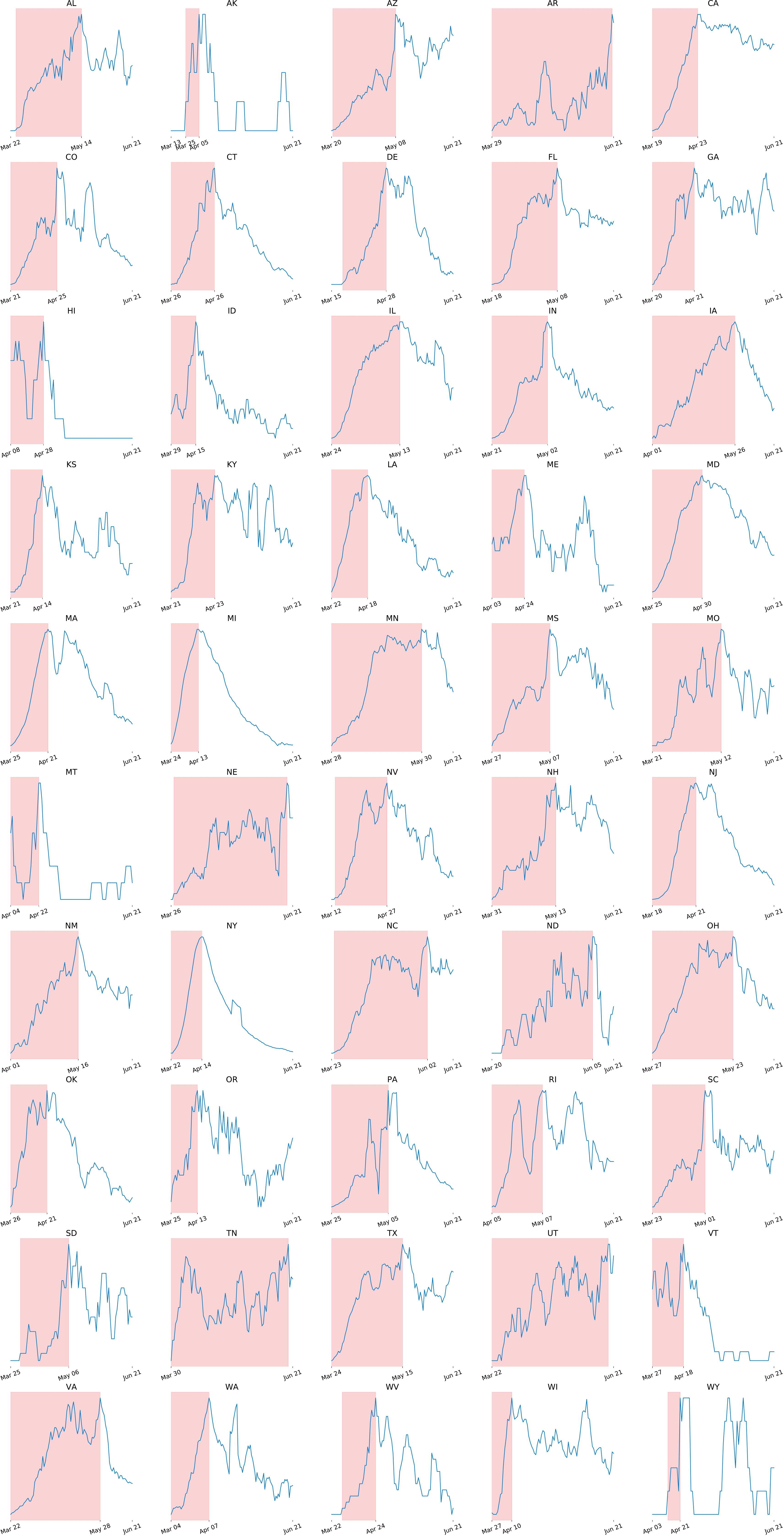}
    \caption{Uptrend region, in red, for deaths}
    \label{fig:supplementary_deaths_uptrend}
\end{figure}

\begin{figure}
    \centering
    \includegraphics[width=0.65\textwidth]{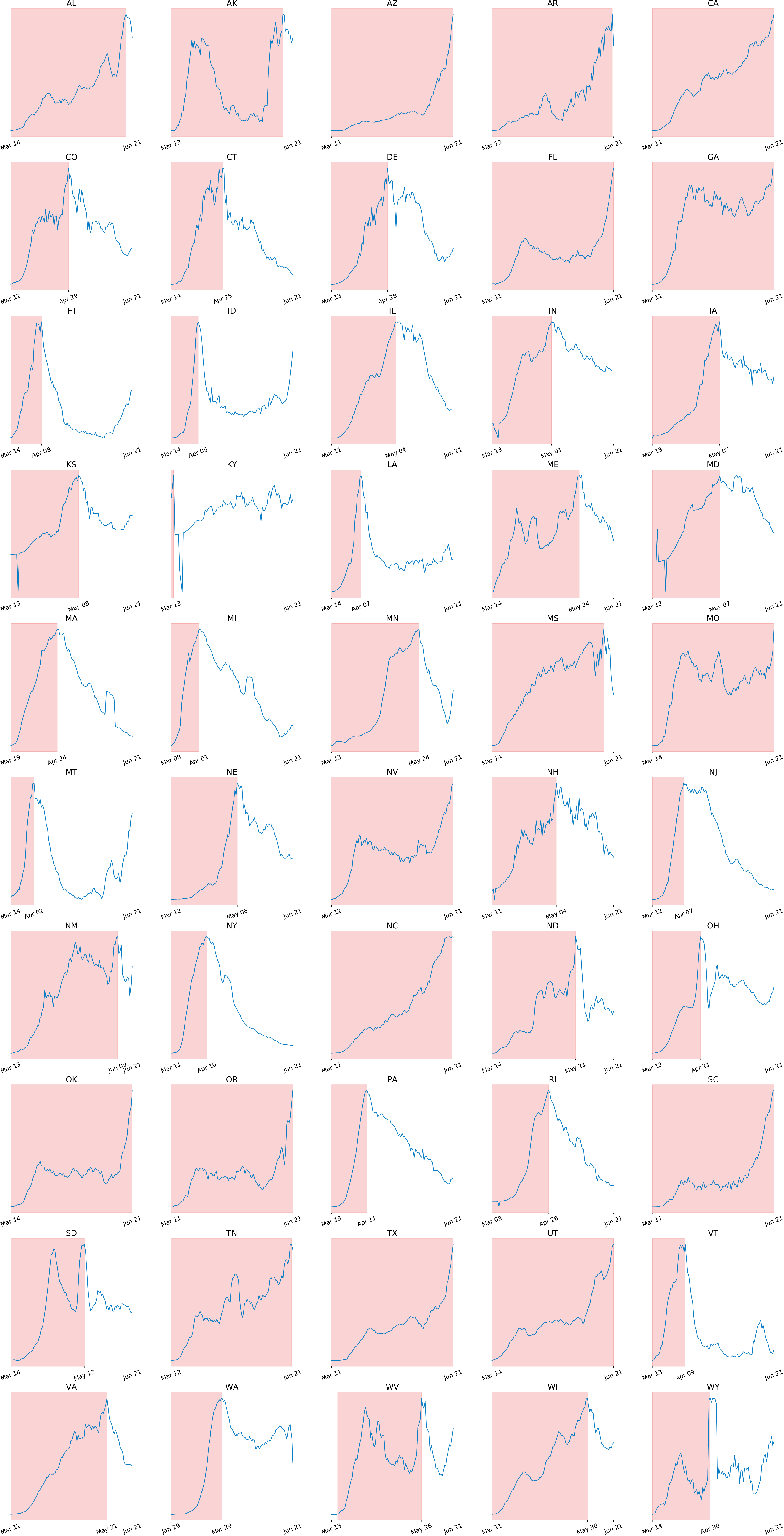}
    \caption{Uptrend region, in red, for cases}
    \label{fig:supplementary_cases_uptrend}
\end{figure}

\begin{figure}
    \centering
    \includegraphics[width=0.65\textwidth]{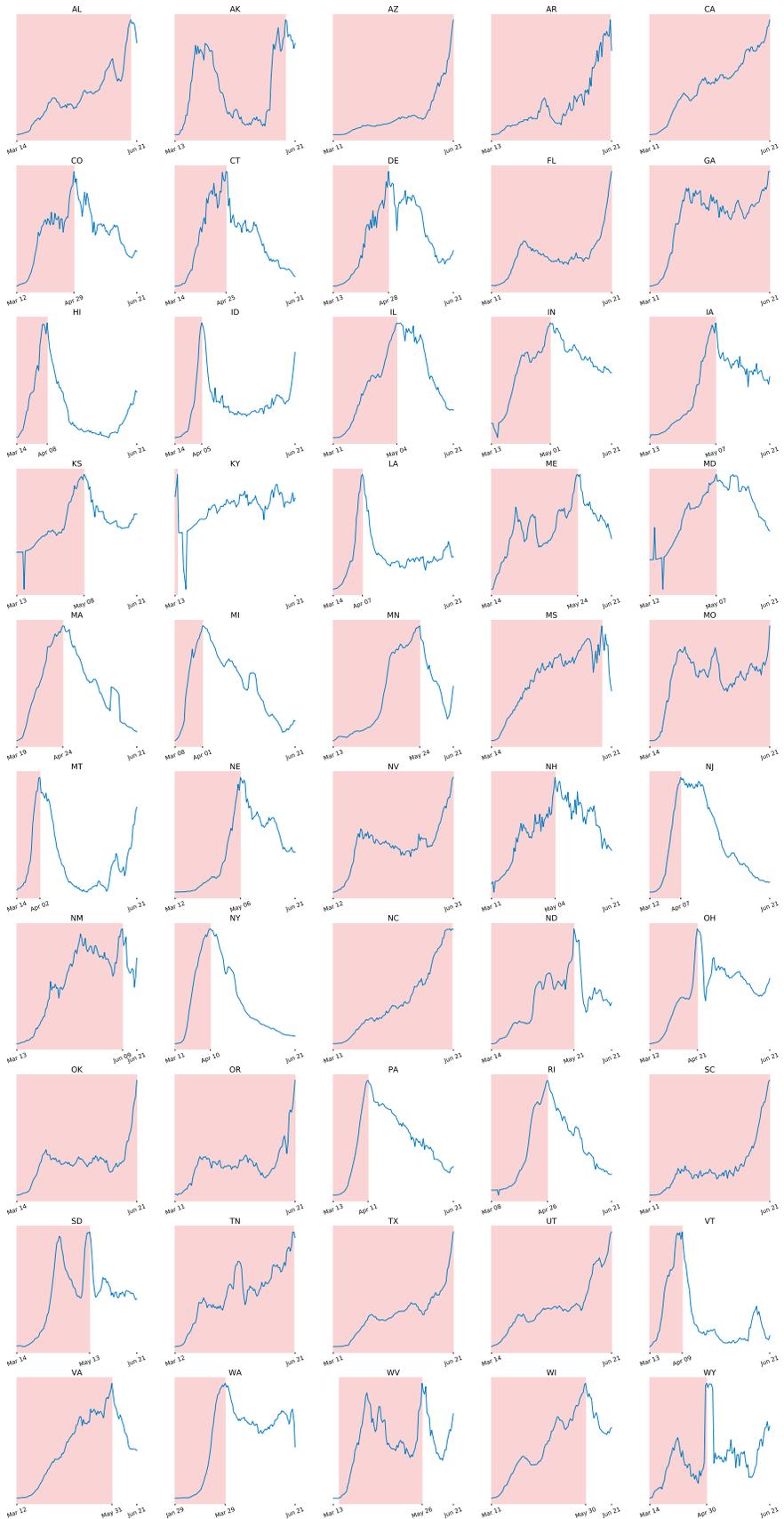}
    \caption{Uptrend region, in red, for ILI}
    \label{fig:supplementary_ili_uptrend}
\end{figure}

\begin{figure}
    \centering
    \includegraphics[width=0.65\textwidth]{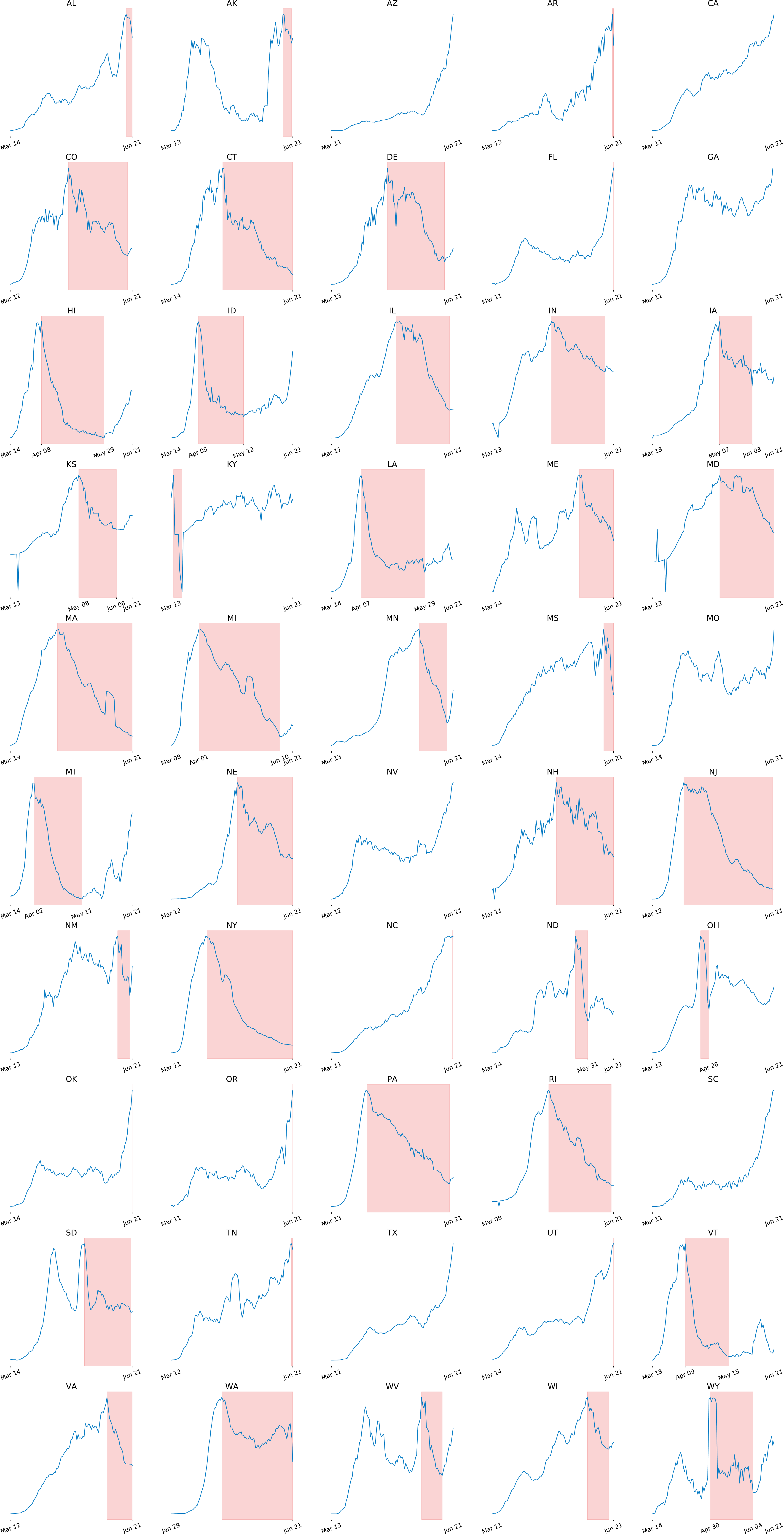}
    \caption{Downtrend region, in red, for cases}
    \label{fig:supplementary_cases_downtrend}
\end{figure}

\begin{figure}
    \centering
    \includegraphics[width=0.65\textwidth]{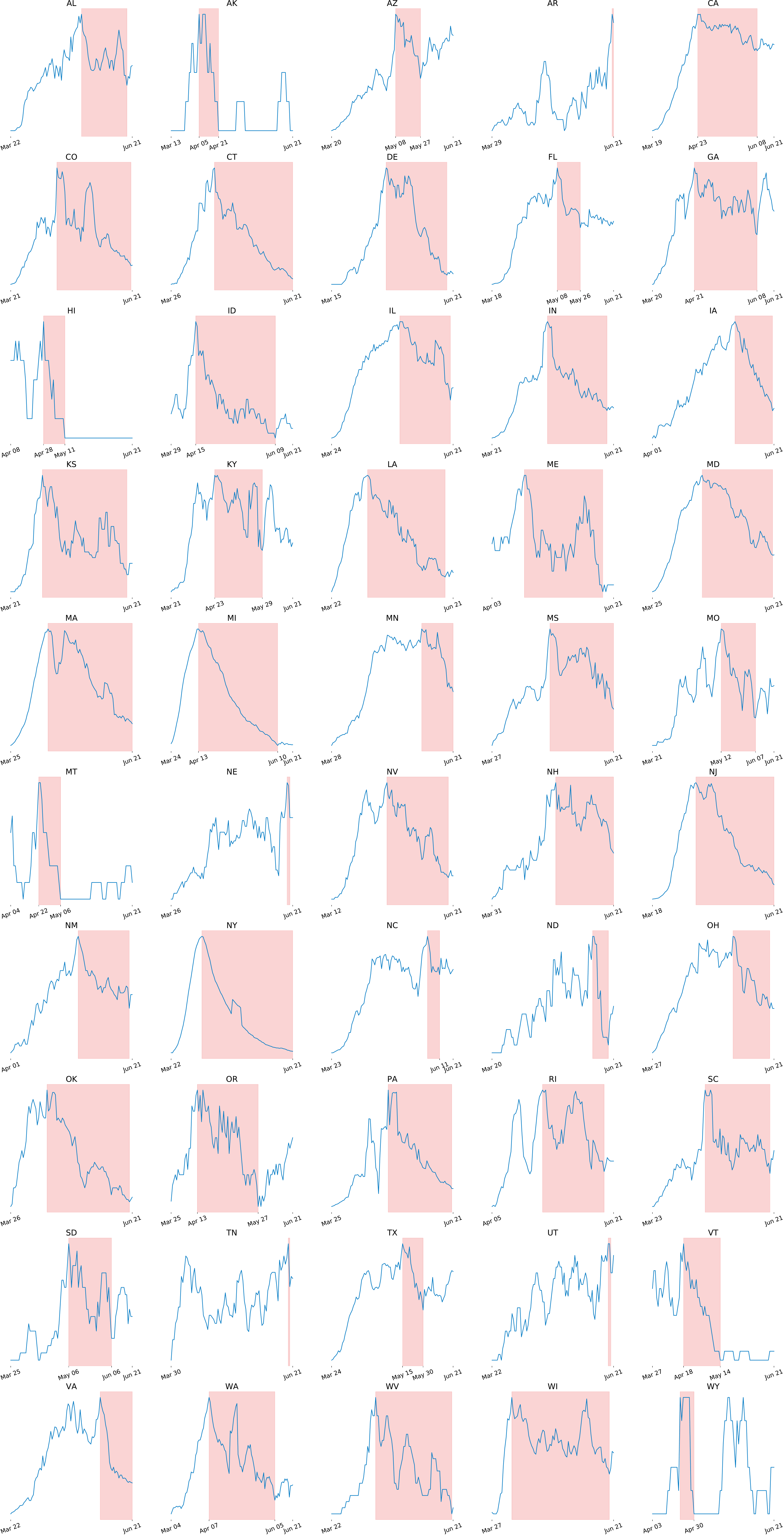}
    \caption{Downtrend region, in red, for deaths}
    \label{fig:supplementary_deaths_downtrend}
\end{figure}

\begin{figure}
    \centering
    \includegraphics[width=0.65\textwidth]{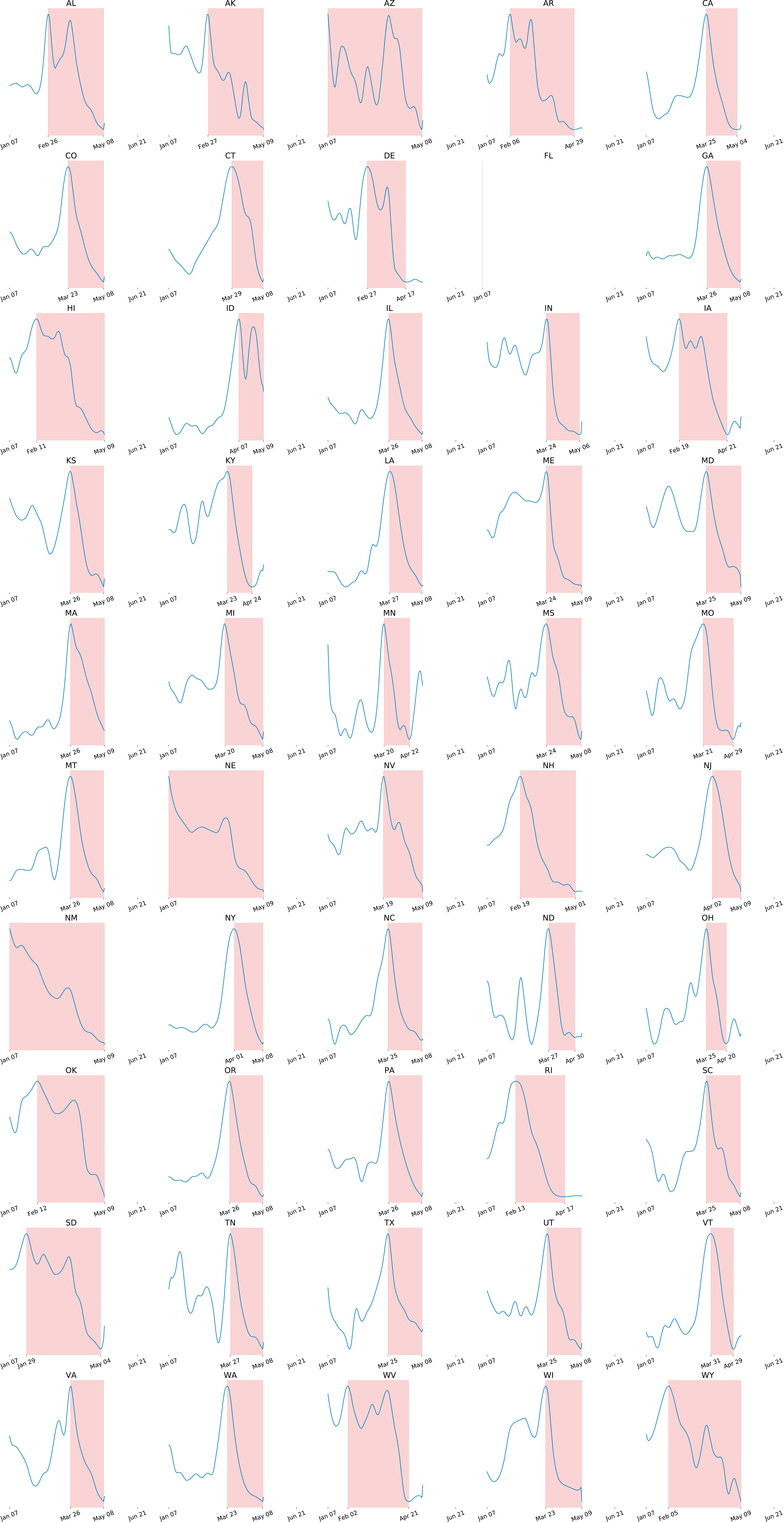}
    \caption{Downtrend region, in red, for ILI}
    \label{fig:supplementary_ili_downtrend}
\end{figure}

A 7-day trailing smooth, calculated as the mean over non-$nan$ values, was first applied to all variables. We next defined a time window for the analysis based off of the response variable's uptrend or downtrend. We chose the uptrend to span the last-occurring global minimum and the first-occurring global maximum (for Excess ILI data, which exhibit multiple uptrends and downtrends, we chose the uptrend to start at the local minimum temporally closest to the global maximum). The uptrend regions are shown in Figures \ref{fig:supplementary_cases_uptrend}, \ref{fig:supplementary_deaths_uptrend}, and \ref{fig:supplementary_ili_uptrend}. We repeated this approach for the downtrend phase, which we chose to span the last-occurring global maximum and the first-occurring global minimum following the downtrend start. See Figures \ref{fig:supplementary_cases_downtrend}, \ref{fig:supplementary_deaths_downtrend}, and \ref{fig:supplementary_ili_downtrend}, which show in red how the downtrend region for confirmed cases, deaths, and Excess ILI yielded by this approach.\vspace{12pt} 

\begin{figure}[t!]
    \centering
    \begin{subfigure}[t]{0.6\textwidth}
    \centering
        \hspace*{2.3cm}\includegraphics[width=\textwidth]{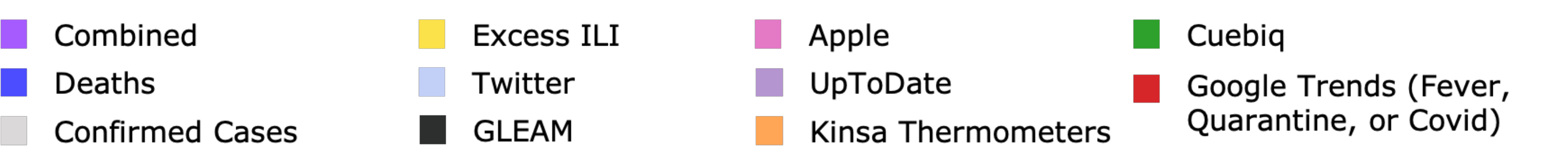}
        \label{fig:legend}
    \end{subfigure}
    \newline
    \begin{subfigure}[t]{1.0\textwidth}
        \includegraphics[width=\textwidth]{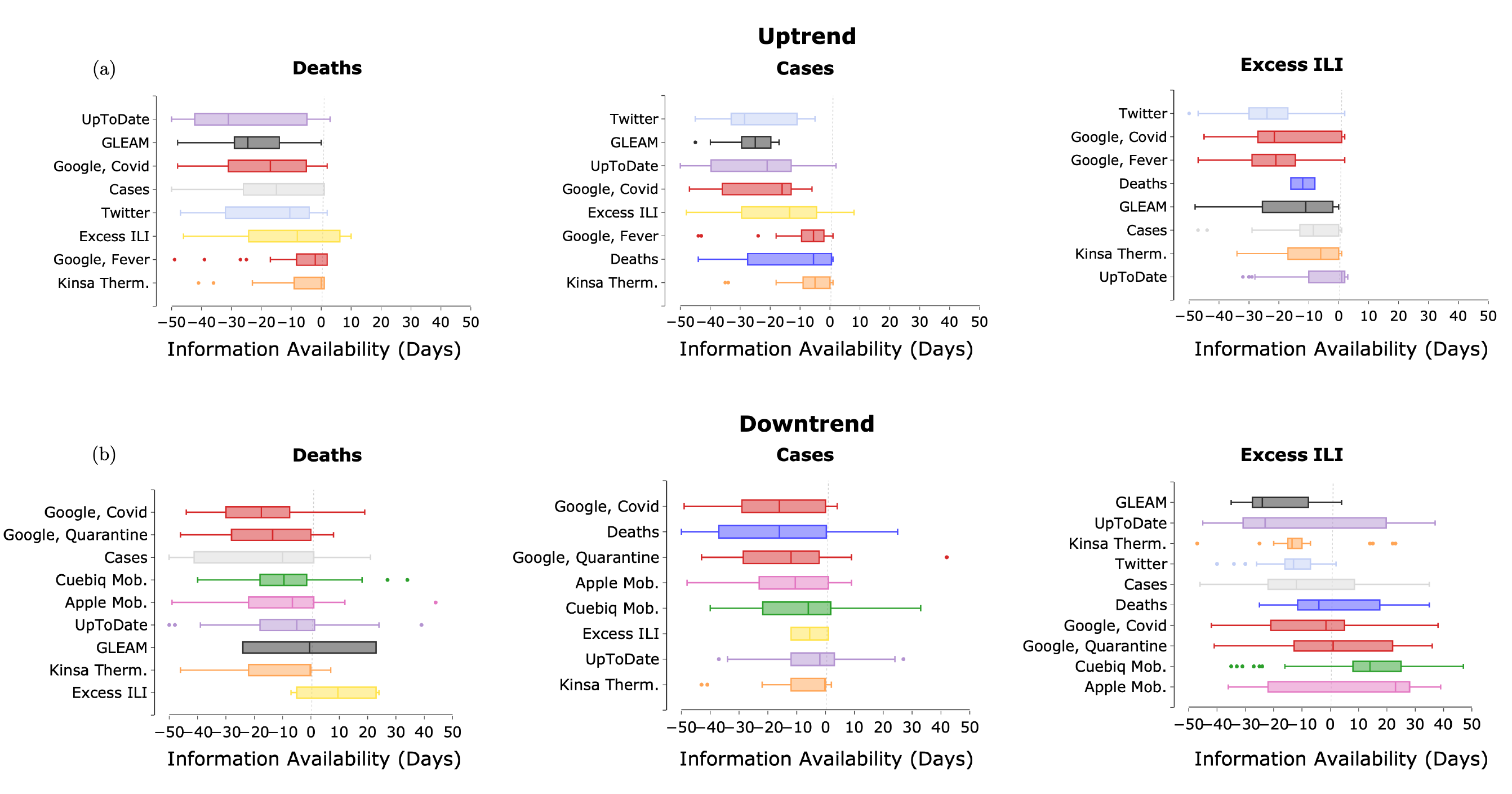}
    \end{subfigure} ~
    \caption{Lagged correlation results for pairwise comparisons between COVID-19 activity proxies and gold standards for US states with available data. (a) is illustrative of the difference in ``uptrends'' between input and response variables. (b) analogously illustrates differences in ``downtrends''. Deaths, cases, and Excess ILI are also included as input variables as a reasonableness check of COVID-19's trajectory. Note that the ``Combined'' time series is not relevant for this analysis. Violin plots cascade by median value, where each plot is shifted according to its deviation from real-time availability. Differences between input and response variables exceeding 50 days are omitted. Negative values indicate an input variable preceding the response variable.}
  \label{fig:pearson_box_plots}
\end{figure}

We then calculated the Pearson correlation coefficient - $r$ - between each outcome-proxy pairing over the uptrend or downtrend time window; this is the ``no lag'' correlation. Next, Pearson's $r$ corresponding to lag $l$ was calculated by shifting the proxy time series back or forward by $l$ days and regressing against the unchanged outcome time series. We calculated the regression for all lags in the range $l \in [0, \pm50]$, stopping when the proxy time series contained a $nan$ value (data unavailable). The optimal lag is, then, the positive or negative lag (in days) corresponding to whichever $r$ value had the greatest magnitude. The calculation of the optimal lag for each proxy-outcome pairing in all states with available data was used to produce Figure \ref{fig:pearson_box_plots}. As there is a delay in real-time reporting for the proxies - 10 days for Excess ILI, 3 days for UpToDate, 2 days for Google Trends and Twitter, and 1 day for all others (GLEAM has a 0 day delay) - each was transformed accordingly. Without these delays, there is presumed synchronicity between signal and response.\vspace{12pt}  

We found the results of the lagged correlation analysis to marginally align with those arising from the event detection analysis. Twitter and Google Searches were among the earliest uptrend signals (median earliness of 2-3 weeks across outcomes); UpToDate, unlike in the event detection analysis, emerged as an early signal of uptrend for deaths (4.5 weeks). Google Searches and some form of mobility data could be seen as early downtrend signals (median earliness of 2 weeks across outcomes).\vspace{12pt} 

For most proxies, the correspondence between analyses did not hold. We attributed this, as well as other inconsistencies and inaccuracies, to the sensitivity of the lagged correlation analysis to several parameters. One such parameter is the definition of uptrend and downtrend initiation. As is shown in \ref{fig:supplementary_deaths_downtrend}, it can be non-trivial to define the start of these trends (see RI, for example, which has multiple peaks and valleys). The lagged correlation analysis is more susceptible to ``false starts'' than the event detection analysis as it uses arbitrarily-selected time series extrema - not exponentiation - as proxies for growth and decay. These false starts can lead to artificial inflating or deflating of Pearson's $r$. Artificial significance of signals can also stem from the maximum lag cutoff (50 days, here). This is especially true for the downtrend, where permitting higher lags can inadvertently lead to correlation of a COVID-19 outcome downtrend with a COVID-19 proxy uptrend. Finally, the length of the uptrend and downtrend (time window) can profoundly affect the lagged correlation analysis' performance. Smaller time windows will allow for more possible lags at the expense of shorter uptrends or downtrends, and vice versa for larger time windows. This can lead to situations where a downtrend in deaths seemingly predates a downtrend in cases (see Figure \ref{fig:pearson_box_plots}b); this may also be attributable to colinearity in cases and deaths. Ultimately, these limitations make the lagged correlation analysis ineffective as an early warning approach.\vspace{12pt} 


